\documentclass[12pt]{article}
\usepackage{jheppub}
\usepackage{ytableau}
\usepackage{comment}
\usepackage[vcentermath]{youngtab}
\usepackage{tikz}

\newcommand\be{\begin{equation}}
\newcommand\ee{\end{equation}}

\newcommand\Tr{\mathrm{Tr}}

\preprint{
RUP-25-21
}

\title{Interface line operators in $\mathcal{N}=4$ SYM theories and supersymmetric indices}
\abstract{
We study configurations of two $\mathcal{N}=4$ super Yang-Mills theories of unitary gauge groups 
connected by the BPS interfaces involving line operators. 
We find strong evidence of S-duality of the configurations as precise matching of the line defect half-indices 
which enumerate the BPS local operators at the junctions of the interfaces and line operators. 
The interface line defect half-indices are expressible as intriguing closed-form formulae 
involving the $q$-binomial coefficients and the principal specializations of the Schur functions. 
}
\author[a]{Yasuyuki Hatsuda}
\author[b]{and Tadashi Okazaki}

\emailAdd{yhatsuda@rikkyo.ac.jp, tokazaki@seu.edu.cn}
\affiliation[a]{Department of Physics, Rikkyo University, Toshima, Tokyo 171-8501, Japan}
\affiliation[b]{
Shing-Tung Yau Center and School of Physics of Southeast University,\\
Yifu Architecture Building, No.2 Sipailou, Xuanwu district, Nanjing, Jiangsu, 210096, China}
\begin{document}
%%%%%%%%%%%%%%%%%%%%%%%%%%%%%%%%%%%%%%%%%%%%
%%%%%%%%%%%%%%%%%%%%%%%%%%%%%%%%%%%%%%%%%%%%
\maketitle

%%%%%%%%%%%%%%%%%%%%%%%%%%%%%%%%%%%%%%%%%%%%%%%%%
%%%%%%%%%%%%%%%%%%%%%%%%%%%%%%%%%%%%%%%%%%%%%%%%%
\section{Introduction and summary}
\label{sec_intro}
%%%%%%%%%%%%%%%%%%%%%%%%%%%%%%%%%%%%%%%%%%%%%%%%%
%%%%%%%%%%%%%%%%%%%%%%%%%%%%%%%%%%%%%%%%%%%%%%%%%

%previous work
In the previous work \cite{Hatsuda:2025yzp} we investigated the line defect half-indices which  
encode the BPS spectra of the configurations involving the line operators in the presence of the basic half-BPS boundary conditions 
in 4d $\mathcal{N}=4$ super Yang-Mills (SYM) theories. 
In the presence of the Neumann boundary condition \cite{Gaiotto:2008sa}, the half-indices can be generalized by introducing the electric Wilson lines 
transforming in certain representations of the preserved gauge group on the boundary. 
On the other hand, the S-dual configurations involve the Nahm pole boundary conditions \cite{Nahm:1979yw,Diaconescu:1996rk,Gaiotto:2008sa}, 
for which one can introduce the magnetic 't Hooft lines \cite{tHooft:1977nqb,Kapustin:2005py,Kapustin:2006pk,Gaiotto:2011nm,Witten:2011zz}. 
We demonstrated that dual pairs of the line defect half-indices match as powerful tests of S-duality of the boundary line operators. 
In addition, we found that the matrix integrals for the line defect half-indices of the Neumann boundary conditions involving the Wilson lines 
are intimately tied to the Macdonald polynomials \cite{MR1354144,MR1976581,MR1314036,MR1354956} 
which show up in a framework of the double affine Hecke algebra (DAHA) \cite{MR1185831,MR1314036,MR1358032,MR1354956,MR1613515,MR1768938,MR1715325,MR1792347,MR1411136}. 

%This paper
The aim of this work is to generalize the results to the half-BPS interfaces involving the line operators for $\mathcal{N}=4$ SYM theories. 
The basic interfaces can be constructed by introducing an NS5-brane or D5-brane \cite{Gaiotto:2008ak} in the Hanany-Witten setup of Type IIB string theory \cite{Hanany:1996ie} 
in such a way that two stacks of D3-branes terminate on the $5$-brane from both sides. 
A remarkable feature of the interfaces is that they contain non-trivial 3d degrees of freedom which are coupled to the 4d gauge theories, 
as opposed to the basic half-BPS boundary conditions. 
The NS5-type interface and D5-type interface are expected to be S-dual. 
In fact, the interface half-indices agree with each other \cite{Gaiotto:2019jvo}. 
Here we further introduce the line operators at the interfaces. 
The electric Wilson lines transforming in the fundamental, symmetric and antisymmetric representations in 4d $\mathcal{N}=4$ SYM theories 
can be realized by introducing a fundamental string \cite{Maldacena:1998im,Rey:1998ik}, 
a D3-brane \cite{Drukker:2005kx,Gomis:2006im,Gomis:2006sb,Rodriguez-Gomez:2006fmx,Yamaguchi:2007ps} 
and a D5-brane \cite{Yamaguchi:2006tq,Gomis:2006sb,Rodriguez-Gomez:2006fmx,Hartnoll:2006hr} respectively. 
In the presence of the interface the Wilson lines can form the gauge invariants with the 3d and 4d degrees of freedom 
and they can be analogously constructed in Type IIB string theory as discussed for 3d $\mathcal{N}=4$ gauge theories \cite{Assel:2015oxa}. 
When we introduce the magnetic lines at the interface, they require the 3d matter fields to have singularities as well as the bulk fields. 
They result in the shifted spins of the 3d charged operators just like an effect of the vortex lines. 
In other words, the magnetic lines at the interfaces will intrinsically behave as the vortex lines in the 3d gauge theories. 
In fact, we confirm that various Wilson line defect half-indices for the interfaces agree with 
what can be obtained from the half-indices of the dual interfaces by applying the prescription for the 3d supersymmetric indices 
in the presence of the vortex lines \cite{Drukker:2012sr} (also see \cite{Okazaki:2019ony,Hayashi:2025guk}). 
The matching pairs have been checked by comparing the terms with high powers of $q$ in the expansions of the half-indices. 
Compared with the boundary line operators, 
i.e. the one-sided half-space configurations of $\mathcal{N}=4$ SYM theories obeying the basic BPS boundary conditions and involving the line operators, 
there are further distinct properties of the line operators at the interfaces
\begin{itemize}
\item 
Both electric and magnetic lines can be non-trivial for each type of the interfaces 
because some gauge group is left unbroken. 

\item 
Line operators can be non-trivial even for $U(1)$ gauge theories since the interfaces contain the 3d charged matter fields. 
\end{itemize}

There are several remarks on the computations of the line defect half-indices for the interfaces. 
As discussed in \cite{Hatsuda:2025yzp}, the line defect half-indices of the basic half-BPS Neumann boundary conditions with the Wilson lines 
are given by the matrix integral involving the weight functions for the inner products of the Macdonald functions \cite{MR1354144,MR1976581,MR1314036,MR1354956}. 
In the half-BPS limit they reduce to the integral involving the measure of the Hall-Littlewood functions 
so that the line defect correlators can be expressed in terms of the Hall-Littlewood functions and the Kostka-Foulkes polynomials. 

%NS5-type interface Wilson and super-Macdonald, HL
Analogously, the line defect half-indices of the NS5-type interface are given by the matrix integrals which contain 
the generalized forms of the weight functions for the Macdonald functions. 
We present a combinatorial sum formula for the line defect half-indices of the antisymmetric Wilson lines. 
In the Higgs limit, the line defect half-indices of the NS5-type interface reduce to the forms which generalize the weight functions for the Hall-Littlewood functions. 
We find closed-form expressions for the Higgs limits of the line defect half-indices of the antisymmetric and symmetric Wilson lines 
in terms of the $q$-binomial coefficients. 

%D5-type interface Wilson and specialization of Schur
On the other hand, for the D5-type interface with the Wilson line we conjecture the closed-form expressions for the line defect half-indices 
as the infinite series that generalizes the ``vortex expansions'' \cite{Gaiotto:2019jvo,Hatsuda:2024uwt,Hatsuda:2025mvj}. 
In particular, for the Wilson line transforming as the irreducible representation, 
the normalized line defect half-indices in the Higgs limit are given by the principal specializations of the Schur functions. 
The results exhibit an explicit one-to-one correspondence between the BPS operators at the junction of the D5-type interface and the Wilson line and the column-strict plane partitions. 

%%%%%%%%%%%%%%%%%%%%%%%%%%%%%%%%%%%%%%%%%%%%%%%%%
\subsection{Structure}
%%%%%%%%%%%%%%%%%%%%%%%%%%%%%%%%%%%%%%%%%%%%%%%%%
In section \ref{sec_InterfaceLine} we review the BPS configuration consisting of line operators at the interfaces in $\mathcal{N}=4$ super Yang Mills theories of unitary gauge groups. 
They can be realized as brane configurations in Type IIB string theory. 
The BPS spectrum of the configuration is encoded by the line defect half-indices. 
In section \ref{sec_u1u1} we analyze the $U(1)|U(1)$ interface-lines. 
The line defect half-indices are exactly computed by picking up residues at towers of poles. 
In section \ref{sec_uNu1} we study the $U(N)|U(1)$ interface-lines with $N>1$. 
In this case the proposed dualities of interface-lines involve the Wilson lines in the symmetric representation of $U(N)$ associaed with the non-minuscule module.  
We also present the exact expressions for the line defect half-indices for the D5-type interface by summing over the residues at simple poles. 
In section \ref{sec_uNuN} the $U(N)|U(N)$ interface-lines with $N>1$ are examined. 
We find strong evidence of the dualities of the interface-lines associated with the minuscule representation 
by checking matching pairs of the line defect half-indices. 
In section \ref{sec_uNuM} we examine the $U(N)|U(M)$ interface-lines with unequal ranks $N>M>1$. 
We find matching pairs of the one-point functions of the interface-lines 
and propose the generalized vortex expansions for the one-point function of the D5-type interface Wilson line in arbitrary representation. 
The dualities of the interface-lines associated with the minuscule representations are generalized. 
Also we provide several closed-form expressions for the line defect half-indices the for the $U(N)|U(M)$ interface-lines 
in terms of the principal specializations of the Schur functions. 
In Appendix \ref{app_expansion} we show the $q$-series expansions of the line defect half-indices. 

%%%%%%%%%%%%%%%%%%%%%%%%%%%%%%%%%%%%%%%%%%%%%%%%%
\subsection{Future works}
%%%%%%%%%%%%%%%%%%%%%%%%%%%%%%%%%%%%%%%%%%%%%%%%%
There are several open questions which we hope to report in future work. 
\begin{itemize}

\item In this work we focus on the line operators at the interfaces between unitary gauge groups. 
The configurations can be generalized as the line operators at the interfaces of type BCD between orthogonal and symplectic gauge groups 
(see \cite{Witten:1998xy,Hanany:2000fq,Feng:2000eq,Evans:1997hk,Gaiotto:2008ak} for the relevant brane constructions with orientifolds). 
As different choices of line operators correspond to distinct physical theories \cite{Aharony:2013hda}, 
an introduction of the line operators at the interface will generalize the half-indices for the interface of BCD type previously studied in \cite{Hatsuda:2024lcc} 
as well as the line defect indices \cite{Hatsuda:2025jze} for the gauge theories of BCD type. 

\item In the absence of the line defects the interface half-indices can be expressed as the infinite series, the vortex expansions \cite{Gaiotto:2019jvo,Hatsuda:2024uwt,Hatsuda:2025mvj}. 
While we find the generalized vortex expansions for the D5-type interfaces with the Wilson lines, 
it would be nice to give proof with some extra work. 
Also it would be intriguing to explore similar formulae for the NS5-type interfaces with the Wilson lines. 

\item Several aspects of the holographic dual descriptions of the line operators at the interfaces or boundaries for $\mathcal{N}=4$ SYM theories have been investigated  
e.g. in \cite{Nagasaki:2011ue,Estes:2012nx,Nagasaki:2013hwa,Karch:2022rvr,Kristjansen:2024dnm}. 
We expect that our results are useful to extract the spectra of the gravity duals 
(see e.g. \cite{Gang:2012yr,Drukker:2015spa,Hatsuda:2023iwi,Hatsuda:2023iof,Hatsuda:2025jze} 
for the relevant analyses of the line defect indices). 

\item The giant graviton-like expansions of the line defect indices and half-indices have been studied e.g. in  \cite{Imamura:2024lkw,Imamura:2024pgp,Beccaria:2024oif,Hatsuda:2024uwt,Imamura:2024zvw,Hatsuda:2024lcc,Imamura:2025fqa}. 
In particular, our formulae for the line defect half-indices of the D5-type interfaces with the Wilson lines 
will be useful to examine the giant graviton-like expansions of the line defect half-indices 
as they can allow for higher order expansion coefficients. 

\end{itemize}

%%%%%%%%%%%%%%%%%%%%%%%%%%%%%%%%%%%%%%%%
%%%%%%%%%%%%%%%%%%%%%%%%%%%%%%%%%%%%%%%%
\section{Interface line operators}
\label{sec_InterfaceLine}
%%%%%%%%%%%%%%%%%%%%%%%%%%%%%%%%%%%%%%%%
%%%%%%%%%%%%%%%%%%%%%%%%%%%%%%%%%%%%%%%%

%%%%%%%%%%%%%%%%%%%%%%%%%%%%%%%%%%%%%%%%
\subsection{Brane construction}
%%%%%%%%%%%%%%%%%%%%%%%%%%%%%%%%%%%%%%%%
We consider the supersymmetric configurations 
consisting of line operators at the interfaces in $\mathcal{N}=4$ SYM theories of unitary gauge groups 
which can be realized as brane configurations in Type IIB string theory. 
Let us briefly review the brane setup. 

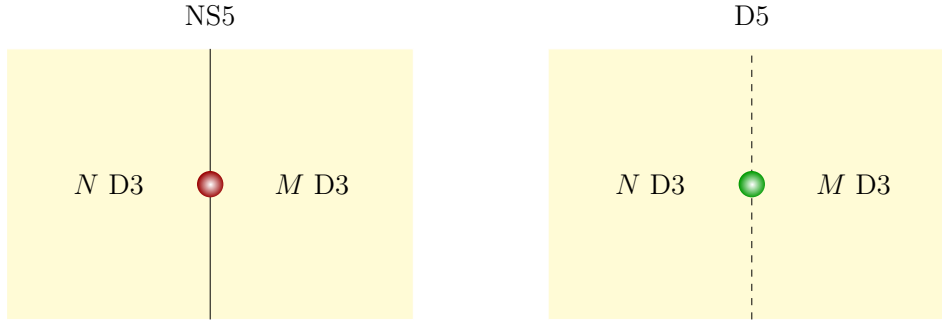
\begin{figure}
\usetikzlibrary{shapes}
\centering
\scalebox{0.9}{
\begin{tikzpicture}
\filldraw [fill=yellow!20!white,draw=white] (-9,0) -- (-3,0) -- (-3,4) -- (-9,4);
\draw (-6,0) -- (-6,4); 
\node at (-7.5,2) {$N$ D3};
\node at (-4.5,2) {$M$ D3};
\node[circle,outer color=red!60!black,inner color=white,minimum width=0.25cm] (radial) at (-6,2) {};
\node at (-6,4.5) {NS5};
\filldraw [fill=yellow!20!white,draw=white] (-1,0) -- (5,0) -- (5,4) -- (-1,4);
\draw (2,0)[dashed] -- (2,4); 
\node at (0.5,2) {$N$ D3};
\node at (3.5,2) {$M$ D3};
\node[circle,outer color=green!60!black,inner color=white,minimum width=0.25cm] (radial) at (2,2) {};
\node at (2,4.5) {D5}; 
\end{tikzpicture}
}
\caption{The NS5-type interface in 4d $U(N)\times U(M)$ bulk theory (left). 
The D5-type interface (right). The horizontal and vertical directions are the directions $6$ and $2$ respectively. 
The line operators are drawn as circles sitting at $x^6=0$, the location of the interfaces. }
\label{fig_uN|uM}
\end{figure}

Consider two stacks of $N$ D3-branes and $M$ D3-branes, both of which extend along the directions $0126$ 
and terminate at $x^6=0$ on a single NS5-brane that spans along the directions $012345$ 
or a D5-brane along the directions $012789$ from both sides (see Figure \ref{fig_uN|uM}). 
From the perspective of 4d $\mathcal{N}=4$ SYM theories on the stacks of D3-branes, 
the $5$-branes can be viewed as the half-BPS interface or domain wall preserving $8$ supercharges. 
The rotational symmetry groups $SO(3)_{345}$ and $SO(3)_{789}$ in he brane configuration 
correspond to the $SU(2)_C$ and $SU(2)_H$ R-symmetry factors of the 3d $\mathcal{N}=4$ supersymmetry. 
For the NS5-type interface, the full $U(N)\times U(M)$ gauge group is preserved, 
whereas the D5-type interface breaks it down to its subgroup $U(\min (N,M))$ and involves the rank $|N-M|$ Nahm pole. 
In both interfaces there exist additional 3d matter fields. 
For the NS5-type interface, one finds the 3d hypermultiplet transforming in the bifundamental representation of $U(N)\times U(M)$ gauge group 
arising from the open strings across the NS5-brane. 
For the D5-type interface, the 3-5 open strings suspended between the D3- and D5-branes give rise to the massless spectrum 
as 3d matter fields at the interface transforming as the fundamental representation of $U(\min (N,M))$ gauge group. 

%Wilson
In addition, let us introduce fundamental strings stretched along the directions $05$. 
They realize Wilson line operators at the interface $x^6=0$. 
They break the $SU(2)_{C}$ R-symmetry down to $U(1)_C$ and preserve $4$ supercharges. 
When $k$ fundamental strings end on D3-branes at one end 
and on another D5-brane along the directions $034789$, which we call $\widetilde{\textrm{D5}}$, 
one finds the Wilson line transforming in the rank-$k$ antisymmetric representation \cite{Yamaguchi:2006tq,Gomis:2006sb,Rodriguez-Gomez:2006fmx,Hartnoll:2006hr}. 
Instead, when $k$ fundamental strings are suspended between D3-branes and another D5-brane along the directions $012789$, 
one obtains the Wilson line in the rank-$k$ symmetric representation \cite{Assel:2015oxa}.\footnote{In the absence of the interface or boundary, the Wilson lines in the symmetric representation in 4d $\mathcal{N}=4$ SYM theories can be realized by a probe D3-brane \cite{Drukker:2005kx,Gomis:2006im}. }  
More generally, the Wilson line operator in the representation associated with the Young diagram $\lambda$ $=$ $(\lambda_1,\cdots, \lambda_r)$ 
can be realized as a configuration of $r$ coincident D5-branes $(\textrm{D5}_1,\cdots, \textrm{D5}_r)$ \cite{Gomis:2006sb}
where $\textrm{D5}_i$ is the $i$-th D5-brane with $\lambda_i$ units of fundamental string charge dissolved in it. 
They can be also constructed as a configuration of $\lambda_1$ coincident $\widetilde{\textrm{D5}}$-branes $(\widetilde{\textrm{D5}}_1,\cdots, \widetilde{\textrm{D5}}_{\lambda_1})$. 
The $j$-th $\widetilde{\textrm{D5}}$-brane $\widetilde{\textrm{D5}}_j$, $j=1,\cdots, \lambda_1$ carries $\lambda'_{j}$ units of fundamental string charge 
where $\lambda'$ $=$ $(\lambda_1',\cdots,\lambda_{\lambda_1}')$ is the transpose of $\lambda$. 

%vortex
Similarly, we consider magnetic lines at the interface $x^6=0$ by adding D1-branes which extend along the directions $09$. 
Such D1-branes can be suspended along the direction $9$ between the D3-branes and $\widetilde{\textrm{NS5}}$-branes that extend along the directions $034578$. 
The configuration breaks the $SU(2)_H$ R-symmetry down to $U(1)_H$. 
The magnetic lines in $\mathcal{N}=4$ SYM theory, which are known as the 't Hooft lines \cite{tHooft:1977nqb,Kapustin:2005py} 
are characterized by magnetic charge $B$ as an element of the cocharacter of gauge group. 
In the description of such magnetic lines, the bulk fields are required to obey singular boundary conditions along the line operators. 
When we introduce the magnetic lines at the interface in $\mathcal{N}=4$ SYM theories, a little caution is needed. 
The magnetic charge $B$ corresponds to an insertion of flux through the two-dimensional $12$ plane at $x^6=0$ rather than the three-dimensional space 
and the 3d charged matter fields at the interface are also required to obey the singular boundary conditions. 
From this aspect, they can be viewed as the \textit{vortex lines}, the magnetic lines in 3d gauge theories. 
%monopole bubbling
Unlike the Wilson lines, the magnetic charge $B$ of the vortex lines may be altered 
due to the non-perturbative screening, which is known as the monopole bubbling effect \cite{Kapustin:2006pk}. 
In the brane configuration the magnetic charge $B$ encodes the number of the D1-branes. 
The monopole bubbling occurs when the location of the D1-branes coincide with 
that of pairs of the $\widetilde{\textrm{NS5}}$-branes along the directions $12$ \cite{Brennan:2018yuj}. 
When one considers the magnetic lines associated with the minuscule representation \cite{MR2109105} 
whose weights belong to a single Weyl group orbit, they are free from the monopole bubbling. 
For the theory of unitary gauge group, the magnetic charges $B$ associated with the minuscule representations take the form $(1^k,0^{N-k})$, 
for which the brane configuration contains $k$ D1-branes stretched between D3-branes and a single $\widetilde{\textrm{NS5}}$-brane. 

The brane configuration is summarized as follows: 
\begin{align}
\begin{array}{c|cccccccccc}
&0&1&2&3&4&5&6&7&8&9 \\ \hline 
\textrm{D3}&\circ&\circ&\circ&&&&\circ&&& \\
\textrm{NS5}&\circ&\circ&\circ&\circ&\circ&\circ&&&& \\
\textrm{D5}&\circ&\circ&\circ&&&&&\circ&\circ&\circ \\ \hline
\textrm{F1}&\circ&&&&&\circ&&&& \\
\widetilde{\textrm{D5}}&\circ&&&\circ&\circ&&&\circ&\circ&\circ \\  
\textrm{D1}&\circ&&&&&&&&&\circ \\
\widetilde{\textrm{NS5}}&\circ&&&\circ&\circ&\circ&&\circ&\circ& \\
\end{array}
\end{align}
where $\circ$ indicates the directions along which the branes extend. 

%S-duality
The $S$ transformation of the $SL(2,\mathbb{Z})$ action of Type IIB string theory 
swaps NS5-branes with D5-branes and fundamental strings with D1-branes and vice versa. 
This leads to the following intriguing conjectural dualities: 
\begin{align}
\label{duality_conj}
\begin{array}{ccc}
\textrm{Wilson lines}&\leftrightarrow&\textrm{vortex lines}\\
\textrm{NS5-type interface}&&\textrm{D5-type interface} \\
&& \\
\textrm{vortex lines}&\leftrightarrow&\textrm{Wilson lines}\\
\textrm{NS5-type interface}&&\textrm{D5-type interface} \\
\end{array}
\end{align}
In the following sections, we compute the line defect half-indices which encodes the spectra of the BPS operators at the junctions of the line operators and the interfaces. 
We find strong evidence of the dualities (\ref{duality_conj}) by finding matching pairs of the line defect half-indices.

%%%%%%%%%%%%%%%%%%%%%%%%%%%%%%%%%%%%%%%%
\subsection{Line defect half-indices}
%%%%%%%%%%%%%%%%%%%%%%%%%%%%%%%%%%%%%%%%
%DEF
The half-index of 4d $\mathcal{N}=4$ SYM theory with gauge group $G$ obeying the half-BPS boundary condition 
associated with the codimension-$1$ defect $\mathcal{B}$ 
can be defined as a trace over the cohomology $\mathcal{H}$ of the preserved supercharges 
that belong to the subalgebra of the 3d $\mathcal{N}=4$ superalgebra\footnote{See \cite{Gaiotto:2019jvo,Okazaki:2019ony} for the convention. Also see \cite{Dimofte:2011py,Gang:2012yr,Gang:2012ff} for other conventions. }
\begin{align}
\label{DEF_hind}
\mathbb{II}_{\mathcal{B}}^{\textrm{4d $G$}}(t,z;q)
&=\Tr (-1)^F q^{J+\frac{H+C}{4}} t^{H-C}z^f. 
\end{align}
Here $F$ is the Fermion number operator, 
$J$ spin, $C$, $H$ the Cartan generators of the $SU(2)_C$ and $SU(2)_H$ R-symmetry 
and $f$ the Cartan generators of the other global symmetries. 
Let us define 
\begin{align}
\label{qpoch_def}
(a;q)_{0}&:=1,\qquad
(a;q)_{n}:=\prod_{k=0}^{n-1}(1-aq^{k}),\qquad 
(q)_{n}:=\prod_{k=1}^{n}(1-q^{k}),
\nonumber \\
(a;q)_{\infty}&:=\prod_{k=0}^{\infty}(1-aq^{k}),\qquad 
(q)_{\infty}:=\prod_{k=1}^{\infty} (1-q^k), 
\end{align}
with $a, q \in \mathbb{C}$ and $|q|<1$. 
We also introduce the following notations: 
\begin{align}
(x^{\pm};q)_{n}&:=(x;q)_{n}(x^{-1};q)_{n}, \\
(x_1,\cdots,x_k;q)_{n}&:=(x_1;q)_{n}\cdots (x_k;q)_{n}. 
\end{align}

For example, the half-index of the NS5-type $U(N)|U(M)$ interface can be evaluated from the matrix integral of the form \cite{Gaiotto:2019jvo}
\begin{align}
\label{uNuMN}
\mathbb{II}_{\mathcal{N}}^{U(N)|U(M)}(t;q)
&=\frac{1}{N!}\frac{(q)_{\infty}^N}{(q^{\frac12}t^{-2};q)_{\infty}^N}
\oint 
\prod_{i=1}^N
\frac{ds_i^{(1)}}{2\pi is_i^{(1)}}
\prod_{i<j}
\frac{(s_i^{(1)\pm}s_j^{(1)\mp};q)_{\infty}}
{(q^{\frac12}t^{-2}s_i^{(1)\pm}s_j^{(1)\mp};q)_{\infty}}
\nonumber\\
&\times 
\frac{1}{M!}\frac{(q)_{\infty}^M}{(q^{\frac12}t^{-2};q)_{\infty}^M}
\oint 
\prod_{i=1}^M
\frac{ds_i^{(2)}}{2\pi is_i^{(2)}}
\prod_{i<j}
\frac{(s_i^{(2)\mp}s_j^{(2)\pm};q)_{\infty}}
{(q^{\frac12}t^{-2}s_i^{(2)\pm}s_j^{(2)\mp};q)_{\infty}}
\nonumber\\
&\times 
\prod_{i=1}^{N}
\prod_{j=1}^{M}
\frac{(q^{\frac34}t^{-1}s_i^{(1)\mp}s_j^{(2)\pm};q)_{\infty}}
{(q^{\frac14}ts_i^{(1)\mp}s_j^{(2)\pm};q)_{\infty}}. 
\end{align}
It agrees with the half-index of the D5-type $U(N)|U(M)$ interface of the form \cite{Gaiotto:2019jvo}
\begin{align}
\label{uNuMD}
\mathbb{II}_{\mathcal{D}}^{U(N)|U(M)}(t;q)
&=\frac{1}{N!}
\frac{(q)_{\infty}^{2N}}{(q^{\frac12}t^{\pm2};q)_{\infty}^N}
\oint \prod_{i=1}^N 
\frac{ds_i}{2\pi is_i}
\prod_{i<j}
\frac{(s_i^{\pm}s_j^{\mp};q)_{\infty}(qs_i^{\pm}s_j^{\mp};q)_{\infty}}
{(q^{\frac12}t^2s_i^{\pm}s_j^{\mp};q)_{\infty}(q^{\frac12}t^{-2}s_i^{\pm}s_j^{\mp};q)_{\infty}}
\nonumber\\
&\times 
\prod_{i=1}^{N}
\frac{(q^{\frac34+\frac{N-M}{4}}t^{-1+(N-M)}s_i^{\mp};q)_{\infty}}
{(q^{\frac14+\frac{N-M}{4}}t^{1+(N-M)}s_i^{\pm};q)_{\infty}}
\nonumber\\
&\times 
\prod_{l=1}^{N-M}
\frac{(q^{\frac{l+1}{2}}t^{2(l-1)};q)_{\infty}}
{(q^{\frac{l}{2}}t^{2l};q)_{\infty}}. 
\end{align}

%line defect half-index Wilson
One can generalize the half-indices (\ref{DEF_hind}) by introducing the line operators at the codimension-$1$ defect $\mathcal{B}$. 
The inserted line operators modify the space $\mathcal{H}$ in such a way that 
the half-indices can count the BPS local operators appearing at the junction of the line operators and $\mathcal{B}$ \cite{Cordova:2016uwk}. 
They are referred to as \textit{line defect half-indices} or \textit{line defect correlation functions}. 
When we introduce the BPS Wilson line $W_{\mathcal{R}}$ transforming in the representation $\mathcal{R}$ of gauge group, 
the BPS local operators at the junction of the Wilson lines and the interface should transform in its conjugate representation $\overline{\mathcal{R}}$ to form the gauge invariants. 
Hence the half-indices are modified by including the corresponding character $\chi_{\mathcal{R}}$ of the representation $\mathcal{R}$ in the matrix integral. 
More generally, we can calculate the line defect correlation function of a collection of Wilson lines $W_{\mathcal{R}_k}$ as 
\begin{align}
\label{W_general}
&
\mathbb{II}_{\mathcal{N}/\mathcal{D}}^{U(N)|U(M)}
=\oint \frac{ds}{2\pi is} i_{4d}(s,t;q)i_{3d}(s,t;q)
\nonumber\\
&\xrightarrow{W_{\mathcal{R}_k}}
\langle \prod_{k=1}W_{\mathcal{R}_k}\rangle_{\mathcal{N}/\mathcal{D}}^{U(N)|U(M)}
=\oint \frac{ds}{2\pi is}  i_{4d}(s,t;q)i_{3d}(s,t;q)\prod_{k}\chi_{\mathcal{R}_k}, 
\end{align}
where $i_{4d}(s,t;q)$ stands for contributions associated with the 4d fields and $i_{3d}$ correspond to those from the 3d matter fields. 

%vortex
Adding the BPS gauge vortex lines of magnetic charge $B$ at the interfaces requires the gauge fields and the matter fields 
to have singular configurations with poles and zeros of the order being specified by $B$. 
We propose that the line defect half-indices for the vortex lines associated with the minuscule representations at the interfaces 
can be computed by generalizing the prescription in \cite{Gang:2012yr} for the 't Hooft line defect indices. 
The half-indices of the interfaces contain the indices of 3d matter fields. 
The indices of 3d chiral multiplets can be evaluated from the localization technique \cite{Kim:2009wb,Imamura:2011su,Kapustin:2011jm}. 
In the presence of the vortex line of magnetic charge $B$, the index of the chiral multiplet transforming in the representation $\rho$ 
can be obtained by making use of the spherical harmonics indexed by the effective spin
\begin{align}
\mathsf{s}_{\textrm{eff}}&=\mathsf{s}+\frac12 \rho(B)
\end{align}
for the fields of spin $\mathsf{s}$. 
Accordingly, an insertion of the vortex line acts on the index by shifting the associated gauge fugacity $s$ in the indices of the matter fields. 
\begin{align}
\label{V_general}
&
\mathbb{II}_{\mathcal{N}/\mathcal{D}}^{U(N)|U(M)}
=F(t;q)\oint \frac{ds}{2\pi is} i_{4d}(s,t;q)i_{3d}(s,t;q)
\nonumber\\
&\xrightarrow{V_{B}}
\langle \prod_{k=1}V_{B}\rangle_{\mathcal{N}/\mathcal{D}}^{U(N)|U(M)}
=q^{\frac{\delta}{2}}t^{-2\delta} F(t;q) \oint \frac{ds}{2\pi is} i_{4d}(q^{\alpha(B)}s,t;q)i_{3d}(q^{\frac12\rho(B)}s,t;q). 
\end{align}
Here $F(t;q)$ is the neutral contribution associated with the broken part of the gauge group. 
The factor $i_{4d}(q^{\alpha(B)}s,t;q)$ in the integrand is obtained from the prescription \cite{Gang:2012yr} of the 't Hooft line defect indices, 
while the factor $i_{3d}(q^{\frac12\rho(B)}s,t;q)$ is from the prescription for the 3d supersymmetric indices 
in the presence of the vortex lines \cite{Drukker:2012sr,Okazaki:2019ony,Hayashi:2025guk}. 
The overall factor $q^{\frac{\delta}{2}}t^{-2\delta}$ is an extra contribution from the bare monopole operator of dimension $\delta$. 

%duality
In the following sections we compute the line defect half-indices (\ref{W_general}) and (\ref{V_general}) at the NS5-type and D5-type interfaces. 
We confirm that the two expressions for dual pairs agree with each other 
as a consequence of the dualities (\ref{duality_conj}) under the mirror transformation 
\begin{align}
t&\rightarrow t^{-1}, 
\end{align}
under which the two R-symmetry factors $SU(2)_C$ and $SU(2)_H$ are swapped. 

Note that the half-index (\ref{DEF_hind}) can be viewed as a supersymmetric index 
that generalizes the 3d $\mathcal{N}=4$ full index by involving additional 4d degrees freedom.  
There exist two interesting special fugacity limits, the Higgs limit and Coulomb limit \cite{Razamat:2014pta}
\begin{align}
\label{H_lim}
{\mathbb{II}_{\mathcal{B}}^{\textrm{4d $G$}}}^{(H)}(z;\mathfrak{q})
&=\lim_{
\begin{smallmatrix}
\textrm{$\mathfrak{q}:=q^{1/4}t$: fixed}\\
q\rightarrow 0\\
\end{smallmatrix}
}
\mathbb{II}_{\mathcal{B}}^{\textrm{4d $G$}}(t,z;q),\\
\label{C_lim}
{\mathbb{II}_{\mathcal{B}}^{\textrm{4d $G$}}}^{(C)}(z;\mathfrak{q})
&=\lim_{
\begin{smallmatrix}
\textrm{$\mathfrak{q}:=q^{1/4}t^{-1}$: fixed}\\
q\rightarrow 0\\
\end{smallmatrix}
}
\mathbb{II}_{\mathcal{B}}^{\textrm{4d $G$}}(t,z;q). 
\end{align}
In the Higgs (resp. Coulomb) limit, the indices enumerate the half-BPS local operators parameterizing the Higgs branch (resp. Coulomb branch). 
So these limits can be viewed as the half-BPS limits from the perspective of 3d $\mathcal{N}=4$ supersymmetric field theories. 
They can be also taken for the line defect half-indices at the interfaces. 
They make it easier to find several exact closed-form formulae by means of several techniques, including the character expansions and the Hall-Littlewood expansions, 
as we will see below. 

%normalized
Also we define the normalized line defect half-indices by 
\begin{align}
\langle \prod_{k} \mathcal{W}_{\mathcal{R}_k} \rangle^{U(N)|U(M)}_{\mathcal{N}/\mathcal{D}}(t;q)
&=\frac{\langle \prod_{k} W_{\mathcal{R}_k} \rangle^{U(N)|U(M)}_{\mathcal{N}/\mathcal{D}}(t;q)}
{\mathbb{II}^{U(N)|U(M)}_{\mathcal{N}/\mathcal{D}}(t;q)}, \\
\langle \prod_{k} \mathcal{V}_{B_k} \rangle^{U(N)|U(M)}_{\mathcal{N}/\mathcal{D}}(t;q)
&=\frac{\langle \prod_{k} V_{B_k} \rangle^{U(N)|U(M)}_{\mathcal{N}/\mathcal{D}}(t;q)}
{\mathbb{II}^{U(N)|U(M)}_{\mathcal{N}/\mathcal{D}}(t;q)}. 
\end{align}
They enumerate the effective degrees of freedom due to the inserted interface line operators and simplify the expressions. 
Similarly, we write the Higgs limit and Coulomb limit of the normalized line defect half-indices as
\begin{align}
\label{H_lim_W}
{\langle \prod_{k} \mathcal{W}_{\mathcal{R}_k} \rangle^{U(N)|U(M)}_{\mathcal{N}/\mathcal{D}}}^{(H)}
(\mathfrak{q})
&=\lim_{
\begin{smallmatrix}
\textrm{$\mathfrak{q}:=q^{1/4}t$: fixed}\\
q\rightarrow 0\\
\end{smallmatrix}
}
\langle \prod_{k} \mathcal{W}_{\mathcal{R}_k} \rangle^{U(N)|U(M)}_{\mathcal{N}/\mathcal{D}}(t;q), \\
\label{C_lim_W}
{\langle \prod_{k} \mathcal{W}_{\mathcal{R}_k} \rangle^{U(N)|U(M)}_{\mathcal{N}/\mathcal{D}}}^{(C)}
(\mathfrak{q})
&=\lim_{
\begin{smallmatrix}
\textrm{$\mathfrak{q}:=q^{1/4}t^{-1}$: fixed}\\
q\rightarrow 0\\
\end{smallmatrix}
}
\langle \prod_{k} \mathcal{W}_{\mathcal{R}_k} \rangle^{U(N)|U(M)}_{\mathcal{N}/\mathcal{D}}(t;q), \\
\label{H_lim_V}
{\langle \prod_{k} \mathcal{V}_{B_k} \rangle^{U(N)|U(M)}_{\mathcal{N}/\mathcal{D}}}^{(H)}
(\mathfrak{q})
&=\lim_{
\begin{smallmatrix}
\textrm{$\mathfrak{q}:=q^{1/4}t$: fixed}\\
q\rightarrow 0\\
\end{smallmatrix}
}\langle \prod_{k} \mathcal{V}_{B_k} \rangle^{U(N)|U(M)}_{\mathcal{N}/\mathcal{D}}(t;q), \\
\label{C_lim_V}
{\langle \prod_{k} \mathcal{V}_{B_k} \rangle^{U(N)|U(M)}_{\mathcal{N}/\mathcal{D}}}^{(C)}
(\mathfrak{q})
&=\lim_{
\begin{smallmatrix}
\textrm{$\mathfrak{q}:=q^{1/4}t^{-1}$: fixed}\\
q\rightarrow 0\\
\end{smallmatrix}
}\langle \prod_{k} \mathcal{V}_{B_k} \rangle^{U(N)|U(M)}_{\mathcal{N}/\mathcal{D}}(t;q). 
\end{align}

%%%%%%%%%%%%%%%%%%%%%%%%%%%%%%%%%%%%%%%%
%%%%%%%%%%%%%%%%%%%%%%%%%%%%%%%%%%%%%%%%
\section{$U(1)|U(1)$}
\label{sec_u1u1}
%%%%%%%%%%%%%%%%%%%%%%%%%%%%%%%%%%%%%%%%
%%%%%%%%%%%%%%%%%%%%%%%%%%%%%%%%%%%%%%%%

%%%%%%%%%%%%%%%%%%%%%%%%%%%%%%%%%%%%%%%%
\subsection{Wilson-vortex}
%%%%%%%%%%%%%%%%%%%%%%%%%%%%%%%%%%%%%%%%
%NS5 U(1)|U(1) Wilson
While the line defect indices for the bulk 4d $\mathcal{N}=4$ $U(1)$ gauge theory are trivial, 
those for the $U(1)|U(1)$ interfaces are non-trivial as they support 3d charged matter fields. 

For the NS5-type $U(1)|U(1)$ interface one has $U(1)^{(1)} \times U(1)^{(2)}$ gauge group 
which is coupled to the bifundamental hypermultiplet. 
Let us consider the Wilson line $W_{n^{(1)};n^{(2)}}$ at the interface carrying charges $n^{(1)}$ (resp. $n^{(2)}$)  
under the $U(1)^{(1)}$ (resp. $U(1)^{(2)}$) gauge group. 
The line defect half-index reads
\begin{align}
\label{u1u1N_Wn}
&
\langle W_{n^{(1)};n^{(2)}}\rangle^{U(1)|U(1)}_{\mathcal{N}}(t;q)
\nonumber\\
&=\frac{(q)_{\infty}^2}{(q^{\frac12}t^{-2};q)_{\infty}^2}
\oint \frac{ds^{(1)}}{2\pi is^{(1)}}
{s^{(1)}}^{n^{(1)}}
\oint \frac{ds^{(2)}}{2\pi is^{(2)}}
{s^{(2)}}^{n^{(2)}} 
\frac{(q^{\frac34}t^{-1}s^{(1)\mp}s^{(2)\pm};q)_{\infty}}
{(q^{\frac14}ts^{(1)\pm}s^{(2)\mp};q)_{\infty}}. 
\end{align}
By picking up the residues at poles $s^{(2)}$ $=$ $q^{\frac14+l}ts^{(1)}$, $l=0,1,\cdots$, 
we find that it is non-trivial if and only if $n^{(1)}=-n^{(2)}=n$
\begin{align}
&
\langle W_{n^{(1)};n^{(2)}}\rangle^{U(1)|U(1)}_{\mathcal{N}}(t;q)
\nonumber\\
&=
\begin{cases}
\frac{(q^{\frac12}t^2;q)_{\infty}}{(q^{\frac12}t^{-2};q)_{\infty}}
\sum_{l=0}^{\infty}
\frac{(q^{1+l};q)_{\infty}^2}
{(q^{\frac12+l}t^2;q)_{\infty}^2}
q^{\frac{l}{2}+|n|l+\frac{|n|}{4}}t^{-2l+|n|}&\textrm{for $n^{(1)}=-n^{(2)}=n$}\cr
0&\textrm{otherwise}\cr
\end{cases}. 
\end{align}

%D5 U(1)|U(1) vortex
The NS5-type $U(1)|U(1)$ interface is conjectured to be S-dual to the D5-type $U(1)|U(1)$ interface \cite{Gaiotto:2008ak}. 
In fact, it is shown in \cite{Gaiotto:2019jvo} that the half-indices agree with each other. 
For the D5-type $U(1)|U(1)$ interface there is a whole $U(1)$ gauge group and the charged hypermultiplet. 
As the charged Wilson lines are realized by fundamental strings, they will be dual to the vortex lines described by D1-branes. 
So we expect that the Wilson lines at the NS5-type interface map to the vortex lines at the D5-type $U(1)|U(1)$ interface. 
In the presence of the gauge vortex line with the vortex number $n$ of the $U(1)$ gauge group 
the spins of the BPS local operators of charge $n$ will be shifted by $|n|/2$. 
This leads to the vortex line defect half-index of the D5-type $U(1)|U(1)$ interface with the form
\begin{align}
\label{u1u1D_Vn}
\langle V_n\rangle_{\mathcal{D}}^{U(1)|U(1)}(t;q)
&=q^{\frac{|n|}{4}}t^{-|n|}
\frac{(q)_{\infty}^2}{(q^{\frac12}t^{\pm2};q)_{\infty}}
\oint \frac{ds}{2\pi is}
\frac{(q^{\frac34+\frac{|n|}{2}}t^{-1}s^{\mp};q)_{\infty}}
{(q^{\frac14+\frac{|n|}{2}}ts^{\pm};q)_{\infty}}. 
\end{align}
Here the overall factor $q^{|n|/4}t^{-|n|}$ is a contribution from the bare monopole operator of fixed magnetic flux $n$ at the interface. 
Note that the expression (\ref{u1u1D_Vn}) reduces to the half-index of the D5-type $U(1)|U(1)$ interface in \cite{Gaiotto:2019jvo} when $n=0$. 
The integral (\ref{u1u1D_Vn}) can be computed by summing over the resides at poles $s$ $=$ $q^{\frac14+\frac{|n|}{2}+l}t^{-1}$, $l=0,1,\cdots$
\begin{align}
\langle V_n \rangle_{\mathcal{D}}^{U(1)|U(1)}(t;q)
&=\sum_{l=0}^{\infty}
\frac{(q^{1+l};q)_{\infty}(q^{1+|n|+l};q)_{\infty}}
{(q^{\frac12+l}t^{2};q)_{\infty}(q^{\frac12+|n|+l}t^{2};q)_{\infty}}
q^{\frac{l}{2}+\frac{|n|}{4}}t^{-2l-|n|}. 
\end{align}

As strong evidence of the duality 
between the NS5-type $U(1)|U(1)$ interface Wilson line $W_{n;-n}$ and the D5-type $U(1)|U(1)$ interface vortex line $V_n$, 
we find the matching of the two expressions (\ref{u1u1N_Wn}) and (\ref{u1u1D_Vn}) under the mirror transformation $t$ $\rightarrow$ $t^{-1}$
\begin{align}
\langle W_{n;-n}\rangle^{U(1)|U(1)}_{\mathcal{N}}(t;q)
&=\langle V_n \rangle_{\mathcal{D}}^{U(1)|U(1)}(t^{-1};q). 
\end{align}

Let us consider the half-BPS limits of the defect half-indices. 
In the absence of the line operator, we have \cite{Hatsuda:2024uwt}
\begin{align}
{\mathbb{II}^{U(1)|U(1)}_{\mathcal{N}}}^{(C)}(\mathfrak{q})
&=\frac{1}{(1-\mathfrak{q}^2)^2}, \\
{\mathbb{II}^{U(1)|U(1)}_{\mathcal{N}}}^{(H)}(\mathfrak{q})
&=\frac{1}{1-\mathfrak{q}^2}. 
\end{align}
When we introduce the charged Wilson line operator $W_{n;-n}$ in the NS5-type interface, we find 
\begin{align}
{\langle \mathcal{W}_{n;-n}\rangle^{U(1)|U(1)}_{\mathcal{N}}}^{(C)}(\mathfrak{q})&=0,\\
{\langle \mathcal{W}_{n;-n}\rangle^{U(1)|U(1)}_{\mathcal{N}}}^{(H)}(\mathfrak{q})&=\mathfrak{q}^n. 
\end{align}
The non-trivial line defect half-index appears in the Higgs limit, 
for which one detects the gauge invariant operator of $n$ charged hypermultiplets at the junction of the Wilson line and the NS5-type interface. 

%%%%%%%%%%%%%%%%%%%%%%%%%%%%%%%%%%%%%%%%
\subsection{Vortex-Wilson}
%%%%%%%%%%%%%%%%%%%%%%%%%%%%%%%%%%%%%%%%
%NS5 U(1)|U(1) vortex
Next consider the vortex line at the NS5-type $U(1)|U(1)$ interface. 
Although the vortex line may carry a pair $(n^{(1)},n^{(2)})$ of the vortex numbers associated with the $U(1)^{(1)}$ and $U(1)^{(2)}$ gauge groups, 
they can be related by the screening effect. 
Up to the screening equivalence, the non-trivial vortex lines can be labeled by one of them, say $n^{(1)}$. 
Taking into account the shifted spin due to the vortex line with $n^{(1)}$ $=$ $n$ and $n^{(2)}=0$, 
the line defect half-index is evaluated as
\begin{align}
\label{u1u1N_Vn}
\langle V_{n;0}\rangle_{\mathcal{N}}^{U(1)|U(1)}(t;q)
&=
q^{\frac{|n|}{4}}t^{-|n|}
\frac{(q)_{\infty}^2}{(q^{\frac12}t^{-2};q)_{\infty}^2}
\oint \frac{ds^{(1)}}{2\pi is^{(1)}}
\oint \frac{ds^{(2)}}{2\pi is^{(2)}}
\frac{(q^{\frac34+\frac{|n|}{2}}t^{-1}s^{(1)\mp}s^{(2)\pm};q)_{\infty}}
{(q^{\frac14+\frac{|n|}{2}}ts^{(1)\pm}s^{(2)\mp};q)_{\infty}}. 
\end{align}
Summing over the residues at poles $s^{(2)}$ $=$ $q^{\frac14+\frac{|n|}{2}+l}ts^{(1)}$, $l=0,1,\cdots$, 
it can be evaluated as
\begin{align}
\langle V_{n;0}\rangle_{\mathcal{N}}^{U(1)|U(1)}(t;q)
&=\frac{(q^{\frac12}t^2;q)_{\infty}}{(q^{\frac12}t^{-2};q)_{\infty}}
\sum_{l=0}^{\infty}
\frac{(q^{1+l};q)_{\infty}(q^{1+|n|+l};q)_{\infty}}
{(q^{\frac12+l}t^2;q)_{\infty}(q^{\frac12+|n|+l}t^2;q)_{\infty}}
q^{\frac{l}{2}+\frac{|n|}{4}}t^{-2l-|n|}. 
\end{align}

%D5 U(1)|U(1) Wilson
We expect that 
there exists the Wilson line $W_n$ of the $U(1)$ charge $n$ in the dual D5-type $U(1)|U(1)$ interface. 
It is straightforward to evaluate the line defect half-index 
of the charged Wilson line $W_n$ for the D5-type $U(1)|U(1)$ interface as
\begin{align}
\label{u1u1D_Wn}
\langle W_n\rangle_{\mathcal{D}}^{U(1)|U(1)}(t;q)
&=\frac{(q)_{\infty}^2}{(q^{\frac12}t^{\pm2};q)_{\infty}}
\oint \frac{ds}{2\pi is}
\frac{(q^{\frac34}t^{-1}s^{\mp};q)_{\infty}}
{(q^{\frac14}ts^{\pm};q)_{\infty}}
s^n. 
\end{align}
Picking up the residues at poles, it can be expanded as
\begin{align}
\langle W_n\rangle_{\mathcal{D}}^{U(1)|U(1)}(t;q)
&=\sum_{l=0}^{\infty}
\frac{(q^{1+l};q)_{\infty}^2}{(q^{\frac12+l}t^2;q)_{\infty}^2}
q^{\frac{l}{2}+|n|l+\frac{|n|}{4}}t^{-2l+|n|}
\end{align}

In fact, we find that 
the two expressions (\ref{u1u1N_Vn}) and (\ref{u1u1D_Wn}) agree with each other 
\begin{align}
\langle V_{n;0}\rangle^{U(1)|U(1)}_{\mathcal{N}}(t;q)
&=\langle W_n\rangle_{\mathcal{D}}^{U(1)|U(1)}(t^{-1};q). 
\end{align}
This demonstrates that 
the NS5-type $U(1)|U(1)$ interface vortex line $V_{n;0}$ and the D5-type $U(1)|U(1)$ interface Wilson line $W_n$ are dual. 

In the Coulomb and Higgs limits the normalized interface line defect half-indices (\ref{u1u1N_Vn}) are given by
\begin{align}
{\langle \mathcal{V}_{n;0}\rangle_{\mathcal{N}}^{U(1)|U(1)}}^{(C)}(\mathfrak{q})&=\mathfrak{q}^n,\\
{\langle \mathcal{V}_{n;0}\rangle_{\mathcal{N}}^{U(1)|U(1)}}^{(H)}(\mathfrak{q})&=0. 
\end{align}
Unlike the Wilson line in the NS5-type interface, the Higgs limit is trivial in this case. 
Instead, the vortex line at the NS5-type interface can detect the Coulomb branch operators, 
which are identified with the monopole operators of magnetic charge $n$ at the junction of the vortex line and the NS5-type interface. 

%%%%%%%%%%%%%%%%%%%%%%%%%%%%%%%%%%%%%%%%
%%%%%%%%%%%%%%%%%%%%%%%%%%%%%%%%%%%%%%%%
\section{$U(N)|U(1)$}
\label{sec_uNu1}
%%%%%%%%%%%%%%%%%%%%%%%%%%%%%%%%%%%%%%%%
%%%%%%%%%%%%%%%%%%%%%%%%%%%%%%%%%%%%%%%%

%%%%%%%%%%%%%%%%%%%%%%%%%%%%%%%%%%%%%%%%
\subsection{Wilson-vortex}
%%%%%%%%%%%%%%%%%%%%%%%%%%%%%%%%%%%%%%%%
%NS5 U(N)|U(1) Wilson
Consider the NS5-type interface between $\mathcal{N}=4$ $U(N)$ gauge theory with $N>1$ and $U(1)$ gauge theory. 
It contains the hypermultiplet transforming as the bifundamental representation under $U(N)^{(1)} \times U(1)^{(2)}$ gauge group. 
One can introduce the Wilson line $W_{\mathcal{R}^{(1)}; n^{(2)}}$ 
at the interface transforming as the representation $\mathcal{R}^{(1)}$ under $U(N)^{(1)}$ 
and carrying the charge $n^{(2)}$ of $U(1)^{(2)}$. 
The line defect half-index is given by the matrix integral of the form 
\begin{align}
\label{uNu1N_W}
&
\langle W_{\mathcal{R}^{(1)}; n^{(2)}}\rangle^{U(N)|U(1)}_{\mathcal{N}}(t;q)
\nonumber\\
&=
\frac{1}{N!} \frac{(q)_{\infty}^N}{(q^{\frac12}t^{-2};q)_{\infty}^N}
\oint 
\prod_{i=1}^N
\frac{ds_i^{(1)}}{2\pi is_i^{(1)}}
\prod_{i<j}
\frac{(s_i^{(1)\pm}s_j^{(1)\mp};q)_{\infty}}
{(q^{\frac12}t^{-2}s_i^{(1)\pm}s_j^{(1)\mp};q)_{\infty}}
\chi_{\mathcal{R}^{(1)}}(s^{(1)})
\nonumber\\
&\times 
\frac{(q)_{\infty}}{(q^{\frac12}t^{-2};q)_{\infty}}
\oint \frac{ds^{(2)}}{2\pi is^{(2)}}
{s^{(2)}}^{n^{(2)}}
\prod_{i=1}^N  
\frac{(q^{\frac34}t^{-1}s_i^{(1)\mp}s^{(2)\pm};q)_{\infty}}
{(q^{\frac14}ts_i^{(1)\pm}s^{(2)\mp};q)_{\infty}}, 
\end{align}
where $\chi_{\mathcal{R}^{(1)}}(s^{(1)})$ is the character of the representation $\mathcal{R}^{(1)}$ of $U(N)^{(1)}$. 
We find that it is non-vanishing if and only if 
the Wilson line transforms as $(n)$, the rank-$n$ symmetric representation of $U(N)^{(1)}$ and carries charge $-n$\footnote{For negative $n$ the notation $(n)$ implies the conjugate of the rank-$n$ symmetric representation. }
\begin{align}
\langle W_{\mathcal{R}^{(1)}; n^{(2)}}\rangle^{U(N)|U(1)}_{\mathcal{N}}(t;q)
&=0\qquad \textrm{for $(\mathcal{R}^{(1)},n^{(2)}) \neq ((n),-n)$}. 
\end{align}

%D5 U(N)|U(1) vortex
The S-dual of the NS5-type $U(N)|U(1)$ interface is the D5-type $U(N)|U(1)$ interface \cite{Gaiotto:2008ak}. 
In this case, the $U(N)\times U(1)$ gauge group is broken down to a diagonal subgroup 
so that the interface preserves the $U(1)$ gauge group. 
When $N=2$ there is no singular Nahm pole at the interface, while for $N>2$ there exists a rank-$(N-1)$ Nahm pole. 
The half-index of the D5-type $U(N)|U(1)$ interface takes the form \cite{Gaiotto:2019jvo}
\begin{align}
\label{uNu1D}
\mathbb{II}_{\mathcal{D}}^{U(N)|U(1)}(t;q)
&=\frac{(q)_{\infty}^2}{(q^{\frac12}t^{\pm2};q)_{\infty}}
\oint \frac{ds}{2\pi is}
\frac{(q^{\frac34+\frac{(N-1)}{4}}t^{-1+(N-1)}s^{\mp};q)_{\infty}}
{(q^{\frac14+\frac{(N-1)}{4}}t^{1+(N-1)}s^{\pm};q)_{\infty}}
\nonumber\\
&\times 
\prod_{l=1}^{N-1}\frac{(q^{\frac{l+1}{2}}t^{2(l-1)};q)_{\infty}}
{(q^{\frac{l}{2}}t^{2l};q)_{\infty}}. 
\end{align}
For the D5-type $U(N)|U(1)$ interface 
one can introduce the gauge vortex line $V_n$ of vortex number $n$. 
Due to the singular configuration induced by the vortex line,  
the local operators of gauge charges $n$ living on the interface should receive the shifts of spins by $|n|/2$. 
Accordingly, the corresponding line defect half-index of the D5-type $U(N)|U(1)$ interface vortex $V_n$ is given by
\begin{align}
\label{uNu1D_V}
\langle  V_{n} \rangle_{\mathcal{D}}^{U(N)|U(1)}(t;q)
&=
q^{\frac{|n|}{4}}t^{-|n|}
\frac{(q)_{\infty}^2}{(q^{\frac12}t^{\pm2};q)_{\infty}}
\oint \frac{ds}{2\pi is}
\frac{(q^{\frac{3}{4}+\frac{(N-1)}{4}+\frac{|n|}{2}}t^{-1+(N-1)}s^{\mp};q)_{\infty}}
{(q^{\frac14+\frac{(N-1)}{4}+\frac{|n|}{2}}t^{1+(N-1)}s^{\pm};q)_{\infty}}
\nonumber\\
&\times 
\prod_{l=1}^{N-1}\frac{(q^{\frac{l+1}{2}}t^{2(l-1)};q)_{\infty}}
{(q^{\frac{l}{2}}t^{2l};q)_{\infty}}. 
\end{align}
Picking up the residues at poles $s=q^{\frac{N}{4}+\frac{|n|}{2}+l}t^N$, $l=0,1,\cdots$, we can expand the expression (\ref{uNu1D_V}) as
\begin{align}
\langle  V_{n} \rangle_{\mathcal{D}}^{U(N)|U(1)}(t;q)
&=\sum_{m=0}^{\infty}
\frac{(q^{1+m};q)_{\infty}(q^{\frac{N+1}{2}+|n|+m}t^{2N-2};q)_{\infty}}
{(q^{\frac12+m}t^{2};q)_{\infty}(q^{\frac{N}{2}+|n|+m}t^{2N};q)_{\infty}}
q^{\frac{|n|}{4}+\frac{m}{2}}t^{-|n|-2m}
\nonumber\\
&\times 
\prod_{l=1}^{N-1}\frac{(q^{\frac{l+1}{2}}t^{2(l-1)};q)_{\infty}}
{(q^{\frac{l}{2}}t^{2l};q)_{\infty}}. 
\end{align}

It is expected that the Wilson line in the NS5-type interface maps to the vortex line in the D5-type interface under the action of S-duality. 
In fact, we find that 
\begin{align}
\langle W_{(n); -n}\rangle^{U(N)|U(1)}_{\mathcal{N}}(t;q)
&=\langle  V_{n} \rangle_{\mathcal{D}}^{U(N)|U(1)}(t^{-1};q). 
\end{align}
This indicates that 
the NS5-type interface Wilson line $W_{(n),-n}$ transforming 
as the rank-$n$ symmetric representation of the $U(N)^{(1)}$ gauge group and carrying the electric charge $-n$ of the $U(1)^{(2)}$ gauge group 
is S-dual to the D5-type interface vortex line $V_{n}$ of vortex number $n$. 

%Coulomb Higgs limits
Taking the Coulomb and Higgs limits, the half-index of the NS5-type interface become \cite{Hatsuda:2024uwt}
\begin{align}
{\mathbb{II}^{U(N)|U(1)}_{\mathcal{N}}}^{(C)}(\mathfrak{q})
&=\frac{1}{1-\mathfrak{q}^2}\prod_{n=1}^{N}\frac{1}{1-\mathfrak{q}^{2n}}, \\
{\mathbb{II}^{U(N)|U(1)}_{\mathcal{N}}}^{(H)}(\mathfrak{q})
&=\frac{1}{1-\mathfrak{q}^2}. 
\end{align}
In the presence of the Wilson line operator $W_{(n),-n}$ on the NS5-type interface, 
one finds the normalized defect half-indices
\begin{align}
{\langle \mathcal{W}_{(n);-n}\rangle^{U(N)|U(1)}_{\mathcal{N}}}^{(C)}(\mathfrak{q})&=0,\\
{\langle \mathcal{W}_{(n);-n}\rangle^{U(N)|U(1)}_{\mathcal{N}}}^{(H)}(\mathfrak{q})
&=\mathfrak{q}^n. 
\end{align}
Again the line defect half-indices of the Wilson line $W_{(n);-n}$ at the $U(N)|U(1)$ NS5-type interface are non-trivial in the Higgs limit. 
There exists a single Higgs branch operator consisting of the 3d matter field transforming in the conjugate of the rank-$n$ symmetric representation of $U(N)^{(1)}$ 
and carrying the electric charge $n$ of $U(1)^{(2)}$. 
Note that the expression is independent of rank $N$ of the gauge group. 

%%%%%%%%%%%%%%%%%%%%%%%%%%%%%%%%%%%%%%%%
\subsection{Vortex-Wilson}
%%%%%%%%%%%%%%%%%%%%%%%%%%%%%%%%%%%%%%%%
%NS5 U(N)|U(1) vortex
For the NS5-type $U(N)|U(1)$ interface, 
the dynamical vortex line $V_{k^{(1)};n^{(2)}}$ can be generically labeled by the Abelianized vortex number $k^{(1)}$ 
$=$ $(k_1^{(1)},\cdots,k_N^{(1)})$, a cocharacter of the $U(N)^{(1)}$ gauge group 
as well as the vortex number $n^{(2)}$ of the $U(1)^{(2)}$ gauge group. 
For the vortex line with $k^{(1)}$ $=$ $0$ and $n^{(2)}=n$, 
the spin of the bifundamental hypermultiplet is shifted so that 
the vortex line defect half-index will take the form 
\begin{align}
\label{uNu1N_Vn}
\langle V_{0;n}\rangle_{\mathcal{N}}^{U(N)|U(1)}(t;q)
&=
q^{\frac{N|n|}{4}}t^{-N|n|}
\frac{1}{N!}
\frac{(q)_{\infty}^N}{(q^{\frac12}t^{-2};q)_{\infty}^N}
\oint \prod_{i=1}^N \frac{ds_i}{2\pi is_i}
\prod_{i<j}
\frac{(s_i^{(1)\pm}s_j^{(1)\mp};q)_{\infty}}
{(q^{\frac12}t^{-2}s_i^{(1)\pm}s_{j}^{(1)\mp};q)_{\infty}}
\nonumber\\
&\times 
\frac{(q)_{\infty}}{(q^{\frac12}t^{-2};q)_{\infty}}
\oint \frac{ds^{(2)}}{2\pi is^{(2)}}
\prod_{i=1}^N
\frac{(q^{\frac34+\frac{|n|}{2}}t^{-1}s_i^{(1)\mp}s^{(2)\pm};q)_{\infty}}
{(q^{\frac14+\frac{|n|}{2}}ts_i^{(1)\pm}s^{(2)\mp};q)_{\infty}}. 
\end{align}

%D5 U(N)|U(1) Wilson
In the S-dual D5-type $U(N)|U(1)$ interface one can introduce the Wilson line with charge $n$ of the $U(1)$ gauge group. 
The line defect half-index of the latter configuration is given by 
\begin{align}
\label{uNu1D_Wn}
\langle  W_{n} \rangle_{\mathcal{D}}^{U(N)|U(1)}(t;q)
&=
\frac{(q)_{\infty}^2}{(q^{\frac12}t^{\pm2};q)_{\infty}}
\oint \frac{ds}{2\pi is}
\frac{(q^{\frac{3}{4}+\frac{(N-1)}{4}}t^{-1+(N-1)}s^{\mp};q)_{\infty}}
{(q^{\frac14+\frac{(N-1)}{4}}t^{1+(N-1)}s^{\pm};q)_{\infty}} s^n 
\nonumber\\
&\times 
\prod_{l=1}^{N-1}\frac{(q^{\frac{l+1}{2}}t^{2(l-1)};q)_{\infty}}
{(q^{\frac{l}{2}}t^{2l};q)_{\infty}}. 
\end{align}
Summing over the residues at poles $s=q^{\frac{N}{4}+l}t^{N}$, $l=0,1,\cdots$, we can expanded it as
\begin{align}
\langle  W_{n} \rangle_{\mathcal{D}}^{U(N)|U(1)}(t;q)
&=\sum_{m=0}^{\infty}
\frac{(q^{1+m};q)_{\infty}(q^{\frac{N+1}{2}+m}t^{2N-2};q)_{\infty}}
{(q^{\frac12+m}t^2;q)_{\infty}(q^{\frac{N}{2}+m}t^{2N};q)_{\infty}}
q^{\frac{Nn}{4}+nm+\frac{m}{2}}t^{Nn-2m}
\nonumber\\
&\times 
\prod_{l=1}^{N-1}\frac{(q^{\frac{l+1}{2}}t^{2(l-1)};q)_{\infty}}
{(q^{\frac{l}{2}}t^{2l};q)_{\infty}}. 
\end{align}

In fact, we have confirmed that 
the two expressions (\ref{uNu1N_Vn}) and (\ref{uNu1D_Wn}) match 
\begin{align}
\langle V_{0;n}\rangle_{\mathcal{N}}^{U(N)|U(1)}(t;q)
&=\langle  W_{n} \rangle_{\mathcal{D}}^{U(N)|U(1)}(t^{-1};q). 
\end{align}
This supports the duality between the NS5-type interface vortex line $V_{0;n}$ 
and the D5-type interface Wilson line $W_{n}$. 

In the Coulomb and Higgs limits, the normalized vortex line half-indices (\ref{uNu1N_Vn}) are 
\begin{align}
{\langle \mathcal{V}_{n;0}\rangle_{\mathcal{N}}^{U(N)|U(1)}}^{(C)}(\mathfrak{q})
&=\mathfrak{q}^{Nn},\\
{\langle \mathcal{V}_{n;0}\rangle_{\mathcal{N}}^{U(N)|U(1)}}^{(H)}(\mathfrak{q})&=0. 
\end{align}
Having the vortex line at the NS5-type interface, the Higgs branch operators are eliminated as opposed to the Wilson line. 
The defect half-index depends on the rank $N$ of the gauge group, 
which indicates that the scaling dimension of the Coulomb branch operator consisting of the BPS monopole is proportional to gauge rank $N$. 

%%%%%%%%%%%%%%%%%%%%%%%%%%%%%%%%%%%%%%%%
%%%%%%%%%%%%%%%%%%%%%%%%%%%%%%%%%%%%%%%%
\section{$U(N)|U(N)$}
\label{sec_uNuN}
%%%%%%%%%%%%%%%%%%%%%%%%%%%%%%%%%%%%%%%%
%%%%%%%%%%%%%%%%%%%%%%%%%%%%%%%%%%%%%%%%

%%%%%%%%%%%%%%%%%%%%%%%%%%%%%%%%%%%%%%%%
\subsection{Wilson-vortex}
%%%%%%%%%%%%%%%%%%%%%%%%%%%%%%%%%%%%%%%%
%NS5 U(N)|U(N) Wilson
Consider the NS5-type $U(N)|U(N)$ interface with $N>1$. 
It preserves the non-Abelian $U(N)^{(1)}$ $\times$ $U(N)^{(2)}$ gauge group 
under which the hypermultiplet transforms as the bifundamental representation 
$(\mathbf{N},\overline{\mathbf{N}})$ $\oplus$ $(\overline{\mathbf{N}},\mathbf{N})$. 
The Wilson line $W_{\mathcal{R}^{(1)};\mathcal{R}^{(2)}}$ at the interface 
can transform as a pair $(\mathcal{R}^{(1)},\mathcal{R}^{(2)})$ of the representations 
under the $U(N)^{(1)}$ $\times$ $U(N)^{(2)}$ gauge group. 
The line defect half-index of the Wilson line $W_{\mathcal{R}^{(1)};\mathcal{R}^{(2)}}$ at the NS5-type interface is evaluated as
\begin{align}
\label{uNuNN_W}
&
\langle W_{\mathcal{R}^{(1)};\mathcal{R}^{(2)}} \rangle^{U(N)|U(N)}_{\mathcal{N}}(t;q)
\nonumber\\
&=\frac{1}{N!}\frac{(q)_{\infty}^N}{(q^{\frac12}t^{-2};q)_{\infty}^N}
\oint 
\prod_{i=1}^N
\frac{ds_i^{(1)}}{2\pi is_i^{(1)}}
\prod_{i<j}
\frac{(s_i^{(1)\pm}s_j^{(1)\mp};q)_{\infty}}
{(q^{\frac12}t^{-2}s_i^{(1)\pm}s_j^{(1)\mp};q)_{\infty}}
\chi_{\mathcal{R}^{(1)}}(s^{(1)})
\nonumber\\
&\times 
\frac{1}{N!}\frac{(q)_{\infty}^N}{(q^{\frac12}t^{-2};q)_{\infty}^N}
\oint 
\prod_{i=1}^N
\frac{ds_i^{(2)}}{2\pi is_i^{(2)}}
\prod_{i<j}
\frac{(s_i^{(2)\mp}s_j^{(2)\pm};q)_{\infty}}
{(q^{\frac12}t^{-2}s_i^{(2)\pm}s_j^{(2)\mp};q)_{\infty}}
\chi_{\mathcal{R}^{(2)}}(s^{(2)})
\nonumber\\
&\times 
\prod_{i=1}^{N}
\prod_{j=1}^{N}
\frac{(q^{\frac34}t^{-1}s_i^{(1)\mp}s_j^{(2)\pm};q)_{\infty}}
{(q^{\frac14}ts_i^{(1)\pm}s_j^{(2)\mp};q)_{\infty}}, 
\end{align}
where $\chi_{\mathcal{R}^{(i)}}(s^{(i)})$ are the characters of the representations $\mathcal{R}^{(i)}$ of $U(N)^{(i)}$. 

More generally, when we introduce collections of the Wilson lines 
$W_{\mathcal{R}^{(1)}_i;\mathcal{R}^{(2)}_i}$, with $i=1,\cdots,m$, 
at the NS5-type interface, we have
\begin{align}
\label{uNuNN_Wgeneral}
&
\langle \prod_{i=1}^{m}
W_{\mathcal{R}^{(1)}_i;\mathcal{R}^{(2)}_i} \rangle^{U(N)|U(N)}_{\mathcal{N}}(t;q)
\nonumber\\
&=\frac{1}{N!}\frac{(q)_{\infty}^N}{(q^{\frac12}t^{-2};q)_{\infty}^N}
\oint 
\prod_{i=1}^N
\frac{ds_i^{(1)}}{2\pi is_i^{(1)}}
\prod_{i<j}
\frac{(s_i^{(1)\pm}s_j^{(1)\mp};q)_{\infty}}
{(q^{\frac12}t^{-2}s_i^{(1)\pm}s_j^{(1)\mp};q)_{\infty}}
\prod_{i=1}^m 
\chi_{\mathcal{R}^{(1)}_i}(s^{(1)})
\nonumber\\
&\times 
\frac{1}{N!}\frac{(q)_{\infty}^N}{(q^{\frac12}t^{-2};q)_{\infty}^N}
\oint 
\prod_{i=1}^N
\frac{ds_i^{(2)}}{2\pi is_i^{(2)}}
\prod_{i<j}
\frac{(s_i^{(2)\mp}s_j^{(2)\pm};q)_{\infty}}
{(q^{\frac12}t^{-2}s_i^{(2)\pm}s_j^{(2)\mp};q)_{\infty}}
\prod_{i=1}^m
\chi_{\mathcal{R}^{(2)}_i}(s^{(2)})
\nonumber\\
&\times 
\prod_{i=1}^{N}
\prod_{j=1}^{N}
\frac{(q^{\frac34}t^{-1}s_i^{(1)\mp}s_j^{(2)\pm};q)_{\infty}}
{(q^{\frac14}ts_i^{(1)\pm}s_j^{(2)\mp};q)_{\infty}}. 
\end{align}

%D5 U(N)|U(N) vortex
Under the action of S-duality the NS5-type $U(N)|U(N)$ interface maps to the D5-type $U(N)|U(N)$ interface. 
In the latter configuration the whole $U(N)$ gauge group is left unbroken. 
It contains the hypermultiplet transforming in the fundamental representation of the $U(N)$ gauge group. 
The Wilson line in the NS5-type $U(N)|U(N)$ interface is expected to be dual to the vortex line in the D5-type interface. 
However, when the cocharacter is associated with the non-minuscule module, 
the corresponding line defect half-index may not be simply computed for the fixed vortex number 
due to the monopole bubbling effect \cite{Kapustin:2006pk}. 
Here we propose that 
two types of the vortex lines at the D5-type $U(N)|U(N)$ interface are free from the monopole bubbling 
so that the line defect half-indices can be simply computed by specializing the fugacities. 
The first type is the vortex line $V_{(1^k,0^{N-k})}$ that is labeled by the Abelianized vortex number $(1^k,0^{N-k})$, 
the minuscule weight of the $U(N)$ gauge group
\begin{align}
\label{uNuND_V1^k}
&
\langle  V_{(1^k,0^{N-k})} \rangle_{\mathcal{D}}^{U(N)|U(N)}(t;q)
\nonumber\\
&=
q^{\frac{k}{4}}t^{-k}
\frac{1}{k! (N-k)!}
\frac{(q)_{\infty}^{2N}}{(q^{\frac12}t^{\pm 2};q)_{\infty}^N}
\oint \prod_{i=1}^N \frac{ds_i}{2\pi is_i}
\nonumber\\
&\times 
\prod_{i=1}^k 
\prod_{j=k+1}^{N}
\frac{(q^{\frac12}s_i^{\pm}s_j^{\mp};q)_{\infty}(q^{\frac32}s_i^{\pm}s_j^{\mp};q)_{\infty}}
{(qt^2s_i^{\pm}s_j^{\mp};q)_{\infty}(qt^{-2}s_i^{\pm}s_j^{\mp};q)_{\infty}}
\prod_{
\begin{smallmatrix}
1\le i<j\le k\\
k+1\le i<j\le N\\
\end{smallmatrix}
}
\frac{(s_i^{\pm}s_j^{\mp};q)_{\infty}(qs_i^{\pm}s_j^{\mp};q)_{\infty}}
{(q^{\frac12}t^2 s_i^{\pm}s_j^{\mp};q)_{\infty} (q^{\frac12}t^{-2}s_i^{\pm}s_j^{\mp};q)_{\infty}}
\nonumber\\
&\times 
\prod_{i=1}^{k}
\frac{(q^{\frac54}t^{-1}s_i^{\mp};q)_{\infty}}
{(q^{\frac34}ts_i^{\pm};q)_{\infty}}
\prod_{i=k+1}^{N}
\frac{(q^{\frac34}t^{-1}s_i^{\mp};q)_{\infty}}
{(q^{\frac14}ts_i^{\pm};q)_{\infty}}. 
\end{align}
Another type is the vortex line $V_{(2,1^{N-1})}$ labeled by the Abelianized vortex number $(2,1^{N-1})$
\begin{align}
\label{uNuND_V21^N-1}
&
\langle  V_{(2,1^{N-1})} \rangle_{\mathcal{D}}^{U(N)|U(N)}(t;q)
\nonumber\\
&=
q^{\frac{N+1}{4}}t^{-(N+1)}
\frac{1}{(N-1)!}
\frac{(q)_{\infty}^{2N}}{(q^{\frac12}t^{\pm 2};q)_{\infty}^N}
\oint \prod_{i=1}^N \frac{ds_i}{2\pi is_i}
\nonumber\\
&\times 
\prod_{j=2}^{N}
\frac{(q^{\frac12}s_1^{\pm}s_j^{\mp};q)_{\infty}(q^{\frac32}s_1^{\pm}s_j^{\mp};q)_{\infty}}
{(qt^2s_1^{\pm}s_j^{\mp};q)_{\infty}(qt^{-2}s_1^{\pm}s_j^{\mp};q)_{\infty}}
\prod_{2\le i<j\le N}
\frac{(s_i^{\pm}s_j^{\mp};q)_{\infty}(qs_i^{\pm}s_j^{\mp};q)_{\infty}}
{(q^{\frac12}t^2 s_i^{\pm}s_j^{\mp};q)_{\infty} (q^{\frac12}t^{-2}s_i^{\pm}s_j^{\mp};q)_{\infty}}
\nonumber\\
&\times 
\frac{(q^{\frac74}t^{-1}s_1^{\mp};q)_{\infty}}
{(q^{\frac54}ts_1^{\pm};q)_{\infty}}
\prod_{i=2}^{N}
\frac{(q^{\frac34}t^{-1}s_i^{\mp};q)_{\infty}}
{(q^{\frac14}ts_i^{\pm};q)_{\infty}}. 
\end{align}

As strong evidence of the dual descriptions of the two types of the vortex lines, 
we have confirmed that the two expressions (\ref{uNuND_V1^k}) and (\ref{uNuND_V21^N-1}) agree with 
the Wilson line defect half-indices (\ref{uNuNN_W}) for the NS5-type $U(N)|U(N)$ interface 
under the mirror transformation $t$ $\rightarrow$ $t^{-1}$
\begin{align}
\langle W_{(1^k);\overline{(1^k)}} \rangle^{U(N)|U(N)}_{\mathcal{N}}(t;q)
&=
\langle  V_{(1^k,0^{N-k})} \rangle_{\mathcal{D}}^{U(N)|U(N)}(t^{-1};q), \\
\langle W_{(2,1^{N-1});\overline{(2,1^{N-1})}} \rangle^{U(N)|U(N)}_{\mathcal{N}}(t;q)
&=
\langle  V_{(2,1^{N-1})} \rangle_{\mathcal{D}}^{U(N)|U(N)}(t^{-1};q). 
\end{align}

%Coulomb and Higgs limits
Let us examine the Coulomb and Higgs limits of the line defect half-indices. 
The limits of the interface half-index of the NS5-type $U(N)|U(N)$ interface without any insertion of the line defects are given by \cite{Hatsuda:2024uwt}
\begin{align}
{\mathbb{II}^{U(N)|U(N)}_{\mathcal{N}}}^{(C)}(\mathfrak{q})
&=\prod_{n=1}^N \frac{1}{(1-\mathfrak{q}^{2n})^2}, \\
{\mathbb{II}^{U(N)|U(N)}_{\mathcal{N}}}^{(H)}(\mathfrak{q})
&=\prod_{n=1}^N \frac{1}{1-\mathfrak{q}^{2n}}. 
\end{align}
The Coulomb and Higgs limits of the Wilson line defect half-indices (\ref{uNuNN_Wgeneral}) are given by
\begin{align}
\label{uNuNN_WgeneralC}
{\langle \prod_{i=1}^{m} W_{\mathcal{R}^{(1)}_i;\mathcal{R}^{(2)}_i} \rangle^{U(N)|U(N)}_{\mathcal{N}}}^{(C)}(\mathfrak{q})
&=\langle \prod_{i=1}^{m} W_{\mathcal{R}^{(1)}_i}\rangle^{U(N)}_{\textrm{$\frac12$BPS}}(\mathfrak{q})
\times 
\langle \prod_{i=1}^{m} W_{\mathcal{R}^{(2)}_i}\rangle^{U(N)}_{\textrm{$\frac12$BPS}}(\mathfrak{q}), \\
\label{uNuNN_WgeneralH}
{\langle \prod_{i=1}^{m} W_{\mathcal{R}^{(1)}_i;\mathcal{R}^{(2)}_i} \rangle^{U(N)|U(N)}_{\mathcal{N}}}^{(H)}(\mathfrak{q})
&=\frac{1}{(N!)^2} 
\oint \prod_{I=1}^2\prod_{i=1}^N \frac{ds_i^{(I)}}{2\pi is_i^{(I)}}
\frac{
\prod_{I=1}^2
\prod_{i<j}(1-s_i^{(I)\pm}s_j^{(I)\mp})}
{\prod_{i,j}1-\mathfrak{q}s_i^{(1)\pm}s_j^{(2)\mp}}
\nonumber\\
&\times \prod_{i=1}^m \chi_{\mathcal{R}^{(1)}_i}(s^{(1)})\chi_{\mathcal{R}^{(2)}_i}(s^{(2)}). 
\end{align}
We see that the Coulomb limits (\ref{uNuNN_WgeneralC}) factorize into the half-BPS limits of the Wilson line defect indices for $\mathcal{N}=4$ $U(N)$ SYM theories. 
The expression (\ref{uNuNN_WgeneralH}) for the Higgs limits lead to a natural generalization of the weight function for the Hall-Littlewood function. 

In particular, additional degrees of freedom due to an insertion of the single Wilson line 
$W_{\mathcal{R}^{(1)};\mathcal{R}^{(2)}}$ transforming as a pair $(\mathcal{R}^{(1)},\mathcal{R}^{(2)})$ of the representations 
under the $U(N)^{(1)}\times U(N)^{(2)}$ gauge group at the NS5-type interface can survive in the Higgs limit, 
whereas it vanishes in the Coulomb limit. 
For example, when we introduce the Wilson line operator $W_{\tiny \yng(1);\tiny \overline{\yng(1)}}$ at the NS5-type interface 
which transforms as the bifundamental representation under the $U(N)^{(1)}$ $\times$ $U(N)^{(2)}$ gauge group, we obtain 
\begin{align}
\label{uNuNN_1_C}
{\langle \mathcal{W}_{\tiny \yng(1);\tiny \overline{\yng(1)}}\rangle^{U(N)|U(N)}_{\mathcal{N}}}^{(C)}(\mathfrak{q})&=0,\\
\label{uNuNN_1_H}
{\langle \mathcal{W}_{\tiny \yng(1);\tiny \overline{\yng(1)}}\rangle^{U(N)|U(N)}_{\mathcal{N}}}^{(H)}(\mathfrak{q})
&=\mathfrak{q}(1+\mathfrak{q}^2+\cdots+\mathfrak{q}^{2(N-1)}). 
\end{align}
While the Coulomb limit of the defect half-index vanishes, 
the Higgs limit encodes the BPS spectrum of the gauge invariant operators 
formed by the Higgs branch operators and the Wilson line operator $W_{\tiny \yng(1);\tiny \overline{\yng(1)}}$ at the interface. 
%antisym
Let us consider the Wilson line operators transforming in the higher-rank conjugate representations of the gauge groups $U(N)^{(1)}$ and $U(N)^{(2)}$ at the NS5-type interface. 
For the Wilson line $W_{(1^k);\overline{(1^k)}}$ in the rank-$k$ antisymmetric representations, we find 
\begin{align}
\label{uNuNN_W1^k_C}
{\langle \mathcal{W}_{(1^k);\overline{(1^k)}}\rangle^{U(N)|U(N)}_{\mathcal{N}}}^{(C)}(\mathfrak{q})&=0,\\
\label{uNuNN_W1^k_H}
{\langle \mathcal{W}_{(1^k);\overline{(1^k)}}\rangle^{U(N)|U(N)}_{\mathcal{N}}}^{(H)}(\mathfrak{q})
&=\mathfrak{q}^{k}
\left(
\begin{matrix}
N\\
k\\
\end{matrix}
\right)_{\mathfrak{q}^2}, 
\end{align}
where
\begin{align}
\left(
\begin{matrix}
m\\
r\\
\end{matrix}
\right)_{\mathfrak{q}}
&=\frac{(\mathfrak{q})_{m}}{(\mathfrak{q})_r(\mathfrak{q})_{m-r}}
=\frac{(1-\mathfrak{q}^m)(1-\mathfrak{q}^{m-1})\cdots (1-\mathfrak{q}^{m-r+1})}{(1-\mathfrak{q})(1-\mathfrak{q}^2)\cdots (1-\mathfrak{q}^r)}
\end{align}
are the $q$-binomial coefficients. 
%adjoint
For the adjoint representation we get
\begin{align}
{\langle \mathcal{W}_{(2,1^{N-1});\overline{(2,1^{N-1})}}\rangle^{U(N)|U(N)}_{\mathcal{N}}}^{(C)}(\mathfrak{q})&=0,\\
\label{uNuNN_Wadj_H}
{\langle \mathcal{W}_{(2,1^{N-1});\overline{(2,1^{N-1})}}\rangle^{U(N)|U(N)}_{\mathcal{N}}}^{(H)}(\mathfrak{q})
&=\mathfrak{q}^{N+1}
\left(
\begin{matrix}
N\\
1\\
\end{matrix}
\right)_{\mathfrak{q}^2}. 
\end{align}
%sym
For the rank-$k$ symmetric representation we obtain
\begin{align}
\label{uNuNN_Wk_C}
{\langle \mathcal{W}_{(k);\overline{(k)}}\rangle^{U(N)|U(N)}_{\mathcal{N}}}^{(C)}(\mathfrak{q})&=0,\\
\label{uNuNN_Wk_H}
{\langle \mathcal{W}_{(k);\overline{(k)}}\rangle^{U(N)|U(N)}_{\mathcal{N}}}^{(H)}(\mathfrak{q})
&=\mathfrak{q}^{k}
\left(
\begin{matrix}
N+k-1\\
k\\
\end{matrix}
\right)_{\mathfrak{q}^2}. 
\end{align}
%relation to the Schur line defect index
More generally, we observe that these normalized line defect half-indices in the Higgs limit are identical 
to the normalized two-point functions of the Wilson lines in the conjugate irreducible representations of 4d $\mathcal{N}=4$ $U(N)$ SYM theory 
in the half-BPS limit \cite{Hatsuda:2023iwi,Hatsuda:2023imp} up to the overall factor\footnote{Here we have defined the variable $\mathfrak{q}$ as the ratio $\mathfrak{q}=q^{1/4}t$, 
whereas the ratio $q^{1/2}t^2$ was used in \cite{Hatsuda:2023iwi,Hatsuda:2023imp}. }
\begin{align}
{\langle \mathcal{W}_{\lambda;\overline{\lambda}}\rangle^{U(N)|U(N)}_{\mathcal{N}}}^{(H)}(\mathfrak{q})
&=\mathfrak{q}^{|\lambda|} \langle \mathcal{W}_{\lambda}\mathcal{W}_{\overline{\lambda}}\rangle^{U(N)}_{\textrm{$\frac12$BPS}}(\mathfrak{q}), 
\end{align}
where $\lambda$ is the Young diagram associated with the irreducible representation of $U(N)$. 
$|\lambda|$ is the number of boxes of $\lambda$. 
See section \ref{sec_uNuM_W} for the proof. 

%one-sided Wilson
One can consider more general configurations as junctions of several line operators at the interfaces. 
For example, given a pair of the Wilson lines transforming in the conjugate representations $\lambda$ and $\overline{\lambda}$ 
under one of the gauge groups, say $U(N)^{(1)}$, we find 
\begin{align}
{\langle \mathcal{W}_{\lambda;\emptyset}\mathcal{W}_{\overline{\lambda};\emptyset} \rangle^{U(N)|U(N)}_{\mathcal{N}}}^{(C)}(\mathfrak{q})
&=\langle \mathcal{W}_{\lambda}\mathcal{W}_{\overline{\lambda}}\rangle^{U(N)}_{\textrm{$\frac12$BPS}}(\mathfrak{q}), \\
{\langle \mathcal{W}_{\lambda;\emptyset}\mathcal{W}_{\overline{\lambda};\emptyset} \rangle^{U(N)|U(N)}_{\mathcal{N}}}^{(H)}(\mathfrak{q})
&=\langle \mathcal{W}_{\lambda}\mathcal{W}_{\overline{\lambda}}\rangle^{U(N)}_{\textrm{$\frac12$BPS}}(\mathfrak{q}). 
\end{align}
The both Coulomb and Higgs limits of the normalized defect half-indices are equal to the normalized two-point function of the Wilson lines of bulk theory in the half-BPS limit \cite{Hatsuda:2023iwi,Hatsuda:2023imp}. 

Next consider the Coulomb and Higgs limits of the vortex line defect half-indices. 
The matrix integral (\ref{uNuND_V1^k}) becomes
\begin{align}
{\langle  \mathcal{V}_{(1^k,0^{N-k})} \rangle_{\mathcal{D}}^{U(N)|U(N)}}^{(C)}(\mathfrak{q})
&=\mathfrak{q}^{k}
\left(
\begin{matrix}
N\\
k\\
\end{matrix}
\right)_{\mathfrak{q}^2}, 
\end{align}
where we have used the inner product of the Hall-Littlewood functions $P_{\mu}(s;\mathfrak{t})$
\begin{align}
\label{HL_inner}
\frac{1}{N!}
\oint \prod_{i=1}^N \frac{ds_i}{2\pi is_i}
\frac{\prod_{i\neq j} 1-\frac{s_i}{s_j}}{\prod_{i,j}1-\mathfrak{t}\frac{s_i}{s_j}}
P_{\mu}(s;\mathfrak{t})P_{\lambda}(s^{-1};\mathfrak{t})
&=\frac{\delta_{\mu\lambda}}{(\mathfrak{t};\mathfrak{t})_{N-l(\mu)}\prod_{j\ge 1}(\mathfrak{t};\mathfrak{t})_{m_j(\mu)}},
\end{align}
where $l(\mu)$ is the length of the Young diagram $\mu$ and $m_j(\mu)$ the multiplicity of $\mu$. 
In fact, this precisely agrees with the expression (\ref{uNuNN_W1^k_H}) for the dual line defect half-index. 
Similarly, the integral (\ref{uNuND_V21^N-1}) reduce to 
\begin{align}
\label{uNuND_V21^N-1C}
\langle  \mathcal{V}_{(2,1^{N-1})} \rangle_{\mathcal{D}}^{U(N)|U(N)}(\mathfrak{q})
&=\mathfrak{q}^{N+1}
\left(
\begin{matrix}
N\\
1\\
\end{matrix}
\right)_{\mathfrak{q}^2}. 
\end{align}
Again this is equivalent to the dual Wilson line defect half-index (\ref{uNuNN_Wadj_H}).

%%%%%%%%%%%%%%%%%%%%%%%%%%%%%%%%%%%%%%%%
\subsection{Vortex-Wilson}
%%%%%%%%%%%%%%%%%%%%%%%%%%%%%%%%%%%%%%%%
%NS5 U(N)|U(N) vortex
One can introduce the vortex lines at the NS5-type $U(N)|U(N)$ interface 
which are labeled by a pair $(k^{(1)},k^{(2)})$ of the Abelianized vortex numbers. 
Again when the cocharacter is associated with the non-minuscule module, 
the configuration may not be stable due to the monopole bubbling effect \cite{Kapustin:2006pk}. 
Here we propose that the defect half-indices for two types of the vortex lines 
associated with the cocharacters $(1^{k},0^{N-k})$ and $(2,1^{N-1})$ can be computed by making the specializations of the fugacities. 
For $k^{(1)}$ $=$ $(1^{k},0^{N-k})$ and $k^{(2)}=0$ we have 
\begin{align}
\label{uNuNN_V1^k}
&
\langle  V_{(1^k,0^{N-k});0} \rangle_{\mathcal{N}}^{U(N)|U(N)}(t;q)
\nonumber\\
&=
q^{\frac{k^2}{4}}t^{-k^2}
\frac{1}{k!(N-k)!}\frac{(q)_{\infty}^N}{(q^{\frac12}t^{-2};q)_{\infty}^N}
\oint\prod_{i=1}^N \frac{ds_i^{(1)}}{2\pi is_i^{(1)}}
\nonumber\\
&\times 
\prod_{i=1}^k
\prod_{j=k+1}^N
\frac{(q^{\frac12}s_i^{(1)\pm}s_j^{(1)\mp};q)_{\infty}}
{(qt^{-2}s_i^{(1)\pm}s_j^{(1)\mp};q)_{\infty}}
\prod_{
\begin{smallmatrix}
1\le i<j\le k\\
k+1\le i<j\le N\\
\end{smallmatrix}
}
\frac{(s_i^{(1)\pm}s_j^{(1)\mp};q)_{\infty}}
{(q^{\frac12}t^{-2}s_i^{(1)\pm}s_j^{(1)\mp};q)_{\infty}}
\nonumber\\
&\times 
\frac{1}{N!}\frac{(q)_{\infty}^N}{(q^{\frac12}t^{-2};q)_{\infty}^N}
\oint \prod_{i=1}^N \frac{ds_i^{(2)}}{2\pi is_i^{(2)}}
\prod_{i<j}
\frac{(s_i^{(2)\pm}s_j^{(2)\mp};q)_{\infty}}
{(q^{\frac12}t^{-2}s_i^{(2)\pm}s_j^{(2)\mp};q)_{\infty}}
\nonumber\\
&\times 
\prod_{i=1}^k \prod_{j=1}^N 
\frac{(q^{\frac54}t^{-1}s_i^{(1)\mp}s_j^{(2)\pm};q)_{\infty}}
{(q^{\frac34}ts_i^{(1)\pm}s_j^{(2)\mp};q)_{\infty}}
\prod_{i=k+1}^{N}
\prod_{j=1}^{N}
\frac{(q^{\frac34}t^{-1}s_i^{(1)\mp}s_j^{(2)\pm};q)_{\infty}}
{(q^{\frac14}ts_i^{(1)\pm}s_j^{(2)\mp};q)_{\infty}}. 
\end{align}
Next consider the other type with $k^{(1)}$ $=$ $(2,1^{N-1})$ and $k^{(2)}=0$. 
The line defect half-index takes the form 
\begin{align}
\label{uNuNN_V21^N-1}
&
\langle  V_{(2,1^{N-1});0} \rangle_{\mathcal{N}}^{U(N)|U(N)}(t;q)
\nonumber\\
&=
q^{\frac{N^2+1}{4}}t^{-(N^2+1)}
\frac{1}{(N-1)!}\frac{(q)_{\infty}^N}{(q^{\frac12}t^{-2};q)_{\infty}^N}
\oint\prod_{i=1}^N \frac{ds_i^{(1)}}{2\pi is_i^{(1)}}
\nonumber\\
&\times 
\prod_{j=2}^N
\frac{(q^{\frac12}s_1^{(1)\pm}s_j^{(1)\mp};q)_{\infty}}
{(qt^{-2}s_1^{(1)\pm}s_j^{(1)\mp};q)_{\infty}}
\prod_{
2\le i<j\le N
}
\frac{(s_i^{(1)\pm}s_j^{(1)\mp};q)_{\infty}}
{(q^{\frac12}t^{-2}s_i^{(1)\pm}s_j^{(1)\mp};q)_{\infty}}
\nonumber\\
&\times 
\frac{1}{N!}\frac{(q)_{\infty}^N}{(q^{\frac12}t^{-2};q)_{\infty}^N}
\oint \prod_{i=1}^N \frac{ds_i^{(2)}}{2\pi is_i^{(2)}}
\prod_{i<j}
\frac{(s_i^{(2)\pm}s_j^{(2)\mp};q)_{\infty}}
{(q^{\frac12}t^{-2}s_i^{(2)\pm}s_j^{(2)\mp};q)_{\infty}}
\nonumber\\
&\times 
\prod_{j=1}^N 
\frac{(q^{\frac74}t^{-1}s_1^{(1)\mp}s_j^{(2)\pm};q)_{\infty}}
{(q^{\frac54}ts_1^{(1)\pm}s_j^{(2)\mp};q)_{\infty}}
\prod_{i=2}^{N}
\prod_{j=1}^{N}
\frac{(q^{\frac54}t^{-1}s_i^{(1)\mp}s_j^{(2)\pm};q)_{\infty}}
{(q^{\frac34}ts_i^{(1)\pm}s_j^{(2)\mp};q)_{\infty}}. 
\end{align}

Furthermore, we expect that the line defect half-indices of the vortex lines 
with pairs of the non-zero equal Abelianized vortex numbers can be obtained in an analogous manner. 
With $k^{(1)}$ $=$ $k^{(2)}$ $=$ $(1^k,0^{N-k})$ 
the vortex line defect half-index is 
\begin{align}
\label{uNuNN_V1^kV1^k}
&
\langle  V_{(1^k,0^{N-k});(1^{k},0^{N-k})} \rangle_{\mathcal{N}}^{U(N)|U(N)}(t;q)
\nonumber\\
&=
\frac{1}{k!(N-k)!}\frac{(q)_{\infty}^N}{(q^{\frac12}t^{-2};q)_{\infty}^N}
\oint\prod_{i=1}^N \frac{ds_i^{(1)}}{2\pi is_i^{(1)}}
\nonumber\\
&\times 
\prod_{i=1}^k
\prod_{j=k+1}^N
\frac{(q^{\frac12}s_i^{(1)\pm}s_j^{(1)\mp};q)_{\infty}}
{(qt^{-2}s_i^{(1)\pm}s_j^{(1)\mp};q)_{\infty}}
\prod_{
\begin{smallmatrix}
1\le i<j\le k\\
k+1\le i<j\le N\\
\end{smallmatrix}
}
\frac{(s_i^{(1)\pm}s_j^{(1)\mp};q)_{\infty}}
{(q^{\frac12}t^{-2}s_i^{(1)\pm}s_j^{(1)\mp};q)_{\infty}}
\nonumber\\
&\times 
\frac{1}{k!(N-k)!}\frac{(q)_{\infty}^N}{(q^{\frac12}t^{-2};q)_{\infty}^N}
\oint \prod_{i=1}^N \frac{ds_i^{(2)}}{2\pi is_i^{(2)}}
\nonumber\\
&\times 
\prod_{i=1}^k
\prod_{j=k+1}^N
\frac{(q^{\frac12}s_i^{(2)\pm}s_j^{(2)\mp};q)_{\infty}}
{(qt^{-2}s_i^{(2)\pm}s_j^{(1)\mp};q)_{\infty}}
\prod_{
\begin{smallmatrix}
1\le i<j\le k\\
k+1\le i<j\le N\\
\end{smallmatrix}
}
\frac{(s_i^{(1)\pm}s_j^{(1)\mp};q)_{\infty}}
{(q^{\frac12}t^{-2}s_i^{(1)\pm}s_j^{(1)\mp};q)_{\infty}}
\nonumber\\
&\times 
\prod_{i=1}^k \prod_{j=k+1}^N 
\frac{(q^{\frac54}t^{-1}s_i^{(1)\mp}s_j^{(2)\pm};q)_{\infty}}
{(q^{\frac34}ts_i^{(1)\pm}s_j^{(2)\mp};q)_{\infty}}
\prod_{i=k+1}^N \prod_{j=1}^k 
\frac{(q^{\frac54}t^{-1}s_i^{(1)\mp}s_j^{(2)\pm};q)_{\infty}}
{(q^{\frac34}ts_i^{(1)\pm}s_j^{(2)\mp};q)_{\infty}}
\nonumber\\
&\times 
\prod_{i=1}^{k}
\prod_{j=1}^{k}
\frac{(q^{\frac34}t^{-1}s_i^{(1)\mp}s_j^{(2)\pm};q)_{\infty}}
{(q^{\frac14}ts_i^{(1)\pm}s_j^{(2)\mp};q)_{\infty}}
\prod_{i=k+1}^{N}
\prod_{j=k+1}^{N}
\frac{(q^{\frac34}t^{-1}s_i^{(1)\mp}s_j^{(2)\pm};q)_{\infty}}
{(q^{\frac14}ts_i^{(1)\pm}s_j^{(2)\mp};q)_{\infty}}. 
\end{align}
For $k^{(1)}$ $=$ $k^{(2)}$ $=$ $(2,1^{N-1})$ 
the vortex line defect half-index is equivalent to that with $k^{(1)}$ $=$ $k^{(2)}$ $=$ $(1,0^{N-1})$, 
which is obtained from (\ref{uNuNN_V1^kV1^k}) by setting $k$ to $1$. 

%D5 U(N)|U(N) Wilson
The vortex lines in the NS5-type $U(N)|U(N)$ interface 
will map to the Wilson lines in the D5-type $U(N)|U(N)$ interface under S-duality. 
As the D5-type $U(N)|U(N)$ interface keeps the $U(N)$ gauge group, 
the dynamical Wilson line $W_{\mathcal{R}}$ is labeled by the representation $\mathcal{R}$ of the $U(N)$. 
The line defect half-indices for the Wilson line operators $W_{\mathcal{R}_i}$, $i=1,\cdots, k$ takes the form 
\begin{align}
\label{uNuND_W}
&
\langle W_{\mathcal{R}_1}\cdots W_{\mathcal{R}_k} \rangle_{\mathcal{D}}^{U(N)|U(N)}(t;q)
\nonumber\\
&=\frac{1}{N!}
\frac{(q)_{\infty}^{2N}}{(q^{\frac12}t^{\pm2};q)_{\infty}^N}
\oint \prod_{i=1}^N 
\frac{ds_i}{2\pi is_i}
\prod_{i<j}
\frac{(s_i^{\pm}s_j^{\mp};q)_{\infty}(qs_i^{\pm}s_j^{\mp};q)_{\infty}}
{(q^{\frac12}t^2 s_i^{\pm}s_j^{\mp};q)_{\infty} (q^{\frac12}t^{-2}s_i^{\pm}s_j^{\mp};q)_{\infty}}
\nonumber\\
&\times 
\prod_{i=1}^N 
\frac{(q^{\frac34}t^{-1}s_i^{\mp};q)_{\infty}}
{(q^{\frac14}ts_i^{\pm};q)_{\infty}}
\prod_{j=1}^{k}
\chi_{\mathcal{R}_j}(s). 
\end{align}

We find that 
the proposed vortex line defect half-indices 
(\ref{uNuNN_V1^k}), (\ref{uNuNN_V21^N-1}) and (\ref{uNuNN_V1^kV1^k}) precisely 
match with the Wilson line defect half-indices (\ref{uNuND_W}) for the D5-type interface upon $t$ $\rightarrow$ $t^{-1}$ as follows: 
\begin{align}
\label{uNuNN_V=uNuND_W1a}
\langle  V_{(1^k,0^{N-k});0} \rangle_{\mathcal{N}}^{U(N)|U(N)}(t;q)
&=
\langle W_{(1^k)} \rangle_{\mathcal{D}}^{U(N)|U(N)}(t^{-1};q), \\
\label{uNuNN_V=uNuND_W1b}
\langle  V_{(2,1^{N-1});0} \rangle_{\mathcal{N}}^{U(N)|U(N)}(t;q)
&=
\langle W_{(2,1^{N-1})} \rangle_{\mathcal{D}}^{U(N)|U(N)}(t^{-1};q), \\
\label{uNuNN_V=uNuND_W2a}
\langle  V_{(1^k,0^{N-k});(1^k,0^{N-k})} \rangle_{\mathcal{N}}^{U(N)|U(N)}(t;q)
&=
\langle W_{(1^k)}W_{\overline{(1^k)}} \rangle_{\mathcal{D}}^{U(N)|U(N)}(t^{-1};q). 
\end{align}

%exact
Here we propose the closed-form expressions for the matrix integral (\ref{uNuND_W}) 
of the one-point function of the Wilson line in the irreducible representation $\mathcal{R}$ for the D5-type $U(N)|U(N)$ interface. 
We find that 
\begin{align}
\label{uNuND_W_expand}
&
\langle W_{\mathcal{R}} \rangle_{\mathcal{D}}^{U(N)|U(N)}(t^{-1};q)
\nonumber\\
&=\sum_{n_1=0}^{\infty}\cdots \sum_{n_N=0}^{\infty}
\frac{
(q^{1+n_1};q)_{\infty}^2
(q^{\frac32+n_1+n_2}t^{-2};q)_{\infty}^2
\cdots
(q^{\frac{N+1}{2}+n_1+\cdots+n_N}t^{-2(N-1)};q)_{\infty}^2
}
{
(q^{\frac12+n_1}t^{-2};q)_{\infty}^2
(q^{1+n_1+n_2}t^{-4};q)_{\infty}^2
\cdots
(q^{\frac{N}{2}+n_1+\cdots+n_N}t^{-2N};q)_{\infty}^2
}
\nonumber\\
&\times 
(q^{\frac12}t^2)^{Nn_1+(N-1)n_2+\cdots+n_N}
\chi_{\mathcal{R}}
\left(
s_i=q^{\sum_{j=1}^i n_j+\frac{2i-1}{4}}t^{-(2i-1)}
\right). 
\end{align}
Here we have flipped a sign of the power of $t$. 
This generalizes the vortex expansions in \cite{Gaiotto:2019jvo,Hatsuda:2024uwt} 
as it reduces to those of the interface half-indices when $\mathcal{R}$ is trivial. 

For example, for the dual pairs (\ref{uNuNN_V=uNuND_W1a}) with $\mathcal{R}$ $=$ $\tiny \yng(1)$ and $(1^N)$, 
the line defect half-indices can be written as
\begin{align}
&
\langle  V_{(1,0^{N-1});0} \rangle_{\mathcal{N}}^{U(N)|U(N)}(t;q)
=\langle W_{\tiny \yng(1)} \rangle_{\mathcal{D}}^{U(N)|U(N)}(t^{-1};q)
\nonumber\\
&=\sum_{n_1=0}^{\infty}\cdots \sum_{n_N=0}^{\infty}
\frac{
(q^{1+n_1};q)_{\infty}^2
(q^{\frac32+n_1+n_2}t^{-2};q)_{\infty}^2
\cdots
(q^{\frac{N+1}{2}+n_1+\cdots+n_N}t^{-2(N-1)};q)_{\infty}^2
}
{
(q^{\frac12+n_1}t^{-2};q)_{\infty}^2
(q^{1+n_1+n_2}t^{-4};q)_{\infty}^2
\cdots
(q^{\frac{N}{2}+n_1+\cdots+n_N}t^{-2N};q)_{\infty}^2
}
\nonumber\\
&\times 
(q^{\frac12}t^2)^{Nn_1+(N-1)n_2+\cdots+n_N}
\Biggl[
q^{n_1+\frac14}t^{-1}
+q^{n_1+n_2+\frac34}t^{-3}
+\cdots
+q^{n_1+\cdots+n_N+\frac{2N-1}{4}}t^{-(2N-1)}
\Biggr]
\end{align}
and 
\begin{align}
&
\langle  V_{(1^{N});0} \rangle_{\mathcal{N}}^{U(N)|U(N)}(t;q)
=\langle W_{(1^N)} \rangle_{\mathcal{D}}^{U(N)|U(N)}(t^{-1};q)
\nonumber\\
&=\sum_{n_1=0}^{\infty}\cdots \sum_{n_N=0}^{\infty}
\frac{
(q^{1+n_1};q)_{\infty}^2
(q^{\frac32+n_1+n_2}t^{-2};q)_{\infty}^2
\cdots
(q^{\frac{N+1}{2}+n_1+\cdots+n_N}t^{-2(N-1)};q)_{\infty}^2
}
{
(q^{\frac12+n_1}t^{-2};q)_{\infty}^2
(q^{1+n_1+n_2}t^{-4};q)_{\infty}^2
\cdots
(q^{\frac{N}{2}+n_1+\cdots+n_N}t^{-2N};q)_{\infty}^2
}
\nonumber\\
&\times 
(q^{\frac12}t^2)^{Nn_1+(N-1)n_2+\cdots+n_N}
q^{Nn_1+(N-1)n_2+\cdots+n_{N}+\frac{N^2}{4}}t^{-N^2}. 
\end{align}

%Coulomb and Higgs limits
Let us look at the Coulomb and Higgs limits of the vortex line defect half-indices for the NS5-type interface. 
The matrix integral (\ref{uNuNN_V1^k}) factorizes into three parts of the integrals in the Coulomb limit, whereas it vanishes in the Higgs limit. 
Making use of the inner product (\ref{HL_inner}) of the Hall-Littlewood functions, 
we find the normalized one-point functions 
\begin{align}
\label{uNuNN_V1^k_C}
{\langle \mathcal{V}_{(1^k,0^{N-k});0} \rangle_{\mathcal{N}}^{U(N)|U(N)}}^{(C)}(\mathfrak{q})
&=\mathfrak{q}^{k^2}
\left(
\begin{matrix}
N\\
k\\
\end{matrix}
\right)_{\mathfrak{q}^2}, \\
{\langle \mathcal{V}_{(1^k,0^{N-k});0} \rangle_{\mathcal{N}}^{U(N)|U(N)}}^{(H)}(\mathfrak{q})
&=0. 
\end{align}
Similarly, the Coulomb limit of the matrix integral (\ref{uNuNN_V21^N-1}) keeps only two parts of the integrals. 
We find
\begin{align}
\label{uNuNN_Vadj_C}
{\langle \mathcal{V}_{(2,1^{N-1});0} \rangle_{\mathcal{N}}^{U(N)|U(N)}}^{(C)}(\mathfrak{q})
&=\mathfrak{q}^{N^2+1}\left(
\begin{matrix}
N\\
1\\
\end{matrix}
\right)_{\mathfrak{q}^2}, \\
{\langle \mathcal{V}_{(2,1^{N-1});0} \rangle_{\mathcal{N}}^{U(N)|U(N)}}^{(H)}(\mathfrak{q})
&=0. 
\end{align}
For the expression (\ref{uNuNN_V1^kV1^k}) for the vortex lines of pairs of the non-trivial vortex numbers $(1^k,0^{N-k})$, 
we find the normalized Coulomb and Higgs limits 
\begin{align}
\label{uNuNN_V1^kV1^k_C}
{\langle  \mathcal{V}_{(1^k,0^{N-k});(1^{k},0^{N-k})} \rangle_{\mathcal{N}}^{U(N)|U(N)}}^{(C)}(\mathfrak{q})
&=\left(
\begin{matrix}
N\\
k\\
\end{matrix}
\right)_{\mathfrak{q}^2}^2, \\ 
\label{uNuNN_V1^kV1^k_H}
{\langle  \mathcal{V}_{(1^k,0^{N-k});(1^{k},0^{N-k})} \rangle_{\mathcal{N}}^{U(N)|U(N)}}^{(H)}(\mathfrak{q})
&=\left(
\begin{matrix}
N\\
k\\
\end{matrix}
\right)_{\mathfrak{q}^2}. 
\end{align}

On the other hand, the Coulomb and Higgs limits of the Wilson line defect half-indices (\ref{uNuND_W}) for the D5-type interface 
reduce to 
\begin{align}
\label{uNuND_W_C}
{\langle W_{\mathcal{R}_1}\cdots W_{\mathcal{R}_k} \rangle_{\mathcal{D}}^{U(N)|U(N)}}^{(C)}(\mathfrak{q})
&={\langle W_{\mathcal{R}_1}\cdots W_{\mathcal{R}_k} \rangle^{U(N)}}_{\textrm{$\frac12$BPS}}(\mathfrak{q}), \\
\label{uNuND_W_H}
{\langle W_{\mathcal{R}_1}\cdots W_{\mathcal{R}_k} \rangle_{\mathcal{D}}^{U(N)|U(N)}}^{(H)}(\mathfrak{q})
&=\frac{1}{N!}\frac{1}{(1-\mathfrak{q}^2)^N}
\oint \prod_{i=1}^N \frac{ds_i}{2\pi is_i}
\prod_{i<j}
\frac{1-s_i^{\pm}s_j^{\mp}}{1-\mathfrak{q}^2 s_i^{\pm}s_j^{\mp}}
\nonumber\\
&\times 
\prod_{i=1}^N \frac{1}{1-\mathfrak{q}s_i^{\pm}}
\prod_{j=1}^{k}
\chi_{\mathcal{R}_j}(s). 
\end{align}
We see that the Coulomb limits (\ref{uNuND_W_C}) are simply identified with the half-BPS limits of the line defect indices for $\mathcal{N}=4$ $U(N)$ SYM theories. 

For the one-point function the Coulomb limits always vanish, just like the line defect indices. 
However, the Higgs limits of the one-point functions (\ref{uNuND_W_H}) is non-trivial. 
According to the closed-form formula (\ref{uNuND_W_expand}), 
the normalized Higgs limits are simply given by
\begin{align}
\label{uNuND_W1pt_H}
{\langle \mathcal{W}_{\mathcal{R}} \rangle_{\mathcal{D}}^{U(N)|U(N)}}^{(H)}(\mathfrak{q})
&=\chi_{\mathcal{R}}(s_1=\mathfrak{q},s_2=\mathfrak{q}^3,\cdots,s_N=\mathfrak{q}^{2N-1}). 
\end{align}
For example, when $\mathcal{R}$ $=$ $(1)$, $(1^k)$ and $(k)$, the expression (\ref{uNuND_W1pt_H}) reduces to 
\begin{align}
\label{uNuND_W1ptfund_H}
{\langle \mathcal{W}_{(1)} \rangle_{\mathcal{D}}^{U(N)|U(N)}}^{(H)}(\mathfrak{q})
&=\mathfrak{q}(1+\mathfrak{q}^2+\cdots+\mathfrak{q}^{2(N-1)}), \\
\label{uNuND_W1ptasym_H}
{\langle \mathcal{W}_{(1^k)} \rangle_{\mathcal{D}}^{U(N)|U(N)}}^{(H)}(\mathfrak{q})
&=\mathfrak{q}^{k^2}
\left(
\begin{matrix}
N\\
k\\
\end{matrix}
\right)_{\mathfrak{q}^2}, \\
\label{uNuND_W1ptsym_H}
{\langle \mathcal{W}_{(k)} \rangle_{\mathcal{D}}^{U(N)|U(N)}}^{(H)}(\mathfrak{q})
&=\mathfrak{q}^{k}
\left(
\begin{matrix}
N+k-1\\
k\\
\end{matrix}
\right)_{\mathfrak{q}^2}. 
\end{align}
The expressions (\ref{uNuND_W1ptfund_H}) and (\ref{uNuND_W1ptasym_H}) are compatible with the conjectured duality relation (\ref{uNuNN_V=uNuND_W1a})
as they are equal to the Coulomb limits (\ref{uNuNN_V1^k_C}) with $k=1$ and general $k$. 
Note that the expressions (\ref{uNuND_W1ptfund_H}), (\ref{uNuND_W1ptasym_H}) and (\ref{uNuND_W1ptsym_H}) 
agree with the normalized line defect half-indices (\ref{uNuNN_1_H}), (\ref{uNuNN_W1^k_H}) and (\ref{uNuNN_Wk_H}) 
in the Higgs limits for the Wilson lines at the NS5-type interface up to the overall factor of $\mathfrak{q}$. 

For the two-point functions, the both Coulomb and Higgs limits can be non-trivial. 
For example, for a pair of the Wilson lines in the rank-$k$ antisymmetric representations, 
they are given by 
\begin{align}
\label{uNuND_W2ptasym_C}
{\langle \mathcal{W}_{(1^k)}\mathcal{W}_{\overline{(1^k)}} \rangle_{\mathcal{D}}^{U(N)|U(N)}}^{(C)}(\mathfrak{q})
&=\left(
\begin{matrix}
N\\
k\\
\end{matrix}
\right)_{\mathfrak{q}^2}, \\
\label{uNuND_W2ptasym_H}
{\langle \mathcal{W}_{(1^k)}\mathcal{W}_{\overline{(1^k)}} \rangle_{\mathcal{D}}^{U(N)|U(N)}}^{(H)}(\mathfrak{q})
&=\left(
\begin{matrix}
N\\
k\\
\end{matrix}
\right)_{\mathfrak{q}^2}^2. 
\end{align}
In fact, the expressions (\ref{uNuND_W2ptasym_C}) and (\ref{uNuND_W2ptasym_H}) 
agree with the dual vortex line defect half-indices (\ref{uNuNN_V1^kV1^k_H}) and  (\ref{uNuNN_V1^kV1^k_C}) respectively.

%%%%%%%%%%%%%%%%%%%%%%%%%%%%%%%%%%%%%%%%
\section{$U(N)|U(M)$}
\label{sec_uNuM}
%%%%%%%%%%%%%%%%%%%%%%%%%%%%%%%%%%%%%%%%

%%%%%%%%%%%%%%%%%%%%%%%%%%%%%%%%%%%%%%%%
\subsection{Wilson-vortex}
\label{sec_uNuM_W}
%%%%%%%%%%%%%%%%%%%%%%%%%%%%%%%%%%%%%%%%
%NS5 U(N)|U(M) Wilson
Let us consider the NS5-type $U(N)|U(M)$ interface with unequal ranks $N>M>1$. 
It contains $U(N)^{(1)}$ $\times$ $U(M)^{(2)}$ gauge group 
that is coupled to the hypermultiplet transforming in the bifundamental representation $(\mathbf{N},\overline{\mathbf{M}})$ $\oplus$ $(\overline{\mathbf{N}},\mathbf{M})$. 
The Wilson line $W_{\mathcal{R}^{(1)};\mathcal{R}^{(2)}}$ 
at the NS5-type interface transforms as a pair $(\mathcal{R}^{(1)},\mathcal{R}^{(2)})$ 
of the representations under the $U(N)^{(1)}$ $\times$ $U(M)^{(2)}$ gauge group. 
The line defect half-index of a junction of the Wilson lines $W_{\mathcal{R}^{(1)}_i;\mathcal{R}^{(2)}_i}$, with $i=1,\cdots, m$, 
can be evaluated by introducing the corresponding characters 
in the integrand of (\ref{uNuMN})
\begin{align}
\label{uNuMN_W}
&
\langle \prod_{i=1}^{m} W_{\mathcal{R}^{(1)}_i;\mathcal{R}^{(2)}_i} \rangle^{U(N)|U(M)}_{\mathcal{N}}(t;q)
\nonumber\\
&=\frac{1}{N!}\frac{(q)_{\infty}^N}{(q^{\frac12}t^{-2};q)_{\infty}^N}
\oint 
\prod_{i=1}^N
\frac{ds_i^{(1)}}{2\pi is_i^{(1)}}
\prod_{i<j}
\frac{(s_i^{(1)\pm}s_j^{(1)\mp};q)_{\infty}}
{(q^{\frac12}t^{-2}s_i^{(1)\pm}s_j^{(1)\mp};q)_{\infty}}
\prod_{i=1}^m 
\chi_{\mathcal{R}^{(1)}_i}(s^{(1)})
\nonumber\\
&\times 
\frac{1}{M!}\frac{(q)_{\infty}^M}{(q^{\frac12}t^{-2};q)_{\infty}^M}
\oint 
\prod_{i=1}^M
\frac{ds_i^{(2)}}{2\pi is_i^{(2)}}
\prod_{i<j}
\frac{(s_i^{(2)\mp}s_j^{(2)\pm};q)_{\infty}}
{(q^{\frac12}t^{-2}s_i^{(2)\pm}s_j^{(2)\mp};q)_{\infty}}
\prod_{i=1}^m
\chi_{\mathcal{R}^{(2)}_i}(s^{(2)})
\nonumber\\
&\times 
\prod_{i=1}^{N}
\prod_{j=1}^{M}
\frac{(q^{\frac34}t^{-1}s_i^{(1)\mp}s_j^{(2)\pm};q)_{\infty}}
{(q^{\frac14}ts_i^{(1)\mp}s_j^{(2)\pm};q)_{\infty}}. 
\end{align}
It is very hard to evaluate this integral exactly.
Interestingly, for $m=1$, $\mathcal{R}^{(1)}_1=(1^{k})$ and $\mathcal{R}^{(2)}_1=\overline{(1^k)}$, we can use the procedure in \cite{Hatsuda:2025mvj} by using Macdonald polynomials. Then we find the following combinatorial sum formula:
\begin{align}
\langle W_{(1^{k});\overline{(1^k)}}\rangle^{U(N)|U(M)}_{\mathcal{N}}(t;q)
&=\frac{(q)_\infty^2}{(q^{\frac{1}{2}}t^{-2};q^{\frac{1}{2}}t^{-2})_N (q^{\frac{N}{2}+1}t^{-2N};q)_\infty (q^{\frac{1}{2}}t^{-2};q^{\frac{1}{2}}t^{-2})_M (q^{\frac{M}{2}+1}t^{-2M};q)_\infty} \nonumber \\
&\quad\times\sum_{\ell(\lambda) \leq M} (q^{\frac{1}{2}} t^2)^{|\lambda|+\frac{k}{2}}\sum_{\mu \in V_M^k(\lambda)}\varphi_{\mu/\lambda}'(q, q^{\frac{1}{2}}t^{-2}) \psi_{\mu/\lambda}'(q,q^{\frac{1}{2}}t^{-2})\nonumber \\
&\quad \times \prod_{i=1}^{\ell(\mu)} \frac{(q^{\frac{N-i+1}{2}} t^{-2N+2i-2};q)_{\mu_i}(q^{\frac{M-i+1}{2}} t^{-2M+2i-2};q)_{\mu_i}}{(q^{\frac{N-i}{2}+1} t^{-2N+2i};q)_{\mu_i}(q^{\frac{M-i}{2}+1} t^{-2M+2i};q)_{\mu_i}}
\end{align}
where $\ell(\lambda)$ is the length of a partition $\lambda$, and
\begin{align}
V_{M}^k(\lambda)&=\{ \mu \vdash |\lambda|+k \; | \; \ell(\mu) \leq M \;\; \text{and} \;\; \mu/\lambda\;\; \text{is a vertical strip} \},\label{eq:Vnr}\\
\psi_{\mu/\lambda}'(\mathsf{q},\mathsf{t})&=\psi_{\mu'/\lambda'}(\mathsf{t},\mathsf{q}),\nonumber \\
\psi_{\mu/\lambda}(\mathsf{q},\mathsf{t})&=\prod_{1\leq i \leq j \leq \ell(\lambda)} \frac{(\mathsf{t}^{j-i+1}\mathsf{q}^{\lambda_i-\lambda_j};\mathsf{q})_{\mu_i-\lambda_i}(\mathsf{t}^{j-i}\mathsf{q}^{\lambda_i-\mu_{j+1}+1};\mathsf{q})_{\mu_i-\lambda_i}}{(\mathsf{t}^{j-i}\mathsf{q}^{\lambda_i-\lambda_j+1};\mathsf{q})_{\mu_i-\lambda_i}(\mathsf{t}^{j-i+1}\mathsf{q}^{\lambda_i-\mu_{j+1}};\mathsf{q})_{\mu_i-\lambda_i}},\\
\varphi_{\mu/\lambda}'(\mathsf{q},\mathsf{t})&=\varphi_{\mu'/\lambda'}(\mathsf{t},\mathsf{q}),\nonumber \\
\varphi_{\mu/\lambda}(\mathsf{q},\mathsf{t})&=\prod_{1\leq i \leq j \leq \ell(\mu)} \frac{(\mathsf{t}^{j-i+1}\mathsf{q}^{\mu_i-\mu_j};\mathsf{q})_{\mu_j-\lambda_j}(\mathsf{t}^{j-i}\mathsf{q}^{\lambda_i-\mu_{j+1}+1};\mathsf{q})_{\mu_{j+1}-\lambda_{j+1}}}{(\mathsf{t}^{j-i}\mathsf{q}^{\mu_i-\mu_j+1};\mathsf{q})_{\mu_j-\lambda_j}(\mathsf{t}^{j-i+1}\mathsf{q}^{\lambda_i-\mu_{j+1}};\mathsf{q})_{\mu_{j+1}-\lambda_{j+1}}}.
\end{align}
See \cite{Hatsuda:2025mvj} for more detail.

%D5 U(N)|U(M) vortex
The NS5-type $U(N)|U(M)$ interface is S-dual to the D5-type $U(N)|U(M)$ interface. 
For this interface the $U(N)$ gauge group is broken down to the block-diagonal $U(M)$ subgroup  
and therefore the whole $U(M)$ gauge group is preserved. 
For $N-M=1$ there is no Nahm pole, 
while for $N-M>1$  the interface involves a regular Nahm pole of rank $N-M$ in the $U(N-M)$ subgroup of the $U(N)$ group. 
The diagonal $U(1)$ subgroup in $U(N-M)$ is identified with the global symmetry at the interface. 
For the D5-type $U(N)|U(M)$ interface we can introduce the vortex line labeled by the cocharacter $(k_1,\cdots,k_M)$ of the $U(M)$ gauge group. 
We expect that the vortex lines with 
the Abelianized vortex number $(1^k,0^{M-k})$ or $(2,1^{M-1})$ are stable 
as they are free from the monopole bubbling effect \cite{Kapustin:2006pk}. 
Taking into account the shifted spins of the matter fields due to the vortex configuration, 
the half-index (\ref{uNuMD}) will be generalized as 
\begin{align}
\label{uNuMD_V1^k}
&
\langle  V_{(1^k,0^{M-k})} \rangle_{\mathcal{D}}^{U(N)|U(M)}(t;q)
\nonumber\\
&=
q^{\frac{k}{4}}t^{-k}
\frac{1}{k! (M-k)!}
\frac{(q)_{\infty}^{2M}}{(q^{\frac12}t^{\pm 2};q)_{\infty}^M}
\oint \prod_{i=1}^M \frac{ds_i}{2\pi is_i}
\nonumber\\
&\times 
\prod_{i=1}^k 
\prod_{j=k+1}^{M}
\frac{(q^{\frac12}s_i^{\pm}s_j^{\mp};q)_{\infty}(q^{\frac32}s_i^{\pm}s_j^{\mp};q)_{\infty}}
{(qt^2s_i^{\pm}s_j^{\mp};q)_{\infty}(qt^{-2}s_i^{\pm}s_j^{\mp};q)_{\infty}}
\prod_{
\begin{smallmatrix}
1\le i<j\le k\\
k+1\le i<j\le M\\
\end{smallmatrix}
}
\frac{(s_i^{\pm}s_j^{\mp};q)_{\infty}(qs_i^{\pm}s_j^{\mp};q)_{\infty}}
{(q^{\frac12}t^2 s_i^{\pm}s_j^{\mp};q)_{\infty} (q^{\frac12}t^{-2}s_i^{\pm}s_j^{\mp};q)_{\infty}}
\nonumber\\
&\times 
\prod_{i=1}^{k}
\frac{(q^{\frac{3}{4}+\frac{(N-M)}{4}+\frac12}t^{-1+(N-M)}s_i^{\mp};q)_{\infty}}
{(q^{\frac14+\frac{(N-M)}{4}+\frac12}t^{1+(N-M)}s_i^{\pm};q)_{\infty}}
\prod_{i=k+1}^{M}
\frac{(q^{\frac34+\frac{(N-M)}{4}}t^{-1+(N-M)}s_i^{\mp};q)_{\infty}}
{(q^{\frac14+\frac{(N-M)}{4}}t^{1+(N-M)}s_i^{\pm};q)_{\infty}}
\nonumber\\
&\times 
\prod_{l=1}^{N-M}
\frac{(q^{\frac{l+1}{2}}t^{2(l-1)};q)_{\infty}}
{(q^{\frac{l}{2}}t^{2l};q)_{\infty}}
\end{align}
in the presence of the vortex line with $(1^k,0^{M-k})$ 
at the D5-type $U(N)|U(M)$ interface. 
Similarly, we have
\begin{align}
\label{uNuMD_V21^M-1}
&
\langle  V_{(2,1^{M-1})} \rangle_{\mathcal{D}}^{U(N)|U(M)}(t;q)
\nonumber\\
&=
q^{\frac{M+1}{4}}t^{-(M+1)}
\frac{1}{(M-1)!}
\frac{(q)_{\infty}^{2M}}{(q^{\frac12}t^{\pm 2};q)_{\infty}^M}
\oint \prod_{i=1}^M \frac{ds_i}{2\pi is_i}
\nonumber\\
&\times 
\prod_{j=2}^{M}
\frac{(q^{\frac12}s_1^{\pm}s_j^{\mp};q)_{\infty}(q^{\frac32}s_1^{\pm}s_j^{\mp};q)_{\infty}}
{(qt^2s_1^{\pm}s_j^{\mp};q)_{\infty}(qt^{-2}s_1^{\pm}s_j^{\mp};q)_{\infty}}
\prod_{
2\le i<j\le M
}
\frac{(s_i^{\pm}s_j^{\mp};q)_{\infty}(qs_i^{\pm}s_j^{\mp};q)_{\infty}}
{(q^{\frac12}t^2 s_i^{\pm}s_j^{\mp};q)_{\infty} (q^{\frac12}t^{-2}s_i^{\pm}s_j^{\mp};q)_{\infty}}
\nonumber\\
&\times 
\frac{(q^{\frac{3}{4}+\frac{(N-M)}{4}+1}t^{-1+(N-M)}s_1^{\mp};q)_{\infty}}
{(q^{\frac14+\frac{(N-M)}{4}+1}t^{1+(N-M)}s_1^{\pm};q)_{\infty}}
\prod_{i=2}^{M}
\frac{(q^{\frac34+\frac{(N-M)}{4}+\frac12}t^{-1+(N-M)}s_i^{\mp};q)_{\infty}}
{(q^{\frac14+\frac{(N-M)}{4}+\frac12}t^{1+(N-M)}s_i^{\pm};q)_{\infty}}
\nonumber\\
&\times 
\prod_{l=1}^{N-M}
\frac{(q^{\frac{l+1}{2}}t^{2(l-1)};q)_{\infty}}
{(q^{\frac{l}{2}}t^{2l};q)_{\infty}}
\end{align}
when we introduce the vortex line with $(2,1^{M-1})$ 
at the D5-type $U(N)|U(M)$ interface. 

Indeed, we find that 
the expressions (\ref{uNuMD_V1^k}) and (\ref{uNuMD_V21^M-1}) coincide with 
the Wilson line defect half-index (\ref{uNuMN_W}) with $m=1$ for the NS5-type $U(N)|U(M)$ interface
\begin{align}
\langle W_{(1^{k});\overline{(1^k)}}\rangle^{U(N)|U(M)}_{\mathcal{N}}(t;q)
&=\langle  V_{(1^k,0^{M-k})} \rangle_{\mathcal{D}}^{U(N)|U(M)}(t^{-1};q), \\
\langle W_{(2,1^{M-1});\overline{(2,1^{M-1})}}\rangle^{U(N)|U(M)}_{\mathcal{N}}(t;q)
&=\langle  V_{(2,1^{M-1})} \rangle_{\mathcal{D}}^{U(N)|U(M)}(t^{-1};q). 
\end{align}

In the Coulomb and Higgs limits, the Wilson line defect half-indices (\ref{uNuMN_W}) reduce to 
\begin{align}
\label{uNuMN_WC}
{\langle \prod_{i=1}^{m} W_{\mathcal{R}^{(1)}_i;\mathcal{R}^{(2)}_i} \rangle^{U(N)|U(M)}_{\mathcal{N}}}^{(C)}(\mathfrak{q})
&=\langle \prod_{i=1}^m W_{\mathcal{R}^{(1)}_i}\rangle^{U(N)}_{\textrm{$\frac12$BPS}}(\mathfrak{q})
\times \langle \prod_{i=1}^m W_{\mathcal{R}^{(2)}_i}\rangle^{U(M)}_{\textrm{$\frac12$BPS}}(\mathfrak{q}), \\
\label{uNuMN_WH}
{\langle \prod_{i=1}^{m} W_{\mathcal{R}^{(1)}_i;\mathcal{R}^{(2)}_i} \rangle^{U(N)|U(M)}_{\mathcal{N}}}^{(H)}(\mathfrak{q})
&=\frac{1}{N!}\oint \prod_{i=1}^{N}\frac{ds_i^{(1)}}{2\pi is_i^{(1)}}
\prod_{i<j}
(1-s_i^{(1)\pm}s_j^{(1)\mp})
\nonumber\\
&\times \frac{1}{M!}\oint \prod_{i=1}^{M}\frac{ds_i^{(2)}}{2\pi is_i^{(2)}}
\prod_{i<j}
(1-s_i^{(2)\pm}s_j^{(2)\mp})
\nonumber\\
&\times \prod_{i=1}^N \prod_{j=1}^{M}
\frac{1}{1-\mathfrak{q}s_i^{(1)\pm}s_j^{(2)\mp}}
\prod_{i=1}^m \chi_{\mathcal{R}^{(1)}_i}(s^{(1)})
\prod_{i=1}^m \chi_{\mathcal{R}^{(2)}_i}(s^{(2)}). 
\end{align}
It is straightforward to compute the Coulomb limit 
as the matrix integral (\ref{uNuMN_WC}) results in the line defect correlator for $U(N)$ SYM theory and that for $U(M)$ SYM theory in the half-BPS limit \cite{Hatsuda:2023iwi,Hatsuda:2023imp}. 
The Higgs limit (\ref{uNuMN_WH}) involves the factors which naturally generalize the weight function for the Hall-Littlewood function. 

It follows from (\ref{uNuMN_WC}) that the line defect half-index of the Wilson line $W_{\mathcal{R}^{(1)};\mathcal{R}^{(2)}}$ vanishes in the Coulomb limit. 
However, it is non-trivial in the Higgs limit. 
For example, for $(\mathcal{R}^{(1)},\mathcal{R}^{(2)})$ $=$ $(\tiny \yng(1),\tiny \overline{\yng(1)})$, 
the normalized line defect half-index is given by
\begin{align}
\label{uNuMN_W1H}
{\langle \mathcal{W}_{\tiny \yng(1);\tiny \overline{\yng(1)}} \rangle^{U(N)|U(M)}_{\mathcal{N}}}^{(H)}(\mathfrak{q})
&=\mathfrak{q}(1+\mathfrak{q}^2+\cdots+\mathfrak{q}^{2(M-1)}). 
\end{align}
For $(\mathcal{R}^{(1)},\mathcal{R}^{(2)})$ $=$ $(\tiny (1^k),\tiny \overline{(1^k)})$ with $k\le M$ 
we find
\begin{align}
\label{uNuMN_W1^kH}
{\langle \mathcal{W}_{(1^k);\tiny \overline{(1^k)}} \rangle^{U(N)|U(M)}_{\mathcal{N}}}^{(H)}(\mathfrak{q})
&=\mathfrak{q}^k \left(
\begin{matrix}
M\\
k\\
\end{matrix}
\right)_{\mathfrak{q}^2}. 
\end{align}
For $(\mathcal{R}^{(1)},\mathcal{R}^{(2)})$ $=$ $((2,1^{M-1});(2,1^{M-1}))$ we have
\begin{align}
\label{uNuMN_WadjH}
{\langle \mathcal{W}_{(2,1^{M-1});(2,1^{M-1})}\rangle_{\mathcal{N}}^{U(N)|U(M)}}^{(H)}
&=\mathfrak{q}^{M+1} \left(
\begin{matrix}
M\\
1\\
\end{matrix}
\right)_{\mathfrak{q}^2}
\end{align}
Also we obtain for $(\mathcal{R}^{(1)},\mathcal{R}^{(2)})$ $=$ $(\tiny (k),\tiny \overline{(k)})$ 
\begin{align}
{\langle \mathcal{W}_{(k);\tiny \overline{(k)}} \rangle^{U(N)|U(M)}_{\mathcal{N}}}^{(H)}(\mathfrak{q})
&=\mathfrak{q}^k \left(
\begin{matrix}
M+k-1\\
k\\
\end{matrix}
\right)_{\mathfrak{q}^2}. 
\end{align}
It follows that the line defect half-index of the Wilson line $W_{\lambda;\overline{\lambda}}$ in the Higgs limit is proportional to 
the two-point function of the Wilson lines $W_{\lambda}$ and $W_{\overline{\lambda}}$ in $\mathcal{N}=4$ $U(M)$ gauge theory in the half-BPS limit \cite{Hatsuda:2023iwi,Hatsuda:2023imp} 
\begin{align}
\label{Wilson-equality}
{\langle \mathcal{W}_{\lambda;\overline{\lambda}} \rangle^{U(N)|U(M)}_{\mathcal{N}}}^{(H)}(\mathfrak{q})
&=\mathfrak{q}^{|\lambda|}\langle \mathcal{W}_{\lambda}\mathcal{W}_{\overline{\lambda}}\rangle^{U(M)}_{\textrm{$\frac12$BPS}}(\mathfrak{q}). 
\end{align}
Let us prove it. We start with
\begin{align}
{\langle W_{\lambda;\overline{\lambda}} \rangle^{U(N)|U(M)}_{\mathcal{N}}}^{(H)}(\mathfrak{q})
&=\frac{1}{N!}\oint \prod_{i=1}^{N}\frac{ds_i^{(1)}}{2\pi is_i^{(1)}}
\prod_{i<j}
(1-s_i^{(1)\pm}s_j^{(1)\mp})
\nonumber\\
&\times \frac{1}{M!}\oint \prod_{i=1}^{M}\frac{ds_i^{(2)}}{2\pi is_i^{(2)}}
\prod_{i<j}
(1-s_i^{(2)\pm}s_j^{(2)\mp})
\nonumber\\
&\times \prod_{i=1}^N \prod_{j=1}^{M}
\frac{1}{1-\mathfrak{q}s_i^{(1)\pm}s_j^{(2)\mp}}
\chi_{\lambda}(s^{(1)})
\chi_{\lambda}(s^{(2)-1})
\end{align}
Using the Cauchy identity:
\begin{align}
\prod_{i=1}^N \prod_{j=1}^{M}
\frac{1}{1-\mathfrak{q}x_i/y_j}
=\sum_{\ell(\mu) \leq M} \mathfrak{q}^{|\mu|} \chi_{\mu}(x)
\chi_{\mu}(y^{-1}),
%\prod_{i=1}^N \prod_{j=1}^{M}
%\frac{1}{1-\mathfrak{q}s_j^{(2)}/s_i^{(1)}}
%&=\sum_{\ell(\nu) \leq M} \mathfrak{q}^{|\nu|} \chi_{\nu}(s^{(1)-1})
%\chi_{\nu}(s^{(2)}),
\end{align}
and the Littlewood-Richardson coefficients:
\begin{align}
\chi_{\lambda}(x)\chi_{\mu}(x)=\sum_{\ell(\nu)\leq M} N_{\lambda, \mu}^{\nu} \chi_{\nu}(x),
\end{align}
we can perform the integral exactly. We find the following sum formula:
\begin{align}
{\langle W_{\lambda;\overline{\lambda}} \rangle^{U(N)|U(M)}_{\mathcal{N}}}^{(H)}(\mathfrak{q})
=\sum_{\ell(\mu) \leq M} \sum_{\ell(\nu) \leq M} \mathfrak{q}^{|\mu|+|\nu|} (N_{\lambda, \mu}^{\nu})^2.
\end{align}
Next, we consider the matrix integral
\begin{align}
\langle W_{\lambda}W_{\overline{\lambda}}\rangle^{U(M)}_{\textrm{$\frac12$BPS}}(\mathfrak{q})
=\frac{1}{M!} \oint \prod_{i=1}^M \frac{ds_i}{2\pi i s_i} \prod_{i<j} (1-s^{\pm} s^{\mp}) \prod_{i,j=1}^M \frac{1}{1-\mathfrak{q}^2 s_i s_j^{-1}}\chi_{\lambda}(s) \chi_{\lambda}(s^{-1}).
\end{align}
Using the Cauchy identity and the Littlewood-Richardson coefficients again, we obtain
\begin{align}
\langle W_{\lambda}W_{\overline{\lambda}}\rangle^{U(M)}_{\textrm{$\frac12$BPS}}(\mathfrak{q})
=\sum_{\ell(\mu) \leq M} \sum_{\ell(\nu) \leq M} \mathfrak{q}^{2|\mu|} (N_{\lambda, \mu}^{\nu})^2.
\end{align}
Noticing that the Littlewood-Richardson coefficients $N_{\lambda, \mu}^{\nu}$ are vanishing except for $|\nu|=|\lambda|+|\mu|$,
we arrive at the equality \eqref{Wilson-equality}.

When we consider junctions of Wilson lines, the Coulomb limits do not vanish in general 
though they can be simply expressed in terms of the line defect correlators for $U(N)$ and $U(M)$ SYM theories in the half-BPS limit \cite{Hatsuda:2023iwi,Hatsuda:2023imp}. 

Next consider the Coulomb and Higgs limits of the vortex line defect half-indices. 
Making use of the orthogonality (\ref{HL_inner}) of the Hall-Littlewood functions, 
we obtain from the matrix integral (\ref{uNuMD_V1^k}) the normalized line defect half-indices
\begin{align}
\label{uNuMD_V1^kC}
{\langle \mathcal{V}_{(1^k,0^{M-k})}\rangle_{\mathcal{D}}^{U(N)|U(M)}}^{(C)}
&=\mathfrak{q}^k \left(
\begin{matrix}
M\\
k\\
\end{matrix}
\right)_{\mathfrak{q}^2}, \\
\label{uNuMD_V1^kH}
{\langle \mathcal{V}_{(1^k,0^{M-k})}\rangle_{\mathcal{D}}^{U(N)|U(M)}}^{(H)}
&=0. 
\end{align}
Similarly, the normalized vortex line defect half-indices are obtained form (\ref{uNuMD_V21^M-1}) 
\begin{align}
\label{uNuMD_V21^M-1C}
{\langle \mathcal{V}_{(2,1^{M-1})}\rangle_{\mathcal{D}}^{U(N)|U(M)}}^{(C)}
&=\mathfrak{q}^{M+1} \left(
\begin{matrix}
M\\
1\\
\end{matrix}
\right)_{\mathfrak{q}^2}, \\
\label{uNuMD_V21^M-1H}
{\langle \mathcal{V}_{(2,1^{M-1})}\rangle_{\mathcal{D}}^{U(N)|U(M)}}^{(H)}
&=0. 
\end{align}
In fact, the expressions (\ref{uNuMD_V1^kC}) and (\ref{uNuMD_V21^M-1C}) precisely agree with the dual ones (\ref{uNuMN_W1^kH}) and (\ref{uNuMN_WadjH}) respectively.

%%%%%%%%%%%%%%%%%%%%%%%%%%%%%%%%%%%%%%%%
\subsection{Vortex-Wilson}
%%%%%%%%%%%%%%%%%%%%%%%%%%%%%%%%%%%%%%%%
%NS5 U(N)|U(M) vortex
Next consider 
the vortex lines in the NS5-type $U(N)|U(M)$ interface. 
In this case, the vortex lines may be characterized by pairs $(k^{(1)},k^{(2)})$ 
of the Abelianized vortex numbers associated with the $U(N)^{(1)}$ and $U(M)^{(2)}$ gauge groups. 
However, the vortex numbers may be lowered according to the monopole bubbling effect \cite{Kapustin:2006pk}. 
Here we consider the cases with $k^{(1)}=0$ and $k^{(2)}$ being associated with the minuscule weights, 
i.e. $(1^k,0^{M-k})$ and $(2,1^{M-1})$. 
We expect that the half-index (\ref{uNuMN}) is decorated by the vortex line with 
$k^{(1)}=0$ and $k^{(2)}=(1^k,0^{M-k})$ as
\begin{align}
\label{uNuMN_V1^k}
&
\langle V_{0;(1^k,0^{M-k})} \rangle_{\mathcal{N}}^{U(N)|U(M)}(t;q)
\nonumber\\
&=
q^{\frac{k^2+(N-M)k}{4}}t^{-2(k^2+(N-M)k)}
\frac{1}{N!}\frac{(q)_{\infty}^N}{(q^{\frac12}t^{-2};q)_{\infty}^N}
\oint 
\prod_{i=1}^N
\frac{ds_i^{(1)}}{2\pi is_i^{(1)}}
\prod_{i<j}
\frac{(s_i^{(1)\pm}s_j^{(1)\mp};q)_{\infty}}
{(q^{\frac12}t^{-2}s_i^{(1)\pm}s_j^{(1)\mp};q)_{\infty}}
\nonumber\\
&\times 
\frac{1}{k!(M-k)!}\frac{(q)_{\infty}^M}{(q^{\frac12}t^{-2};q)_{\infty}^M}
\oint 
\prod_{i=1}^M
\frac{ds_i^{(2)}}{2\pi is_i^{(2)}}
\nonumber\\
&\times 
\prod_{i=1}^k
\prod_{j=k+1}^N
\frac{(q^{\frac12}s_i^{(1)\pm}s_j^{(1)\mp};q)_{\infty}}
{(qt^{-2}s_i^{(1)\pm}s_j^{(1)\mp};q)_{\infty}}
\prod_{
\begin{smallmatrix}
1\le i<j\le k\\
k+1\le i<j\le M\\
\end{smallmatrix}
}
\frac{(s_i^{(2)\pm}s_j^{(2)\mp};q)_{\infty}}
{(q^{\frac12}t^{-2}s_i^{(2)\pm}s_j^{(2)\mp};q)_{\infty}}
\nonumber\\
&\times
\prod_{i=1}^{N}
\prod_{j=1}^{k}
\frac{(q^{\frac54}t^{-1}s_i^{(1)\mp}s_j^{(2)\pm};q)_{\infty}}
{(q^{\frac34}ts_i^{(1)\mp}s_j^{(2)\pm};q)_{\infty}}
\prod_{i=1}^{N}
\prod_{j=k+1}^{M}
\frac{(q^{\frac34}t^{-1}s_i^{(1)\mp}s_j^{(2)\pm};q)_{\infty}}
{(q^{\frac14}ts_i^{(1)\mp}s_j^{(2)\pm};q)_{\infty}}. 
\end{align}
When $k^{(1)}=0$ and $k^{(2)}$ $=$ $(2,1^{M-1})$ we have 
\begin{align}
\label{uNuMN_V21^M-1}
&
\langle  V_{0;(2,1^{M-1})} \rangle_{\mathcal{N}}^{U(N)|U(M)}(t;q)
\nonumber\\
&=
q^{\frac{M^2+1}{4}}t^{-(M^2+1)}
\frac{1}{N!}\frac{(q)_{\infty}^N}{(q^{\frac12}t^{-2};q)_{\infty}^N}
\oint \prod_{i=1}^N \frac{ds_i^{(1)}}{2\pi is_i^{(1)}}
\prod_{i<j}
\frac{(s_i^{(1)\pm}s_j^{(1)\mp};q)_{\infty}}
{(q^{\frac12}t^{-2}s_i^{(1)\pm}s_j^{(1)\mp};q)_{\infty}}
\nonumber\\
&\times 
\frac{1}{(M-1)!}\frac{(q)_{\infty}^M}{(q^{\frac12}t^{-2};q)_{\infty}^M}
\oint\prod_{i=1}^M \frac{ds_i^{(2)}}{2\pi is_i^{(2)}}
\nonumber\\
&\times 
\prod_{j=2}^M
\frac{(q^{\frac12}s_1^{(2)\pm}s_j^{(2)\mp};q)_{\infty}}
{(qt^{-2}s_1^{(2)\pm}s_j^{(2)\mp};q)_{\infty}}
\prod_{
2\le i<j\le M
}
\frac{(s_i^{(2)\pm}s_j^{(2)\mp};q)_{\infty}}
{(q^{\frac12}t^{-2}s_i^{(2)\pm}s_j^{(2)\mp};q)_{\infty}}
\nonumber\\
&\times 
\prod_{j=1}^N 
\frac{(q^{\frac74}t^{-1}s_1^{(1)\mp}s_j^{(2)\pm};q)_{\infty}}
{(q^{\frac54}ts_1^{(1)\pm}s_j^{(2)\mp};q)_{\infty}}
\prod_{i=2}^{N}
\prod_{j=1}^{N}
\frac{(q^{\frac54}t^{-1}s_i^{(1)\mp}s_j^{(2)\pm};q)_{\infty}}
{(q^{\frac34}ts_i^{(1)\pm}s_j^{(2)\mp};q)_{\infty}}. 
\end{align}

%D5 U(N)|U(M) Wilson
In the S-dual configuration we have the D5-type $U(N)|U(M)$ interface involving 
the Wilson line $W_{\mathcal{R}}$ transforming as the representation $\mathcal{R}$ of the preserved $U(M)$ gauge group. 
When we introduce a collection of the Wilson lines $W_{\mathcal{R}_i}$, with $i=1,\cdots, m$, 
the line defect half-index takes the form 
\begin{align}
\label{uNuMD_W}
&
\langle \prod_{i=1}^m 
W_{\mathcal{R}_i} \rangle_{\mathcal{D}}^{U(N)|U(M)}(t;q)
\nonumber\\
&=\frac{1}{M!}
\frac{(q)_{\infty}^{2M}}{(q^{\frac12}t^{\pm2};q)_{\infty}^M}
\oint \prod_{i=1}^M 
\frac{ds_i}{2\pi is_i}
\prod_{i<j}
\frac{(s_i^{\pm}s_j^{\mp};q)_{\infty}(qs_i^{\pm}s_j^{\mp};q)_{\infty}}
{(q^{\frac12}t^2s_i^{\pm}s_j^{\mp};q)_{\infty}(q^{\frac12}t^{-2}s_i^{\pm}s_j^{\mp};q)_{\infty}}
\nonumber\\
&\times 
\prod_{i=1}^{M}
\frac{(q^{\frac34+\frac{N-M}{4}}t^{-1+(N-M)}s_i^{\mp};q)_{\infty}}
{(q^{\frac14+\frac{N-M}{4}}t^{1+(N-M)}s_i^{\pm};q)_{\infty}}
\prod_{i=1}^{m} \chi_{\mathcal{R}_i}(s)
\nonumber\\
&\times 
\prod_{l=1}^{N-M}
\frac{(q^{\frac{l+1}{2}}t^{2(l-1)};q)_{\infty}}
{(q^{\frac{l}{2}}t^{2l};q)_{\infty}}. 
\end{align}

We have checked that 
the vortex line defect half-indices (\ref{uNuMN_V1^k}) and (\ref{uNuMN_V21^M-1}) 
for the NS5-type interface agree with the Wilson line defect half-indices (\ref{uNuMD_W}) for the D5-type interface
\begin{align}
\langle V_{0;(1^k,0^{M-k})} \rangle_{\mathcal{N}}^{U(N)|U(M)}(t;q)&=
\langle W_{(1^{k})} \rangle_{\mathcal{D}}^{U(N)|U(M)}(t^{-1};q), \\
\langle V_{0;(2,1^{M-1})} \rangle_{\mathcal{N}}^{U(N)|U(M)}(t;q)&=
\langle W_{(2,1^{M-1})} \rangle_{\mathcal{D}}^{U(N)|U(M)}(t^{-1};q). 
\end{align}

In addition, we propose the closed-form expressions for the matrix integral (\ref{uNuMD_W}) with the form
\begin{align}
\label{uNuMD_W_expand}
&
\langle W_{\mathcal{R}} \rangle_{\mathcal{D}}^{U(N)|U(M)}(t^{-1};q)
\nonumber\\
&=\sum_{n_1=0}^{\infty}
\cdots 
\sum_{n_M=0}^{\infty}
\frac{(q^{1+n_1};q)_{\infty}(q^{1+\frac{N-M}{2}+n_1}t^{-2(N-M)};q)_{\infty}}
{(q^{\frac12+n_1}t^{-2};q)_{\infty}(q^{\frac12+\frac{N-M}{2}+n_1}t^{-2(N-M+1)};q)_{\infty}}
\nonumber\\
&\times 
\frac{(q^{\frac32+n_1+n_2}t^{-2};q)_{\infty}(q^{\frac32+\frac{N-M}{2}+n_1+n_2}t^{-2(N-M+1)};q)_{\infty}}
{(q^{1+n_1+n_2}t^{-4};q)_{\infty}(q^{1+\frac{N-M}{2}+n_1+n_2}t^{-2(N-M+2)};q)_{\infty}}\times \cdots 
\nonumber\\
&\times 
\frac{(q^{\frac{M+1}{2}+n_1+\cdots +n_M}t^{-2(M-1)};q)_{\infty}
(q^{\frac{N+1}{2}+n_1+\cdots+n_M}t^{-2(N-1)};q)_{\infty}}
{(q^{\frac{M}{2}+n_1+\cdots +n_M}t^{-2M};q)_{\infty}
q^{\frac{N}{2}+n_1+\cdots+n_M}t^{-2N};q)_{\infty}
}
(q^{\frac12}t^2)^{Mn_1+(M-1)n_2+\cdots+n_M}
\nonumber\\
&\times 
\chi_{\mathcal{R}}
\left(
s_i=q^{\sum_{j=1}^i n_j+\frac{N-M+2i-1}{4}}t^{-(N-M+2i-1)}
\right)
\prod_{l=1}^{N-M}\frac{(q^{\frac{l+1}{2}}t^{2(l-1)};q)_{\infty}}
{(q^{\frac{l}{2}}t^{2l};q)_{\infty}}. 
\end{align}
When $N=M$, this reduces to (\ref{uNuND_W_expand}). 
One can view this as a generalization of the vortex expansion of the half-index of the D5-type $U(N)|U(M)$ interface 
in \cite{Gaiotto:2019jvo,Hatsuda:2024uwt}. 
Interestingly, the effect of an insertion of the Wilson line in the representation $\mathcal{R}$ 
is simply manipulated by adding the specialized character $\chi_{\mathcal{R}}(s)$ in the summand of the vortex expansion. 

For example, for the fundamental Wilson line in the D5-type $U(N)|U(M)$ interface 
which is dual to the vortex line with $k^{(1)}=0$ and $k^{(2)}=(1,0^{M-1})$ in the NS5-type $U(N)|U(M)$ interface, 
we have
\begin{align}
&
\langle V_{0;(1,0^{M-1})} \rangle_{\mathcal{N}}^{U(N)|U(M)}(t;q)
=\langle W_{\tiny \yng(1)} \rangle_{\mathcal{D}}^{U(N)|U(M)}(t^{-1};q)
\nonumber\\
&=\sum_{n_1=0}^{\infty}
\cdots 
\sum_{n_M=0}^{\infty}
\frac{(q^{1+n_1};q)_{\infty}(q^{1+\frac{N-M}{2}+n_1}t^{-2(N-M)};q)_{\infty}}
{(q^{\frac12+n_1}t^{-2};q)_{\infty}(q^{\frac12+\frac{N-M}{2}+n_1}t^{-2(N-M+1)};q)_{\infty}}
\nonumber\\
&\times 
\frac{(q^{\frac32+n_1+n_2}t^{-2};q)_{\infty}(q^{\frac32+\frac{N-M}{2}+n_1+n_2}t^{-2(N-M+1)};q)_{\infty}}
{(q^{1+n_1+n_2}t^{-4};q)_{\infty}(q^{1+\frac{N-M}{2}+n_1+n_2}t^{-2(N-M+2)};q)_{\infty}}\times \cdots 
\nonumber\\
&\times 
\frac{(q^{\frac{M+1}{2}+n_1+\cdots +n_M}t^{-2(M-1)};q)_{\infty}
(q^{\frac{N+1}{2}+n_1+\cdots+n_M}t^{-2(N-1)};q)_{\infty}}
{(q^{\frac{M}{2}+n_1+\cdots +n_M}t^{-2M};q)_{\infty}
q^{\frac{N}{2}+n_1+\cdots+n_M}t^{-2N};q)_{\infty}
}
(q^{\frac12}t^2)^{Mn_1+(M-1)n_2+\cdots+n_M}
\nonumber\\
&\times 
\left[
q^{n_1+\frac{N-M+1}{4}}t^{-(N-M+1)}
+q^{n_1+n_2+\frac{N-M+3}{4}}t^{-(N-M+3)}
+\cdots
+q^{n_1+\cdots+n_M+\frac{N+M-1}{4}}t^{-(N+M-1)}
\right]
\nonumber\\
&\times 
\prod_{l=1}^{N-M}\frac{(q^{\frac{l+1}{2}}t^{2(l-1)};q)_{\infty}}
{(q^{\frac{l}{2}}t^{2l};q)_{\infty}}. 
\end{align}

%Coulomb Higgs limits
Let us examine the Coulomb and Higgs limits of the line defect half-indices. 
For the vortex line defect half-indices (\ref{uNuMN_V1^k}) and (\ref{uNuMN_V21^M-1}) the Higgs limits vanish. 
In the Coulomb limit they reduce to
\begin{align}
\label{uNuMD_W1^kH}
{\langle \mathcal{V}_{0;(1^k,0^{M-k})} \rangle_{\mathcal{N}}^{U(N)|U(M)}}^{(C)}(\mathfrak{q})
&=\mathfrak{q}^{k^2+(N-M)k}
\left(
\begin{matrix}
M\\
k\\
\end{matrix}
\right)_{\mathfrak{q}^2}, \\
{\langle \mathcal{V}_{0;(2,1^{M-1})} \rangle_{\mathcal{N}}^{U(N)|U(M)}}^{(C)}(\mathfrak{q})
&=\mathfrak{q}^{M^2+1}
\left(
\begin{matrix}
M\\
1\\
\end{matrix}
\right)_{\mathfrak{q}^2}. 
\end{align}

On the other hand, the matrix integral (\ref{uNuMD_W}) for the Wilson line defect half-index becomes 
\begin{align}
\label{uNuMD_WC}
{\langle \prod_{i=1}^m W_{\mathcal{R}_i} \rangle_{\mathcal{D}}^{U(N)|U(M)}}^{(C)}(\mathfrak{q})
&=\langle \prod_{i=1}^m W_{\mathcal{R}_i}\rangle^{U(M)}_{\textrm{$\frac12$BPS}}(\mathfrak{q}), \\
\label{uNuMD_WH}
{\langle \prod_{i=1}^m W_{\mathcal{R}_i} \rangle_{\mathcal{D}}^{U(N)|U(M)}}^{(H)}(\mathfrak{q})
&=\frac{1}{M!}\frac{1}{(1-\mathfrak{q}^2)^M}\oint \prod_{i=1}^M
\frac{ds_i}{2\pi is_i}\prod_{i<j}\frac{1-s_i^{\pm}s_j^{\mp}}{1-\mathfrak{q}^2s_i^{\pm}s_j^{\mp}}
\nonumber\\
&\times 
\prod_{i=1}^M \frac{1}{1-\mathfrak{q}^{N-M+1}s_i^{\pm}}
\prod_{i=1}^m \chi_{\mathcal{R}_i}(s)
\prod_{l=1}^{N-M}\frac{1}{1-\mathfrak{q}^{2l}}. 
\end{align}
in the Coulomb and Higgs limits. 

When we introduce a single Wilson line transforming in the representation $\mathcal{R}$, 
the Coulomb limits vanish just like the line defect indices according to (\ref{uNuMD_WC}). 
On the other hand, the normalized one-point functions in the Higgs limit are non-trivial. 
It follows from (\ref{uNuMD_W_expand}) that they are given by
\begin{align}
\label{uNuMD_WH1pt}
{\langle \mathcal{W}_{\mathcal{R}} \rangle_{\mathcal{D}}^{U(N)|U(M)}}^{(H)}(\mathfrak{q})
&=\chi_{\mathcal{R}}(\mathfrak{q}^{N-M+1},\mathfrak{q}^{N-M+3},\cdots, \mathfrak{q}^{N+M-1}). 
\end{align}
This can be viewed as a specialization of the formula (\ref{uNuMD_W_expand}). 
The result (\ref{uNuMD_WH1pt}) implies that 
the BPS spectrum of the Higgs branch operators formed by the 3d bosonic fields at the junction of the D5-type interface and Wilson line 
in the irreducible representation $\mathcal{R}$ is generated by the principal specializations of the Schur functions $s_{\lambda}(s)$ labeled by the associated Young diagram $\lambda$.  
The line defect half-index (\ref{uNuMD_WH1pt}) can be explicitly written as \cite{MR1110850}
\begin{align}
{\langle \mathcal{W}_{\mathcal{R}} \rangle_{\mathcal{D}}^{U(N)|U(M)}}^{(H)}(\mathfrak{q})
&=\mathfrak{q}^{\sum_{i\ge 1}(2i+N-M-1)\lambda_i}
\prod_{(i,j)\in \lambda}
\frac{1-\mathfrak{q}^{2M+2j-2i}}{1-\mathfrak{q}^{2h(i,j)}}, 
\end{align}
where
\begin{align}
h(i,j)&=\lambda_i+\lambda'_j-i-j+1
\end{align}
is the Hook length of boxes $(i,j)$ in $\lambda$. 
According to the combinatorial interpretation \cite{MR1354144,MR325407} of the principal specializations of the Schur functions, 
the expression (\ref{uNuMD_WH1pt}) implies that 
the degeneracy of the $n$-th excited Higgs branch operators appearing at the junction of the D5-type interface and Wilson line 
in the irreducible representation labeled by the Young diagram $\lambda$ is one to one with 
the number of column-strict plane partitions of $n$ of shape $\lambda$ and parts\footnote{
The column-strict plane partitions are plane partitions whose parts along columns are strictly decreasing. 
The non-zero entries of the plane partition are called the parts. The sum $n$ $=$ $\sum_{i,j} n_{ij}$ of all entries is called the norm of the plane partition (see e.g. \cite{MR325407} for the details).} being chosen from $0,1,\cdots,M-1$. 
The lowest excited Higgs branch operator that forms the gauge invariant together with the Wilson line carries the charge
\begin{align}
\Delta&=2n(\lambda)+(N-M+1)|\lambda|, 
\end{align}
where 
\begin{align}
n(\lambda)&=\sum_{i}(i-1)\lambda_i. 
\end{align}
The total number of the gauge invariants is obtained by taking the limit $\mathfrak{q}\rightarrow 1$
\begin{align}
\prod_{{i,j}\in \lambda} 
\frac{M+j-i}{h(i,j)}, 
\end{align}
which is equal to the total number of the column-strict plane partitions.\footnote{It is also equal to the total number of semi-standard Young tableaux (SSYT) of shape $\lambda$ with entries from $1,2,\cdots, M$. }

The correspondence is summarized as
\begin{align}
\begin{array}{|c|c|c|} \hline 
\textrm{$U(N|M)$ D5-type interface Wilson line $W_{\lambda}$}&\textrm{column-strict plane partitions}\\ \hline 
\textrm{gauge rank $M$}&\textrm{highest value of parts $M-1$} \\
\textrm{charge $\Delta+2(n-1)$ of the Higgs branch operators}&\textrm{norm $n$} \\
\textrm{irreducible representation $\lambda$}&\textrm{shape $\lambda$} \\ \hline
\end{array}
\end{align}

For example, for $N=5$, $M=3$ and $\lambda$ $=$ $\tiny \yng(3,1)$ we have 
the lowest charge $\Delta$ $=$ $2\cdot 1+(5-3+1)\cdot 4$ $=$ $14$. 
The Higgs limit of the line defect half-index is evaluated as
\begin{align}
\label{u5u3D_W31H}
{\langle \mathcal{W}_{\tiny \yng(3,1)} \rangle_{\mathcal{D}}^{U(5)|U(3)}}^{(H)}(\mathfrak{q})
&=\mathfrak{q}^{14}+2\mathfrak{q}^{16}+3\mathfrak{q}^{18}+3\mathfrak{q}^{20}+3\mathfrak{q}^{22}+2\mathfrak{q}^{24}+\mathfrak{q}^{26}. 
\end{align}
The terms $\mathfrak{q}^{14+2(n-1)}$ with $n=1,\cdots, 7$ in the expression (\ref{u5u3D_W31H}) correspond to the following 
column-strict plane partitions of $n$ of shape $\tiny \yng(3,1)$ and parts being chosen from $0,1$ and $2$: 
\begin{align}
n&=1\quad :\young(100,0), \\
n&=2\quad :\young(110,0), \quad \young(200,0), \\
n&=3\quad :\young(111,0), \quad \young(210,0), \quad \young(200,1), \\
n&=4\quad :\young(211,0), \quad \young(210,1), \quad \young(220,0), \\
n&=5\quad :\young(221,0), \quad \young(211,1), \quad \young(220,1), \\
n&=6\quad :\young(222,0), \quad \young(221,1), \\
n&=7\quad :\young(222,1).
\end{align}

Note that when $\lambda$ $=$ $(1^k)$, the principal specialization of the Schur function is given by 
the $q$-binomial coefficient
\begin{align}
s_{(1^k)}(q^a,q^{a+1},\cdots, q^{a+(N-1)})
&=q^{\frac{k(k-1)}{2}+ka}
\left(
\begin{matrix}
N\\
k\\
\end{matrix}
\right)_{q}. 
\end{align}
So we obtain from the expression (\ref{uNuMD_WH1pt}) 
\begin{align}
{\langle \mathcal{W}_{(1^k)} \rangle_{\mathcal{D}}^{U(N)|U(M)}}^{(H)}(\mathfrak{q})
&=\mathfrak{q}^{k^2+(N-M)k}
\left(
\begin{matrix}
M\\
k\\
\end{matrix}
\right)_{\mathfrak{q}^2}, 
\end{align}
which precisely agrees with the vortex line defect half-index (\ref{uNuMD_W1^kH}) of the S-dual configuration.

%%%%%%%%%%%%%%%%%%%%%%%%%%%%%%%%%%%
\acknowledgments{
The work of Y.H. was supported in part by JSPS KAKENHI Grant Nos. 22K03641 and 23K25790.
The work of T.O. was supported by the Startup Funding no.\ 4007012317 of the Southeast University. 
}
%%%%%%%%%%%%%%%%%%%%%%%%%%%%%%%%%%%

\appendix

%%%%%%%%%%%%%%%%%%%%%%%%%%%%%%%%%%%%%%%%%%%%%%%
%%%%%%%%%%%%%%%%%%%%%%%%%%%%%%%%%%%%%%%%%%%%%%%
\section{Series expansions}
\label{app_expansion}
%%%%%%%%%%%%%%%%%%%%%%%%%%%%%%%%%%%%%%%%%%%%%%%
%%%%%%%%%%%%%%%%%%%%%%%%%%%%%%%%%%%%%%%%%%%%%%%
We exhibit first several terms in the $q$-series expansions. 
The matching of the indices are checked at least up to the terms with $q^5$ unless the order is not indicated.

%%%%%%%%%%%%%%%%%%%%%%%%%%%%%%%%%%%%%%%%%%%%%%%
\subsection{$U(2)|U(1)$}
%%%%%%%%%%%%%%%%%%%%%%%%%%%%%%%%%%%%%%%%%%%%%%%
For the Wilson lines at the NS5-type interface and the vortex lines at the D5-type interface we have 
\begin{align}
\begin{array}{c|c}
\textrm{line defect half-indices}&\textrm{expansions}\\ \hline
\langle W_{\tiny \yng(1);-1} \rangle_{\mathcal{N}}^{U(2)|U(1)}(t;q)
&
tq^{\frac14}+(2t^{-1}+t^3)q^{\frac34}+(3t^{-3}-t+t^5)q^{\frac54}+(4t^{-5}-t^{-1}+t^7)q^{\frac74}\\ 
=\langle V_{(1,0)}\rangle_{\mathcal{D}}^{U(2)|U(1)}(t^{-1};q)
&+(5t^{-7}+t+t^9)q^{\frac94}
+(6t^{-9}+t^{-5}-3t^{-1}-t^3+t^{11})q^{\frac{11}{4}}+\cdots \\ \hline 
\langle W_{\tiny \yng(2);-2} \rangle_{\mathcal{N}}^{U(2)|U(1)}(t;q)
&
t^2q^{\frac12}+(2+t^4)q+(3t^{-2}-t^2+t^6)q^{\frac32}+(-1+4t^{-4}+t^8)q^2\\ 
=\langle V_{(2,0)}\rangle_{\mathcal{D}}^{U(2)|U(1)}(t^{-1};q)
&+(5t^{-6}-t^{-2}+t^2+t^{10})q^{\frac52}
+(-2+6t^{-8}-t^{-4}-t^4+t^{12})q^3+\cdots \\ \hline 
\langle W_{\tiny \yng(3);-3} \rangle_{\mathcal{N}}^{U(2)|U(1)}(t;q)
&
t^{3}q^{\frac34}+(2t+t^5)q^{\frac54}+(3t^{-1}-t^3+t^7)q^{\frac74}+(4t^{-3}-t+t^9)q^{\frac94}\\ 
=\langle V_{(3,0)}\rangle_{\mathcal{D}}^{U(2)|U(1)}(t^{-1};q)
&+(5t^{-5}-t^{-1}+t^3+t^{11})q^{\frac{11}{4}}+(6t^{-7}-t^{-3}-2t-t^5+t^{13})q^{\frac{13}{4}}\cdots \\ \hline 
\end{array}
\end{align}
For the vortex lines at the NS5-type interface and the Wilson lines at the D5-type interface we obtain 
\begin{align}
\begin{array}{c|c}
\textrm{line defect half-indices}&\textrm{expansions}\\ \hline
\langle V_{0;1} \rangle_{\mathcal{N}}^{U(2)|U(1)}(t;q)
&
t^{-2}q^{\frac12}+2t^{-4}q+(4t^{-6}-2t^{-2})q^{\frac32}+(1+6t^{-8}-3t^{-4})q^2 \\ 
=\langle W_{1} \rangle_{\mathcal{D}}^{U(2)|U(1)}(t^{-1};q)
&+(9t^{-10}-5t^{-6})q^{\frac52}+(-1+12t^{-12}-6t^{-8}-2t^{-4})q^3+\cdots\\ \hline 
\langle V_{0;2} \rangle_{\mathcal{N}}^{U(2)|U(1)}(t;q)
&
t^{-4}q+2t^{-6}q^{\frac32}+(4t^{-8}-2t^{-4})q^2+(6t^{-10}-3t^{-6})q^{\frac52} \\ 
=\langle W_{2} \rangle_{\mathcal{D}}^{U(2)|U(1)}(t^{-1};q)
&+(9t^{-12}-5t^{-8}-t^{-4})q^3+(12t^{-14}-6t^{-10}-3t^{-6}+t^{-2})q^{\frac72}+\cdots\\ \hline 
\langle V_{0;3} \rangle_{\mathcal{N}}^{U(2)|U(1)}(t;q)
&
t^{-6}q^{\frac32}+2t^{-8}q^2+(4t^{-10}-2t^{-6})q^{\frac52}+(6t^{-12}-3t^{-8})q^3 \\ 
=\langle W_{3} \rangle_{\mathcal{D}}^{U(2)|U(1)}(t^{-1};q)
&+(9t^{-14}-5t^{-10}-t^{-6})q^{\frac72}+(12t^{-16}-6t^{-12}-3t^{-8})q^4+\cdots\\ \hline 
\end{array}
\end{align}

%%%%%%%%%%%%%%%%%%%%%%%%%%%%%%%%%%%%%%%%%%%%%%%
\subsection{$U(3)|U(1)$}
%%%%%%%%%%%%%%%%%%%%%%%%%%%%%%%%%%%%%%%%%%%%%%%
For the Wilson lines at the NS5-type interface and the vortex lines at the D5-type interface we have 
\begin{align}
\begin{array}{c|c}
\textrm{line defect half-indices}&\textrm{expansions}\\ \hline
\langle W_{\tiny \yng(1);-1} \rangle_{\mathcal{N}}^{U(3)|U(1)}(t;q)
&
tq^{\frac14}+(2t^{-1}+t^3)q^{\frac34}+(4t^{-3}-t+t^5)q^{\frac54}+(6t^{-5}-t^{-1}+t^7)q^{\frac74} \\ 
=\langle V_{(1,0)}\rangle_{\mathcal{D}}^{U(2)|U(1)}(t^{-1};q)
&(9t^{-7}-3t^{-3}+t+t^9)q^{\frac94}+(12t^{-9}-2t^{-5}-t^{-1}-t^3+t^{11})q^{\frac{11}{4}}+\cdots \\ \hline 
\langle W_{\tiny \yng(2);-2} \rangle_{\mathcal{N}}^{U(3)|U(1)}(t;q)
&
t^2q^{\frac12}+(2+t^{-4})q+(4t^{-2}-t^{2}+t^6)q^{\frac32}+(-1+6t^{-4}+t^8)q^2\\ 
=\langle V_{(2,0)}\rangle_{\mathcal{D}}^{U(3)|U(1)}(t^{-1};q)
&(9t^{-6}-3t^{-2}+t^2+t^{10})q^{\frac52}
+(-1+12t^{-8}-3t^{-4}-t^4+t^{12})q^3+\cdots \\ \hline 
\langle W_{\tiny \yng(3);-3} \rangle_{\mathcal{N}}^{U(3)|U(1)}(t;q)
&
t^3q^{\frac34}+(2t+t^5)q^{\frac54}+(4t^{-1}-t^3+t^7)q^{\frac74}+(6t^{-3}-t+t^9)q^{\frac94}\\ 
=\langle V_{(3,0)}\rangle_{\mathcal{D}}^{U(3)|U(1)}(t^{-1};q)
&+(9t^{-5}-3t^{-1}+t^3+t^{11})q^{\frac{11}{4}}
+(12t^{-7}-3t^{-3}-t-t^5+t^{13})q^{\frac{13}{4}}+\cdots \\ \hline 
\end{array}
\end{align}
For the vortex lines at the NS5-type interface and the Wilson lines at the D5-type interface we get
\begin{align}
\begin{array}{c|c}
\textrm{line defect half-indices}&\textrm{expansions}\\ \hline
\langle V_{0;1} \rangle_{\mathcal{N}}^{U(3)|U(1)}(t;q)
&
t^{-3}q^{\frac34}+2t^{-5}q^{\frac54}+(4t^{-7}-2t^{-3})q^{\frac74}+(7t^{-9}-3t^{-5}+t^{-1})q^{\frac94} \\ 
=\langle W_{1} \rangle_{\mathcal{D}}^{U(3)|U(1)}(t^{-1};q)
&+(11t^{-11}-6t^{-7})q^{\frac{11}{4}}+(16t^{-13}-9t^{-9}-t^{-5}-t^{-1})q^{\frac{13}{4}}+\cdots\\ \hline 
\langle V_{0;2} \rangle_{\mathcal{N}}^{U(3)|U(1)}(t;q)
&
t^{-6}q^{\frac32}+2t^{-8}q^2+(4t^{-10}-2t^{-6})q^{\frac52}+(7t^{-12}-3t^{-8})q^3 \\ 
=\langle W_{2} \rangle_{\mathcal{D}}^{U(3)|U(1)}(t^{-1};q)
&+(11t^{-14}-6t^{-10}-t^{-6})q^{\frac72}+(16t^{-16}-9t^{-12}-3t^{-8}+t^{-4})q^4+\cdots\\ \hline 
\langle V_{0;3} \rangle_{\mathcal{N}}^{U(3)|U(1)}(t;q)
&
t^{-9}q^{\frac94}+2t^{-11}q^{\frac{11}{4}}+(4t^{-13}-2t^{-9})q^{\frac{13}{4}}+(7t^{-15}-3t^{-11})q^{\frac{15}{4}}\\ 
=\langle W_{3} \rangle_{\mathcal{D}}^{U(3)|U(1)}(t^{-1};q)
&+(11t^{-17}-6t^{-13}-t^{-11})q^{\frac{17}{4}}+(16t^{-19}-9t^{-15}-3t^{-11})q^{\frac{19}{4}}+\cdots\\ \hline 
\end{array}
\end{align}

%%%%%%%%%%%%%%%%%%%%%%%%%%%%%%%%%%%%%%%%%%%%%%%
\subsection{$U(2)|U(2)$}
%%%%%%%%%%%%%%%%%%%%%%%%%%%%%%%%%%%%%%%%%%%%%%%
For the Wilson lines at the NS5-type interface and the vortex lines at the D5-type interface we find
\begin{align}
\begin{array}{c|c}
\textrm{line defect half-indices}&\textrm{expansions}\\ \hline
\langle W_{\tiny \yng(1);\overline{\yng(1)}} \rangle_{\mathcal{N}}^{U(2)|U(2)}(t;q)
&
tq^{\frac14}+(3t^{-1}+2t^2)q^{\frac34}+(6t^{-3}+t+3t^5)q^{\frac54} \\ 
=\langle V_{(1,0)}\rangle_{\mathcal{D}}^{U(2)|U(2)}(t^{-1};q)
&+(10t^{-5}-2t^{-1}+4t^7)q^{\frac74}+(15t^{-7}-4t^{-3}+3t+5t^9)q^{\frac94}+\cdots \\ \hline 
\langle W_{\tiny \yng(1,1);\overline{\yng(1,1)}} \rangle_{\mathcal{N}}^{U(2)|U(2)}(t;q)&
t^2q^{\frac12}+(1+t^4)q+(2t^{-2}+2t^6)q^{\frac32}+(1+2t^{-4}+2t^8)q^2 \\ 
=\langle V_{(1,1)}\rangle_{\mathcal{D}}^{U(2)|U(2)}(t^{-1};q)
&+(3t^{-6}+2t^{-2}+3t^{10})q^{\frac52}+(3t^{-8}+3t^{-4}+t^4+3t^{12})q^3+\cdots\\ \hline 
\langle W_{\tiny \yng(2,1);\overline{\yng(2,1)}} \rangle_{\mathcal{N}}^{U(2)|U(2)}(t;q)&
 t^3q^{\frac34}+(2t+2t^5)q^{\frac54}+(3t^{-1}+t^3+3t^7)q^{\frac74}+(4t^{-3}+t+4t^9)q^{\frac94} \\ 
=\langle V_{(2,1)}\rangle_{\mathcal{D}}^{U(2)|U(2)}(t^{-1};q)
&+(5t^{-5}+2t^{-1}+t^3+5t^{11})q^{\frac{11}{4}}+(6t^{-7}+3t^{-3}+2t^5+6t^{13})q^{\frac{13}{4}}+\cdots \\ \hline 
\end{array}
\end{align}
For the vortex lines at the NS5-type interface and the Wilson lines at the D5-type interface we have 
\begin{align}
\begin{array}{c|c}
\textrm{line defect half-indices}&\textrm{expansions}\\ \hline
\langle V_{0;(1,0)} \rangle_{\mathcal{N}}^{U(2)|U(2)}(t;q)
&
t^{-1}q^{\frac14}+(3t^{-3}+t)q^{\frac34}+(7t^{-5}+t^3)q^{\frac54} \\ 
=\langle W_{\tiny \yng(1)} \rangle_{\mathcal{D}}^{U(2)|U(2)}(t^{-1};q)
&+(13t^{-7}-3t^{-3}+t^5)q^{\frac74}+(22t^{-9}-8t^{-5}+t^{-1}+t^3+t^7)q^{\frac94}+\cdots\\ \hline 
\langle V_{(1,0);(1,0)} \rangle_{\mathcal{N}}^{U(2)|U(2)}(t;q)
&
1+(4t^{-2}+2t^2)q^{\frac12}+(2+10t^{-4}+3t^4)q \\ 
=\langle W_{\tiny \yng(1)}W_{\tiny \overline{\yng(1)}} \rangle_{\mathcal{D}}^{U(2)|U(2)}(t^{-1};q)
&+(20t^{-6}-2t^{-2}+4t^6)q^{\frac32}+(1+35t^{-8}-10t^{-4}+5t^8)q^2+\cdots\\ \hline 
\langle V_{0;(1,1)} \rangle_{\mathcal{N}}^{U(2)|U(2)}(t;q)
&
t^{-4}q+2t^{-6}q^{\frac32}+(5t^{-8}+2t^{-4})q^2+(8t^{-10}-4t^{-6}+t^{-2})q^{\frac52} \\ 
=\langle W_{\tiny \yng(1,1)}\rangle_{\mathcal{D}}^{U(2)|U(2)}(t^{-1};q)
&+(14t^{-12}-8t^{-8}+t^{-4})q^3+\cdots \\ \hline 
\langle V_{(1,1);(1,1)} \rangle_{\mathcal{N}}^{U(2)|U(2)}(t;q)
&
1+(2t^{-2}+t^2)q^{\frac12}+(5t^{-4}+2t^4)q+(8t^{-6}-t^{-2}+2t^6)q^{\frac32} \\ 
=\langle W_{\tiny \yng(1,1)}W_{\tiny\overline{\yng(1,1)}} \rangle_{\mathcal{D}}^{U(2)|U(2)}(t^{-1};q)
&+(14t^{-8}-4t^{-4}+3t^8)q^2+\cdots \\ \hline 
\langle V_{0;(2,1)} \rangle_{\mathcal{N}}^{U(2)|U(2)}(t;q)&
t^{-5}q^{\frac54}+3t^{-7}q^{\frac74}+(7t^{-9}-2t^{-5})q^{\frac94}+(13t^{-11}-6t^{-7}+t^{-3})q^{\frac{11}{4}} \\ 
=\langle W_{\tiny \yng(2,1)} \rangle_{\mathcal{D}}^{U(2)|U(2)}(t^{-1};q)
&+(22t^{-13}-12t^{-9}+t^{-5})q^{\frac{13}{4}}+\cdots \\ \hline 
\end{array}
\end{align}
The line defect half-indices $\langle V_{(1,1);(1,1)} \rangle_{\mathcal{N}}^{U(2)|U(2)}(t;q)$ 
and $\langle W_{\tiny \yng(1,1)}W_{\tiny\overline{\yng(1,1)}} \rangle_{\mathcal{D}}^{U(2)|U(2)}(t^{-1};q)$ 
are equal to the half-index $\mathbb{II}_{\mathcal{N}}^{U(2)|U(2)}(t;q)$ of the $U(2)|U(2)$ interface. 

%%%%%%%%%%%%%%%%%%%%%%%%%%%%%%%%%%%%%%%%%%%%%%%
\subsection{$U(3)|U(3)$}
%%%%%%%%%%%%%%%%%%%%%%%%%%%%%%%%%%%%%%%%%%%%%%%
For the Wilson lines at the NS5-type interface and the vortex lines at the D5-type interface 
we have confirmed the matching pairs up to the terms with $q^4$
\begin{align}
\begin{array}{c|c}
\textrm{line defect half-indices}&\textrm{expansions}\\ \hline
\langle W_{\tiny \yng(1);\overline{\yng(1)}} \rangle_{\mathcal{N}}^{U(3)|U(3)}(t;q)
&
tq^{\frac14}+(3t^{-1}+2t^3)q^{\frac34}+(8t^{-3}+3t+4t^5)q^{\frac54} \\ 
=\langle V_{(1,0,0)}\rangle_{\mathcal{D}}^{U(3)|U(3)}(t^{-1};q)
&+(16t^{-5}+5t^{-1}+4t^3+6t^7)q^{\frac74}+(30t^{-7}+2t^{-3}+4t+3t^5+9t^9)q^{\frac94}+\cdots\\ \hline 
\langle W_{\tiny \yng(1,1);\overline{\yng(1,1)}} \rangle_{\mathcal{N}}^{U(3)|U(3)}(t;q)&
t^2q^{\frac12}+(3+2t^4)q+(7t^{-2}+2t^2+4t^6)q^{\frac32}+(2+13t^{-4}+4t^4+6t^8)q^2 \\ 
=\langle V_{(1,1,0)}\rangle_{\mathcal{D}}^{U(3)|U(3)}(t^{-1};q)
&+(22t^{-6}-t^{-2}+6t^2+2t^6+9t^{10})q^{\frac52}+\cdots\\ \hline 
\langle W_{\tiny \yng(1,1,1);\overline{\yng(1,1,1)}} \rangle_{\mathcal{N}}^{U(3)|U(3)}(t;q)&
t^3q^{\frac34}+(t+t^5)q^{\frac54}+(2t^{-1}+t^7)q^{\frac74}+(3t^{-3}+2t+t^5+3t^9)q^{\frac94} \\ 
=\langle V_{(1,1,1)}\rangle_{\mathcal{D}}^{U(3)|U(3)}(t^{-1};q)
&+(4t^{-5}+2t^{-1}+t^3+4t^{11})q^{\frac{11}{4}}+\cdots\\ \hline 
\langle W_{\tiny \yng(2,1,1);\overline{\yng(2,1,1)}} \rangle_{\mathcal{N}}^{U(3)|U(3)}(t;q)&
t^4q+(2t^2+2t^4)q^{\frac32}+(4+2t^4+4t^8)q^2+(6t^{-2}+4t^2+3t^6+6t^{10})q^{\frac52} \\ 
=\langle V_{(2,1,1)}\rangle_{\mathcal{D}}^{U(3)|U(3)}(t^{-1};q)
&+(5+9t^{-4}+4t^4+2t^8+9t^{12})q^3+\cdots \\ \hline 
\end{array}
\end{align}
For the vortex lines at the NS5-type interface and the Wilson lines at the D5-type interface we have 
\begin{align}
\begin{array}{c|c}
\textrm{line defect half-indices}&\textrm{expansions}\\ \hline
\langle V_{0;(1,0,0)} \rangle_{\mathcal{N}}^{U(3)|U(3)}(t;q)
&
t^{-1}q^{\frac14}+(3t^{-3}+t)q^{\frac34}+(8t^{-5}+t^{-1}+2t^3)q^{\frac54} \\ 
=\langle W_{\tiny \yng(1)} \rangle_{\mathcal{D}}^{U(3)|U(3)}(t^{-1};q)
&+(17t^{-7}+t^{-3}+3t+2t^5)q^{\frac74}+\cdots\\ \hline 
\langle V_{(1,0,0);(1,0,0)} \rangle_{\mathcal{N}}^{U(3)|U(3)}(t;q)
&
1+(4t^{-2}+2t^2)q^{\frac12}+(4+12t^{-4}+4t^4)q \\ 
=\langle W_{\tiny \yng(1)}W_{\tiny \overline{\yng(1)}} \rangle_{\mathcal{D}}^{U(3)|U(3)}(t^{-1};q)
&+(28t^{-6}+7t^{-2}+6t^2+6t^6)q^{\frac32}+\cdots\\ \hline 
\langle V_{0;(1,1,0)} \rangle_{\mathcal{N}}^{U(3)|U(3)}(t;q)
&
t^{-4}q+(3t^{-6}+t^{-2})q^{\frac32}+(1+8t^{-8})q^2\\ 
=\langle W_{\tiny \yng(1,1)}\rangle_{\mathcal{D}}^{U(3)|U(3)}(t^{-1};q)
&+(17t^{-10}-t^{-6}+t^2)q^{\frac52}+(1+33t^{-12}-8t^{-8}+t^{-4}+t^4)q^3+\cdots \\ \hline 
\langle V_{(1,1,0);(1,1,0)} \rangle_{\mathcal{N}}^{U(3)|U(3)}(t;q)
&
1+(4t^{-2}+2t^2)q^{\frac12}+(4+12t^{-4}+4t^4)q\\ 
=\langle W_{\tiny \yng(1,1)}W_{\tiny \overline{\yng(1,1)}}\rangle_{\mathcal{D}}^{U(3)|U(3)}(t^{-1};q)
&+(28t^{-6}+7t^{-2}+6t^2+6t^6)q^{\frac32}+\cdots \\ \hline 
\langle V_{0;(1,1,1)} \rangle_{\mathcal{N}}^{U(3)|U(3)}(t;q)
&
t^{-9}q^{\frac94}+2t^{-11}q^{\frac{11}{4}}+(5t^{-13}-2t^{-9})q^{\frac{13}{4}} \\ 
=\langle W_{\tiny \yng(1,1,1)} \rangle_{\mathcal{D}}^{U(3)|U(3)}(t^{-1};q)
&+(10t^{-15}-4t^{-11}+t^{-7})q^{\frac{15}{4}}+\cdots \\ \hline 
\langle V_{(1,1,1);(1,1,1)} \rangle_{\mathcal{N}}^{U(3)|U(3)}(t;q)
&
1+(2t^{-2}+t^2)q^{\frac12}+(5t^{-4}+2t^4)q\\ 
=\langle W_{\tiny \yng(1,1,1)}W_{\tiny \overline{\yng(1,1,1)}} \rangle_{\mathcal{D}}^{U(3)|U(3)}(t^{-1};q)
&(3+18t^{-8}-2t^{-4}+4t^8)q^2+\cdots \\ \hline 
\langle V_{0;(2,1,1)} \rangle_{\mathcal{N}}^{U(3)|U(3)}(t;q)&
t^{-10}q^{\frac52}+3t^{-12}q^3+(8t^{-14}-2t^{-10})q^{\frac72}+\cdots \\ 
=\langle W_{\tiny \yng(2,1,1)} \rangle_{\mathcal{D}}^{U(3)|U(3)}(t^{-1};q)
& \\ \hline 
\langle V_{(2,1,1);(2,1,1)} \rangle_{\mathcal{N}}^{U(3)|U(3)}(t;q)&
1+(4t^{-2}+2t^2)q^{\frac12}+(4+12t^{-4}+4t^4)q\\ 
=\langle W_{\tiny \yng(2,1,1)}W_{\tiny \overline{\yng(2,1,1)}} \rangle_{\mathcal{D}}^{U(3)|U(3)}(t^{-1};q)
&+(28t^{-6}+7t^{-2}+6t^2+6t^6)q^{\frac32}+\cdots \\ \hline 
\end{array}
\end{align}
The line defect half-indices $\langle V_{(1,1,1);(1,1,1)} \rangle_{\mathcal{N}}^{U(3)|U(3)}(t;q)$ and 
$\langle W_{\tiny \yng(1,1,1)}W_{\tiny\overline{\yng(1,1,1)}} \rangle_{\mathcal{D}}^{U(3)|U(3)}(t^{-1};q)$, 
are equal to the half-index $\mathbb{II}_{\mathcal{N}}^{U(3)|U(3)}(t;q)$ of the $U(3)|U(3)$ interface.

%%%%%%%%%%%%%%%%%%%%%%%%%%%%%%%%%%%%%%%%%%%%%%%
\subsection{$U(3)|U(2)$}
%%%%%%%%%%%%%%%%%%%%%%%%%%%%%%%%%%%%%%%%%%%%%%%
For the Wilson lines at the NS5-type interface and the vortex lines at the D5-type interface we have 
\begin{align}
\begin{array}{c|c}
\textrm{line defect half-indices}&\textrm{expansions}\\ \hline
\langle W_{\tiny \yng(1);\overline{\yng(1)}} \rangle_{\mathcal{N}}^{U(3)|U(2)}(t;q)
&
tq^{\frac14}+(3t^{-1}+2t^3)q^{\frac34}+(7t^{-3}+2t+3t^5)q^{\frac54} \\ 
=\langle V_{(1,0)}\rangle_{\mathcal{D}}^{U(3)|U(2)}(t^{-1};q)
&+(13t^{-5}+t^{-1}+t^3+4t^7)q^{\frac74}+(22t^{-7}-3t^{-3};2t+t^5+5t^9)q^{\frac94}+\cdots\\ \hline 
\langle W_{\tiny \yng(1,1);\overline{\yng(1,1)}} \rangle_{\mathcal{N}}^{U(3)|U(2)}(t;q)&
t^2q^{\frac12}+(2+t^4)q+(4t^{-2}+2t^6)q^{\frac32}+(6t^{-4}+t^4+2t^8)q^2 \\ 
=\langle V_{(1,1)}\rangle_{\mathcal{D}}^{U(3)|U(2)}(t^{-1};q)
&+(9t^{-6}+t^2+3t^{10})q^{\frac52}+(1+12t^{-8}+2t^{-4}+2t^4+t^8+3t^{12})q^3+\cdots\\ \hline 
\langle W_{\tiny \yng(2,1);\overline{\yng(2,1)}} \rangle_{\mathcal{N}}^{U(3)|U(2)}(t;q)
&
t^3q^{\frac34}+(3t+2t^5)q^{\frac54}+(6t^{-1}+2t^3+3t^7)q^{\frac74} \\ 
=\langle V_{(2,1)}\rangle_{\mathcal{D}}^{U(3)|U(2)}(t^{-1};q)
&+(10t^{-3}+t+t^5+4t^9)q^{\frac94}+(15t^{-5}+t^{-1}+2t^3+t^{7}+5t^{11})q^{\frac{11}{4}}+\cdots\\ \hline 
\end{array}
\end{align}
For the vortex lines at the NS5-type interface and the Wilson lines at the D5-type interface we have 
\begin{align}
\begin{array}{c|c}
\textrm{line defect half-indices}&\textrm{expansions}\\ \hline
\langle V_{0;(1,0)} \rangle_{\mathcal{N}}^{U(3)|U(2)}(t;q)
&
t^{-2}q^{\frac12}+(1+3t^{-4})q+(7t^{-6}+t^2)q^{\frac32} \\ 
=\langle W_{\tiny \yng(1)} \rangle_{\mathcal{D}}^{U(3)|U(2)}(t^{-1};q)
&+(14t^{-8}-2t^{-4}+t^4)q^2+(25t^{-10}-7t^{-6}+t^{-2}+t^2+t^6)q^{\frac52}+\cdots\\ \hline 
\langle V_{0;(1,1)} \rangle_{\mathcal{N}}^{U(3)|U(2)}(t;q)
&
t^{-6}q^{\frac32}+2t^{-8}q^2+(5t^{-10}-2t^{-6})q^{\frac52} \\ 
=\langle W_{\tiny \yng(1,1)} \rangle_{\mathcal{D}}^{U(3)|U(2)}(t^{-1};q)
&+(9t^{-12}-4t^{-8}+t^{-4})q^3+(16t^{-14}-9t^{-10}+t^{-6})q^{\frac72}+\cdots\\ \hline 
\langle V_{0;(2,1)} \rangle_{\mathcal{N}}^{U(3)|U(2)}(t;q)
&
t^{-8}q^2+3t^{-10}q^{\frac52}+(7t^{-12}-2t^{-8})q^3 \\ 
=\langle W_{\tiny \yng(2,1)} \rangle_{\mathcal{D}}^{U(3)|U(2)}(t^{-1};q)
&+(14t^{-14}-6t^{-10}+t^{-6})q^{\frac72}+(25t^{-16}-13t^{-12}+t^{-8})q^4+\cdots\\ \hline 
\end{array}
\end{align}

%%%%%%%%%%%%%%%%%%%%%%%%%%%%%%%%%%%%%%%%%%%%%%%
\subsection{$U(4)|U(2)$}
%%%%%%%%%%%%%%%%%%%%%%%%%%%%%%%%%%%%%%%%%%%%%%%
For the Wilson lines at the NS5-type interface and the vortex lines at the D5-type interface 
the matching pairs have been confirmed up to the terms with $q^4$
\begin{align}
\begin{array}{c|c}
\textrm{line defect half-indices}&\textrm{expansions}\\ \hline
\langle W_{\tiny \yng(1);\overline{\yng(1)}} \rangle_{\mathcal{N}}^{U(4)|U(2)}(t;q)
&
tq^{\frac14}+(3t^{-1}+2t^3)q^{\frac34}+(7t^{-3}+2t+3t^5)q^{\frac54} \\ 
=\langle V_{(1,0)}\rangle_{\mathcal{D}}^{U(4)|U(2)}(t^{-1};q)
&+(14t^{-5}+2t^{-1}+t^3+4t^7)q^{\frac74}+(25t^{-7}+3t+t^5+5t^9)q^{\frac94}+\cdots\\ \hline 
\langle W_{\tiny \yng(1,1);\overline{\yng(1,1)}} \rangle_{\mathcal{N}}^{U(4)|U(2)}(t;q)
&
t^2q^{\frac12}+(2+t^4)q+(5t^{-2}+2t^6)q^{\frac32} \\ 
=\langle V_{(1,1)}\rangle_{\mathcal{D}}^{U(4)|U(2)}(t^{-1};q)
&(8t^{-4}+t^4+2t^8)q^2+(2+20t^{-8}-2t^{-4}+2t^4+t^8+3t^{12})q^3+\cdots\\ \hline 
\langle W_{\tiny \yng(2,1);\overline{\yng(2,1)}} \rangle_{\mathcal{N}}^{U(4)|U(2)}(t;q)
&
t^3q^{\frac34}+(3t+2t^5)q^{\frac54}+(7t^{-1}+2t^3+3t^7)q^{\frac74} \\ 
=\langle V_{(2,1)}\rangle_{\mathcal{D}}^{U(4)|U(2)}(t^{-1};q)
&+(13t^{-3}+2t+t^5+4t^9)q^{\frac94}+(22t^{-5}+3t^3+t^7+5t^{11})q^{\frac{11}{4}}+\cdots\\ \hline 
\end{array}
\end{align}
For the vortex lines at the NS5-type interface and the Wilson lines at the D5-type interface we obtain 
\begin{align}
\begin{array}{c|c}
\textrm{line defect half-indices}&\textrm{expansions}\\ \hline
\langle V_{0;(1,0)} \rangle_{\mathcal{N}}^{U(4)|U(2)}(t;q)
&
t^{-3}q^{\frac34}+(3t^{-5}+t^{-1})q^{\frac54}+(7t^{-7}+t)q^{\frac74} \\ 
=\langle W_{\tiny \yng(1)} \rangle_{\mathcal{D}}^{U(4)|U(2)}(t^{-1};q)
&(t^{-3}-2t^5+14t^9)q^{\frac94}+(t^{-5}+t^{-1}+t^3-6t^7+26t^{11})q^{\frac{11}{4}}+\cdots\\ \hline 
\langle V_{0;(1,1)} \rangle_{\mathcal{N}}^{U(4)|U(2)}(t;q)
&
t^{-8}q^2+2t^{-10}q^{\frac52}+(5t^{-12}-2t^{-8})q^3 \\ 
=\langle W_{\tiny \yng(1,1)} \rangle_{\mathcal{D}}^{U(4)|U(2)}(t^{-1};q)
&+(9t^{-14}-4t^{-10}+t^{-6})q^{\frac72}+\cdots\\ \hline 
\langle V_{0;(2,1)} \rangle_{\mathcal{N}}^{U(4)|U(2)}(t;q)
&
t^{-11}q^{\frac{11}{4}}+3t^{-13}q^{\frac{13}{4}}+(7t^{-15}-2t^{-11})q^{\frac{15}{4}}+\cdots \\ 
=\langle W_{\tiny \yng(2,1)} \rangle_{\mathcal{D}}^{U(4)|U(2)}(t^{-1};q)
& \\ \hline 
\end{array}
\end{align}

\bibliographystyle{utphys}
\bibliography{ref}

\def\polhk#1{\setbox0=\hbox{#1}{\ooalign{\hidewidth
  \lower1.5ex\hbox{`}\hidewidth\crcr\unhbox0}}} \def\cprime{$'$}
\providecommand{\href}[2]{#2}\begingroup\raggedright\begin{thebibliography}{10}

\bibitem{Hatsuda:2025yzp}
Y.~Hatsuda and T.~Okazaki, ``{S-duality of boundary lines in $ \mathcal{N} $ =
  4 SYM theories and supersymmetric indices},''
  \href{http://dx.doi.org/10.1007/JHEP08(2025)127}{{\em JHEP} {\bfseries 08}
  (2025) 127}, \href{http://arxiv.org/abs/2505.14962}{{\ttfamily
  arXiv:2505.14962 [hep-th]}}.

\bibitem{Gaiotto:2008sa}
D.~Gaiotto and E.~Witten, ``{Supersymmetric Boundary Conditions in N=4 Super
  Yang-Mills Theory},'' \href{http://dx.doi.org/10.1007/s10955-009-9687-3}{{\em
  J. Statist. Phys.} {\bfseries 135} (2009) 789--855},
\href{http://arxiv.org/abs/0804.2902}{{\ttfamily arXiv:0804.2902 [hep-th]}}.
%%CITATION = ARXIV:0804.2902;%%.

\bibitem{Nahm:1979yw}
W.~Nahm, ``{A Simple Formalism for the BPS Monopole},''
  \href{http://dx.doi.org/10.1016/0370-2693(80)90961-2}{{\em Phys. Lett. B}
  {\bfseries 90} (1980) 413--414}.

\bibitem{Diaconescu:1996rk}
D.-E. Diaconescu, ``{D-branes, monopoles and Nahm equations},''
  \href{http://dx.doi.org/10.1016/S0550-3213(97)00438-0}{{\em Nucl. Phys. B}
  {\bfseries 503} (1997) 220--238},
  \href{http://arxiv.org/abs/hep-th/9608163}{{\ttfamily arXiv:hep-th/9608163}}.

\bibitem{tHooft:1977nqb}
G.~'t~Hooft, ``{On the Phase Transition Towards Permanent Quark Confinement},''
  \href{http://dx.doi.org/10.1016/0550-3213(78)90153-0}{{\em Nucl. Phys. B}
  {\bfseries 138} (1978) 1--25}.

\bibitem{Kapustin:2005py}
A.~Kapustin, ``{Wilson-'t Hooft operators in four-dimensional gauge theories
  and S-duality},'' \href{http://dx.doi.org/10.1103/PhysRevD.74.025005}{{\em
  Phys. Rev. D} {\bfseries 74} (2006) 025005},
  \href{http://arxiv.org/abs/hep-th/0501015}{{\ttfamily arXiv:hep-th/0501015}}.

\bibitem{Kapustin:2006pk}
A.~Kapustin and E.~Witten, ``{Electric-Magnetic Duality And The Geometric
  Langlands Program},''
  \href{http://dx.doi.org/10.4310/CNTP.2007.v1.n1.a1}{{\em Commun. Num. Theor.
  Phys.} {\bfseries 1} (2007) 1--236},
  \href{http://arxiv.org/abs/hep-th/0604151}{{\ttfamily arXiv:hep-th/0604151}}.

\bibitem{Gaiotto:2011nm}
D.~Gaiotto and E.~Witten, ``{Knot Invariants from Four-Dimensional Gauge
  Theory},'' \href{http://dx.doi.org/10.4310/ATMP.2012.v16.n3.a5}{{\em Adv.
  Theor. Math. Phys.} {\bfseries 16} no.~3, (2012) 935--1086},
  \href{http://arxiv.org/abs/1106.4789}{{\ttfamily arXiv:1106.4789 [hep-th]}}.

\bibitem{Witten:2011zz}
E.~Witten, ``{Fivebranes and Knots},''
\href{http://arxiv.org/abs/1101.3216}{{\ttfamily arXiv:1101.3216 [hep-th]}}.
%%CITATION = ARXIV:1101.3216;%%.

\bibitem{MR1354144}
I.~G. Macdonald, {\em Symmetric functions and Hall polynomials}.
\newblock Oxford Mathematical Monographs. The Clarendon Press, Oxford
  University Press, New York, second~ed., 1995.
\newblock With contributions by A. Zelevinsky, Oxford Science Publications.

\bibitem{MR1976581}
I.~G. Macdonald, \href{http://dx.doi.org/10.1017/CBO9780511542824}{{\em Affine
  {H}ecke algebras and orthogonal polynomials}}, vol.~157 of {\em Cambridge
  Tracts in Mathematics}.
\newblock Cambridge University Press, Cambridge, 2003.
\newblock \url{https://doi.org/10.1017/CBO9780511542824}.

\bibitem{MR1314036}
I.~Cherednik, ``Double affine {H}ecke algebras and {M}acdonald's conjectures,''
  \href{http://dx.doi.org/10.2307/2118632}{{\em Ann. of Math. (2)} {\bfseries
  141} no.~1, (1995) 191--216}. \url{https://doi.org/10.2307/2118632}.

\bibitem{MR1354956}
I.~Cherednik, ``Macdonald's evaluation conjectures and difference {F}ourier
  transform,'' \href{http://dx.doi.org/10.1007/BF01231441}{{\em Invent. Math.}
  {\bfseries 122} no.~1, (1995) 119--145}.
  \url{https://doi.org/10.1007/BF01231441}.

\bibitem{MR1185831}
I.~Cherednik, ``Double affine {H}ecke algebras, {K}nizhnik-{Z}amolodchikov
  equations, and {M}acdonald's operators,''
  \href{http://dx.doi.org/10.1155/S1073792892000199}{{\em Internat. Math. Res.
  Notices} no.~9, (1992) 171--180}.
  \url{https://doi.org/10.1155/S1073792892000199}.

\bibitem{MR1358032}
I.~Cherednik, ``Nonsymmetric {M}acdonald polynomials,''
  \href{http://dx.doi.org/10.1155/S1073792895000341}{{\em Internat. Math. Res.
  Notices} no.~10, (1995) 483--515}.
  \url{https://doi.org/10.1155/S1073792895000341}.

\bibitem{MR1613515}
I.~Cherednik, ``Intertwining operators of double affine {H}ecke algebras,''
  \href{http://dx.doi.org/10.1007/s000290050017}{{\em Selecta Math. (N.S.)}
  {\bfseries 3} no.~4, (1997) 459--495}.
  \url{https://doi.org/10.1007/s000290050017}.

\bibitem{MR1768938}
S.~Sahi, \href{http://dx.doi.org/10.1090/conm/254/03963}{``Some properties of
  {K}oornwinder polynomials,''} in {\em {$q$}-series from a contemporary
  perspective ({S}outh {H}adley, {MA}, 1998)}, vol.~254 of {\em Contemp.
  Math.}, pp.~395--411.
\newblock Amer. Math. Soc., Providence, RI, 2000.
\newblock \url{https://doi.org/10.1090/conm/254/03963}.

\bibitem{MR1715325}
S.~Sahi, ``Nonsymmetric {K}oornwinder polynomials and duality,''
  \href{http://dx.doi.org/10.2307/121102}{{\em Ann. of Math. (2)} {\bfseries
  150} no.~1, (1999) 267--282}. \url{https://doi.org/10.2307/121102}.

\bibitem{MR1792347}
J.~V. Stokman, ``Koornwinder polynomials and affine {H}ecke algebras,''
  \href{http://dx.doi.org/10.1155/S1073792800000520}{{\em Internat. Math. Res.
  Notices} no.~19, (2000) 1005--1042}.
  \url{https://doi.org/10.1155/S1073792800000520}.

\bibitem{MR1411136}
J.~F. van Diejen, ``Self-dual {K}oornwinder-{M}acdonald polynomials,''
  \href{http://dx.doi.org/10.1007/s002220050102}{{\em Invent. Math.} {\bfseries
  126} no.~2, (1996) 319--339}. \url{https://doi.org/10.1007/s002220050102}.

\bibitem{Gaiotto:2008ak}
D.~Gaiotto and E.~Witten, ``{S-Duality of Boundary Conditions In N=4 Super
  Yang-Mills Theory},''
  \href{http://dx.doi.org/10.4310/ATMP.2009.v13.n3.a5}{{\em Adv. Theor. Math.
  Phys.} {\bfseries 13} no.~3, (2009) 721--896},
\href{http://arxiv.org/abs/0807.3720}{{\ttfamily arXiv:0807.3720 [hep-th]}}.
%%CITATION = ARXIV:0807.3720;%%.

\bibitem{Hanany:1996ie}
A.~Hanany and E.~Witten, ``{Type IIB superstrings, BPS monopoles, and
  three-dimensional gauge dynamics},''
  \href{http://dx.doi.org/10.1016/S0550-3213(97)00157-0}{{\em Nucl.Phys.}
  {\bfseries B492} (1997) 152--190},
\href{http://arxiv.org/abs/hep-th/9611230}{{\ttfamily arXiv:hep-th/9611230
  [hep-th]}}.
%%CITATION = HEP-TH/9611230;%%.

\bibitem{Gaiotto:2019jvo}
D.~Gaiotto and T.~Okazaki, ``{Dualities of Corner Configurations and
  Supersymmetric Indices},''
  \href{http://dx.doi.org/10.1007/JHEP11(2019)056}{{\em JHEP} {\bfseries 11}
  (2019) 056},
\href{http://arxiv.org/abs/1902.05175}{{\ttfamily arXiv:1902.05175 [hep-th]}}.
%%CITATION = ARXIV:1902.05175;%%.

\bibitem{Maldacena:1998im}
J.~M. Maldacena, ``{Wilson loops in large N field theories},''
  \href{http://dx.doi.org/10.1103/PhysRevLett.80.4859}{{\em Phys. Rev. Lett.}
  {\bfseries 80} (1998) 4859--4862},
  \href{http://arxiv.org/abs/hep-th/9803002}{{\ttfamily arXiv:hep-th/9803002}}.

\bibitem{Rey:1998ik}
S.-J. Rey and J.-T. Yee, ``{Macroscopic strings as heavy quarks in large N
  gauge theory and anti-de Sitter supergravity},''
  \href{http://dx.doi.org/10.1007/s100520100799}{{\em Eur. Phys. J. C}
  {\bfseries 22} (2001) 379--394},
  \href{http://arxiv.org/abs/hep-th/9803001}{{\ttfamily arXiv:hep-th/9803001}}.

\bibitem{Drukker:2005kx}
N.~Drukker and B.~Fiol, ``{All-genus calculation of Wilson loops using
  D-branes},'' \href{http://dx.doi.org/10.1088/1126-6708/2005/02/010}{{\em
  JHEP} {\bfseries 02} (2005) 010},
  \href{http://arxiv.org/abs/hep-th/0501109}{{\ttfamily arXiv:hep-th/0501109}}.

\bibitem{Gomis:2006im}
J.~Gomis and F.~Passerini, ``{Wilson Loops as D3-Branes},''
  \href{http://dx.doi.org/10.1088/1126-6708/2007/01/097}{{\em JHEP} {\bfseries
  01} (2007) 097},
\href{http://arxiv.org/abs/hep-th/0612022}{{\ttfamily arXiv:hep-th/0612022
  [hep-th]}}.
%%CITATION = HEP-TH/0612022;%%.

\bibitem{Gomis:2006sb}
J.~Gomis and F.~Passerini, ``{Holographic Wilson Loops},''
  \href{http://dx.doi.org/10.1088/1126-6708/2006/08/074}{{\em JHEP} {\bfseries
  08} (2006) 074}, \href{http://arxiv.org/abs/hep-th/0604007}{{\ttfamily
  arXiv:hep-th/0604007}}.

\bibitem{Rodriguez-Gomez:2006fmx}
D.~Rodriguez-Gomez, ``{Computing Wilson lines with dielectric branes},''
  \href{http://dx.doi.org/10.1016/j.nuclphysb.2006.06.037}{{\em Nucl. Phys. B}
  {\bfseries 752} (2006) 316--326},
  \href{http://arxiv.org/abs/hep-th/0604031}{{\ttfamily arXiv:hep-th/0604031}}.

\bibitem{Yamaguchi:2007ps}
S.~Yamaguchi, ``{Semi-classical open string corrections and symmetric Wilson
  loops},'' \href{http://dx.doi.org/10.1088/1126-6708/2007/06/073}{{\em JHEP}
  {\bfseries 06} (2007) 073},
  \href{http://arxiv.org/abs/hep-th/0701052}{{\ttfamily arXiv:hep-th/0701052}}.

\bibitem{Yamaguchi:2006tq}
S.~Yamaguchi, ``{Wilson loops of anti-symmetric representation and
  D5-branes},'' \href{http://dx.doi.org/10.1088/1126-6708/2006/05/037}{{\em
  JHEP} {\bfseries 05} (2006) 037},
  \href{http://arxiv.org/abs/hep-th/0603208}{{\ttfamily arXiv:hep-th/0603208}}.

\bibitem{Hartnoll:2006hr}
S.~A. Hartnoll and S.~P. Kumar, ``{Multiply wound Polyakov loops at strong
  coupling},'' \href{http://dx.doi.org/10.1103/PhysRevD.74.026001}{{\em Phys.
  Rev. D} {\bfseries 74} (2006) 026001},
  \href{http://arxiv.org/abs/hep-th/0603190}{{\ttfamily arXiv:hep-th/0603190}}.

\bibitem{Assel:2015oxa}
B.~Assel and J.~Gomis, ``{Mirror Symmetry And Loop Operators},''
  \href{http://dx.doi.org/10.1007/JHEP11(2015)055}{{\em JHEP} {\bfseries 11}
  (2015) 055},
\href{http://arxiv.org/abs/1506.01718}{{\ttfamily arXiv:1506.01718 [hep-th]}}.
%%CITATION = ARXIV:1506.01718;%%.

\bibitem{Drukker:2012sr}
N.~Drukker, T.~Okuda, and F.~Passerini, ``{Exact results for vortex loop
  operators in 3d supersymmetric theories},''
  \href{http://dx.doi.org/10.1007/JHEP07(2014)137}{{\em JHEP} {\bfseries 07}
  (2014) 137}, \href{http://arxiv.org/abs/1211.3409}{{\ttfamily arXiv:1211.3409
  [hep-th]}}.

\bibitem{Okazaki:2019ony}
T.~Okazaki, ``{Mirror symmetry of 3D $\mathcal{N}=4$ gauge theories and
  supersymmetric indices},''
  \href{http://dx.doi.org/10.1103/PhysRevD.100.066031}{{\em Phys. Rev.}
  {\bfseries D100} no.~6, (2019) 066031},
\href{http://arxiv.org/abs/1905.04608}{{\ttfamily arXiv:1905.04608 [hep-th]}}.
%%CITATION = ARXIV:1905.04608;%%.

\bibitem{Hayashi:2025guk}
H.~Hayashi, T.~Nosaka, and T.~Okazaki, ``{Abelian dualities and line defect
  indices for 3d gauge theories},''
  \href{http://arxiv.org/abs/2506.01278}{{\ttfamily arXiv:2506.01278
  [hep-th]}}.

\bibitem{Hatsuda:2024uwt}
Y.~Hatsuda, H.~Lin, and T.~Okazaki, ``{Giant graviton expansions and ETW
  brane},'' \href{http://dx.doi.org/10.1007/JHEP09(2024)181}{{\em JHEP}
  {\bfseries 09} (2024) 181}, \href{http://arxiv.org/abs/2405.14564}{{\ttfamily
  arXiv:2405.14564 [hep-th]}}.

\bibitem{Hatsuda:2025mvj}
Y.~Hatsuda, ``{Deformed Schur indices and Macdonald polynomials},''
  \href{http://arxiv.org/abs/2503.03952}{{\ttfamily arXiv:2503.03952
  [hep-th]}}.

\bibitem{Witten:1998xy}
E.~Witten, ``{Baryons and branes in anti-de Sitter space},''
  \href{http://dx.doi.org/10.1088/1126-6708/1998/07/006}{{\em JHEP} {\bfseries
  07} (1998) 006}, \href{http://arxiv.org/abs/hep-th/9805112}{{\ttfamily
  arXiv:hep-th/9805112}}.

\bibitem{Hanany:2000fq}
A.~Hanany and B.~Kol, ``{On orientifolds, discrete torsion, branes and M
  theory},'' \href{http://dx.doi.org/10.1088/1126-6708/2000/06/013}{{\em JHEP}
  {\bfseries 06} (2000) 013},
  \href{http://arxiv.org/abs/hep-th/0003025}{{\ttfamily arXiv:hep-th/0003025}}.

\bibitem{Feng:2000eq}
B.~Feng and A.~Hanany, ``{Mirror symmetry by O3 planes},''
  \href{http://dx.doi.org/10.1088/1126-6708/2000/11/033}{{\em JHEP} {\bfseries
  11} (2000) 033},
\href{http://arxiv.org/abs/hep-th/0004092}{{\ttfamily arXiv:hep-th/0004092
  [hep-th]}}.
%%CITATION = HEP-TH/0004092;%%.

\bibitem{Evans:1997hk}
N.~J. Evans, C.~V. Johnson, and A.~D. Shapere, ``{Orientifolds, branes, and
  duality of 4-D gauge theories},''
  \href{http://dx.doi.org/10.1016/S0550-3213(97)00384-2}{{\em Nucl. Phys. B}
  {\bfseries 505} (1997) 251--271},
  \href{http://arxiv.org/abs/hep-th/9703210}{{\ttfamily arXiv:hep-th/9703210}}.

\bibitem{Aharony:2013hda}
O.~Aharony, N.~Seiberg, and Y.~Tachikawa, ``{Reading between the lines of
  four-dimensional gauge theories},''
  \href{http://dx.doi.org/10.1007/JHEP08(2013)115}{{\em JHEP} {\bfseries 08}
  (2013) 115}, \href{http://arxiv.org/abs/1305.0318}{{\ttfamily arXiv:1305.0318
  [hep-th]}}.

\bibitem{Hatsuda:2024lcc}
Y.~Hatsuda, H.~Lin, and T.~Okazaki, ``{Orbifold ETW brane and half-indices},''
  \href{http://dx.doi.org/10.1007/JHEP12(2024)227}{{\em JHEP} {\bfseries 12}
  (2025) 227}, \href{http://arxiv.org/abs/2409.16841}{{\ttfamily
  arXiv:2409.16841 [hep-th]}}.

\bibitem{Hatsuda:2025jze}
Y.~Hatsuda, H.~Lin, and T.~Okazaki, ``{$\mathcal{N}=4$ line defect correlators
  of type BCD},'' \href{http://arxiv.org/abs/2502.18110}{{\ttfamily
  arXiv:2502.18110 [hep-th]}}.

\bibitem{Nagasaki:2011ue}
K.~Nagasaki, H.~Tanida, and S.~Yamaguchi, ``{Holographic Interface-Particle
  Potential},'' \href{http://dx.doi.org/10.1007/JHEP01(2012)139}{{\em JHEP}
  {\bfseries 01} (2012) 139}, \href{http://arxiv.org/abs/1109.1927}{{\ttfamily
  arXiv:1109.1927 [hep-th]}}.

\bibitem{Estes:2012nx}
J.~Estes, A.~O'Bannon, E.~Tsatis, and T.~Wrase, ``{Holographic Wilson Loops,
  Dielectric Interfaces, and Topological Insulators},''
  \href{http://dx.doi.org/10.1103/PhysRevD.87.106005}{{\em Phys. Rev. D}
  {\bfseries 87} no.~10, (2013) 106005},
  \href{http://arxiv.org/abs/1210.0534}{{\ttfamily arXiv:1210.0534 [hep-th]}}.

\bibitem{Nagasaki:2013hwa}
K.~Nagasaki and S.~Yamaguchi, ``{\textquoteright{}t Hooft operators on an
  interface and bubbling D5-branes},''
  \href{http://dx.doi.org/10.1103/PhysRevD.89.046002}{{\em Phys. Rev. D}
  {\bfseries 89} no.~4, (2014) 046002},
  \href{http://arxiv.org/abs/1309.3125}{{\ttfamily arXiv:1309.3125 [hep-th]}}.

\bibitem{Karch:2022rvr}
A.~Karch, H.~Sun, and C.~F. Uhlemann, ``{Double holography in string theory},''
  \href{http://dx.doi.org/10.1007/JHEP10(2022)012}{{\em JHEP} {\bfseries 10}
  (2022) 012}, \href{http://arxiv.org/abs/2206.11292}{{\ttfamily
  arXiv:2206.11292 [hep-th]}}.

\bibitem{Kristjansen:2024dnm}
C.~Kristjansen and K.~Zarembo, ``{Integrable Holographic Defect CFTs},''
\newblock 1, 2024.
\newblock \href{http://arxiv.org/abs/2401.17144}{{\ttfamily arXiv:2401.17144
  [hep-th]}}.

\bibitem{Gang:2012yr}
D.~Gang, E.~Koh, and K.~Lee, ``{Line Operator Index on $S^{1}\times S^{3}$},''
  \href{http://dx.doi.org/10.1007/JHEP05(2012)007}{{\em JHEP} {\bfseries 05}
  (2012) 007},
\href{http://arxiv.org/abs/1201.5539}{{\ttfamily arXiv:1201.5539 [hep-th]}}.
%%CITATION = ARXIV:1201.5539;%%.

\bibitem{Drukker:2015spa}
N.~Drukker, ``{The $ \mathcal{N}=4 $ Schur index with Polyakov loops},''
  \href{http://dx.doi.org/10.1007/JHEP12(2015)012}{{\em JHEP} {\bfseries 12}
  (2015) 012}, \href{http://arxiv.org/abs/1510.02480}{{\ttfamily
  arXiv:1510.02480 [hep-th]}}.

\bibitem{Hatsuda:2023iwi}
Y.~Hatsuda and T.~Okazaki, ``{Exact $ \mathcal{N} $ = 2$^{*}$ Schur line defect
  correlators},'' \href{http://dx.doi.org/10.1007/JHEP06(2023)169}{{\em JHEP}
  {\bfseries 06} (2023) 169}, \href{http://arxiv.org/abs/2303.14887}{{\ttfamily
  arXiv:2303.14887 [hep-th]}}.

\bibitem{Hatsuda:2023iof}
Y.~Hatsuda and T.~Okazaki, ``{Excitations of bubbling geometries for line
  defects},'' \href{http://dx.doi.org/10.1103/PhysRevD.109.066013}{{\em Phys.
  Rev. D} {\bfseries 109} no.~6, (2024) 066013},
  \href{http://arxiv.org/abs/2311.13740}{{\ttfamily arXiv:2311.13740
  [hep-th]}}.

\bibitem{Imamura:2024lkw}
Y.~Imamura, ``{Giant Graviton Expansions for the Line Operator Index},''
  \href{http://dx.doi.org/10.1093/ptep/ptae084}{{\em PTEP} {\bfseries 2024}
  no.~6, (2024) 063B03}, \href{http://arxiv.org/abs/2403.11543}{{\ttfamily
  arXiv:2403.11543 [hep-th]}}.

\bibitem{Imamura:2024pgp}
Y.~Imamura and M.~Inoue, ``{Brane expansions for anti-symmetric line operator
  index},'' \href{http://dx.doi.org/10.1007/JHEP08(2024)020}{{\em JHEP}
  {\bfseries 08} (2024) 020}, \href{http://arxiv.org/abs/2404.08302}{{\ttfamily
  arXiv:2404.08302 [hep-th]}}.

\bibitem{Beccaria:2024oif}
M.~Beccaria, ``{Schur line defect correlators and giant graviton expansion},''
  \href{http://dx.doi.org/10.1007/JHEP06(2024)088}{{\em JHEP} {\bfseries 06}
  (2024) 088}, \href{http://arxiv.org/abs/2403.14553}{{\ttfamily
  arXiv:2403.14553 [hep-th]}}.

\bibitem{Imamura:2024zvw}
Y.~Imamura, A.~Sei, and D.~Yokoyama, ``{Giant graviton expansion for general
  Wilson line operator indices},''
  \href{http://dx.doi.org/10.1007/JHEP09(2024)202}{{\em JHEP} {\bfseries 09}
  (2024) 202}, \href{http://arxiv.org/abs/2406.19777}{{\ttfamily
  arXiv:2406.19777 [hep-th]}}.

\bibitem{Imamura:2025fqa}
Y.~Imamura and A.~Sei, ``{D1-brane correction to a line operator index},''
  \href{http://arxiv.org/abs/2506.05788}{{\ttfamily arXiv:2506.05788
  [hep-th]}}.

\bibitem{Brennan:2018yuj}
T.~D. Brennan, A.~Dey, and G.~W. Moore, ``{On \textquoteright{}t Hooft defects,
  monopole bubbling and supersymmetric quantum mechanics},''
  \href{http://dx.doi.org/10.1007/JHEP09(2018)014}{{\em JHEP} {\bfseries 09}
  (2018) 014}, \href{http://arxiv.org/abs/1801.01986}{{\ttfamily
  arXiv:1801.01986 [hep-th]}}.

\bibitem{MR2109105}
N.~Bourbaki, {\em Lie groups and {L}ie algebras. {C}hapters 7--9}.
\newblock Elements of Mathematics (Berlin). Springer-Verlag, Berlin, 2005.
\newblock Translated from the 1975 and 1982 French originals by Andrew
  Pressley.

\bibitem{Dimofte:2011py}
T.~Dimofte, D.~Gaiotto, and S.~Gukov, ``{3-Manifolds and 3d Indices},''
  \href{http://dx.doi.org/10.4310/ATMP.2013.v17.n5.a3}{{\em Adv. Theor. Math.
  Phys.} {\bfseries 17} no.~5, (2013) 975--1076},
\href{http://arxiv.org/abs/1112.5179}{{\ttfamily arXiv:1112.5179 [hep-th]}}.
%%CITATION = ARXIV:1112.5179;%%.

\bibitem{Gang:2012ff}
D.~Gang, E.~Koh, and K.~Lee, ``{Superconformal Index with Duality Domain
  Wall},'' \href{http://dx.doi.org/10.1007/JHEP10(2012)187}{{\em JHEP}
  {\bfseries 10} (2012) 187},
\href{http://arxiv.org/abs/1205.0069}{{\ttfamily arXiv:1205.0069 [hep-th]}}.
%%CITATION = ARXIV:1205.0069;%%.

\bibitem{Cordova:2016uwk}
C.~Cordova, D.~Gaiotto, and S.-H. Shao, ``{Infrared Computations of Defect
  Schur Indices},'' \href{http://dx.doi.org/10.1007/JHEP11(2016)106}{{\em JHEP}
  {\bfseries 11} (2016) 106}, \href{http://arxiv.org/abs/1606.08429}{{\ttfamily
  arXiv:1606.08429 [hep-th]}}.

\bibitem{Kim:2009wb}
S.~Kim, ``{The Complete superconformal index for N=6 Chern-Simons theory},''
  \href{http://dx.doi.org/10.1016/j.nuclphysb.2009.06.025}{{\em Nucl. Phys. B}
  {\bfseries 821} (2009) 241--284},
  \href{http://arxiv.org/abs/0903.4172}{{\ttfamily arXiv:0903.4172 [hep-th]}}.
  [Erratum: Nucl.Phys.B 864, 884 (2012)].

\bibitem{Imamura:2011su}
Y.~Imamura and S.~Yokoyama, ``{Index for three dimensional superconformal field
  theories with general R-charge assignments},''
  \href{http://dx.doi.org/10.1007/JHEP04(2011)007}{{\em JHEP} {\bfseries 04}
  (2011) 007}, \href{http://arxiv.org/abs/1101.0557}{{\ttfamily arXiv:1101.0557
  [hep-th]}}.

\bibitem{Kapustin:2011jm}
A.~Kapustin and B.~Willett, ``{Generalized Superconformal Index for Three
  Dimensional Field Theories},''
\href{http://arxiv.org/abs/1106.2484}{{\ttfamily arXiv:1106.2484 [hep-th]}}.
%%CITATION = ARXIV:1106.2484;%%.

\bibitem{Razamat:2014pta}
S.~S. Razamat and B.~Willett, ``{Down the rabbit hole with theories of class $
  \mathcal{S} $},'' \href{http://dx.doi.org/10.1007/JHEP10(2014)099}{{\em JHEP}
  {\bfseries 10} (2014) 99},
\href{http://arxiv.org/abs/1403.6107}{{\ttfamily arXiv:1403.6107 [hep-th]}}.
%%CITATION = ARXIV:1403.6107;%%.

\bibitem{Hatsuda:2023imp}
Y.~Hatsuda and T.~Okazaki, ``{Large N and large representations of Schur line
  defect correlators},'' \href{http://dx.doi.org/10.1007/JHEP01(2024)096}{{\em
  JHEP} {\bfseries 01} (2024) 096},
  \href{http://arxiv.org/abs/2309.11712}{{\ttfamily arXiv:2309.11712
  [hep-th]}}.

\bibitem{MR1110850}
R.~P. Stanley,
  \href{http://dx.doi.org/10.1111/j.1749-6632.1989.tb16434.x}{``Log-concave and
  unimodal sequences in algebra, combinatorics, and geometry,''} in {\em Graph
  theory and its applications: {E}ast and {W}est ({J}inan, 1986)}, vol.~576 of
  {\em Ann. New York Acad. Sci.}, pp.~500--535.
\newblock New York Acad. Sci., New York, 1989.
\newblock \url{https://doi.org/10.1111/j.1749-6632.1989.tb16434.x}.

\bibitem{MR325407}
R.~P. Stanley, ``Theory and application of plane partitions. {I}, {II},''
  \href{http://dx.doi.org/10.1002/sapm1971503259}{{\em Studies in Appl. Math.}
  {\bfseries 50} (1971) 167--188; ibid. 50 (1971), 259--279}.
  \url{https://doi.org/10.1002/sapm1971503259}.

\end{thebibliography}\endgroup

\end{document}